\documentclass{elsarticle}

\usepackage{graphicx}% Include figure files
\usepackage{dcolumn}% Align table columns on decimal point
\usepackage{bm}% bold math
\usepackage{float}
\usepackage{slashbox}
\usepackage{amsmath,amssymb,amsfonts}
\usepackage{todonotes}
\def\Xint#1{\mathchoice
	{\XXint\displaystyle\textstyle{#1}}%
	{\XXint\textstyle\scriptstyle{#1}}%
	{\XXint\scriptstyle\scriptscriptstyle{#1}}%
	{\XXint\scriptscriptstyle\scriptscriptstyle{#1}}%
	\!\int}
\def\XXint#1#2#3{{\setbox0=\hbox{$#1{#2#3}{\int}$}
		\vcenter{\hbox{$#2#3$}}\kern-.5\wd0}}

\usepackage{hyperref}
\usepackage{color}  
\definecolor{darkGreen}{rgb}{0,0.45,0}
\definecolor{darkBlue}{rgb}{0,0,0.7}
\definecolor{darkRed}{rgb}{0.76, 0.13, 0.28}
\hypersetup{
	colorlinks=true, 
	linktoc=all,    
	linkcolor=darkBlue, 
	citecolor=darkGreen,
	urlcolor = darkRed,
}
\usepackage{comment}

\renewcommand{\d}{{\mathrm{d}}}

\renewcommand{\Re}{\mathrm{R}}
\renewcommand{\Im}{\mathrm{I}}

\usepackage[margin=2.25cm]{geometry}

\newcommand\Tstrut{\rule{0pt}{4.5ex}}         % = `top' strut
\newcommand\Tstrutr{\rule{0pt}{3ex}}         % = `top' strut after h line of regular row
\newcommand\Bstrut{\rule[-3ex]{0pt}{0pt}}   % = `bottom' strut
\newcommand\Bstrutr{\rule[-2ex]{0pt}{0pt}}   % = `bottom' strut after h line of regular row

\begin{document}

%\title{Dynamics of spring- and hinge-mounted membranes in inviscid flow}
\title{Dynamics of tethered membranes in inviscid flow}

\author[1]{Christiana Mavroyiakoumou}
\ead{chrismav@umich.edu}

\author[1]{Silas Alben}
\ead{alben@umich.edu}

% \corref{cor1}\cortext[cor1]{Corresponding authors}

\address[1]{Department of Mathematics, University of Michigan, Ann Arbor, MI 48109, USA}

\date{\today}

\begin{abstract}
We investigate the dynamics of membranes that are held by freely-rotating tethers in fluid flows. The tethered boundary condition allows periodic and chaotic oscillatory motions for certain parameter values. We characterize the oscillations in terms of deflection amplitudes, dominant periods, and numbers of deflection extrema along the membranes across the parameter space of membrane mass density, stretching modulus, pretension, and tether length. We determine the region of instability and the small-amplitude behavior by solving a nonlinear eigenvalue problem. We also consider an infinite periodic membrane model, which yields a regular eigenvalue problem, analytical results, and asymptotic scaling laws. We find qualitative similarities among all three models in terms of the oscillation frequencies and membrane shapes at small and large values of membrane mass, pretension, and tether length/stiffness. 

%effect of changing the  conditions at the leading and trailing edges of a thin extensible membrane in the small-amplitude growth regime and the resulting large-amplitude steady state.  We develop an infinite, periodic membrane model with a periodic array of Hookean springs attached at the bottom surface of the membrane to describe these phenomena by exploiting the fast computations of the associated eigenvalues of the most unstable modes. Scaling laws
%for the real and imaginary parts of the eigenvalues and the dominant wave numbers are estimated in the limits of small and large membrane mass and pretension parameters. We use numerical simulations and scaling arguments, and we compare our predictions with
%previous work.
\end{abstract}
\begin{keyword}
Extensible membranes; Flow–structure interactions; Nonlinear instability
\end{keyword}

\maketitle

%%%%%%%%
\section{Introduction\label{sec:intro}}

There have been many studies of fluid-structure interactions induced by thin flexible bodies. Most of these studies concern flexible beams that are nearly inextensible~\citep{Taneda_JPhysSocJpn_1968,kornecki1976ait,zhang2000flexible,zhu2002simulation,watanabe2002experimental,shelley2005heavy,argentina2005fluid,eloy2007flutter,eloy2008aeroelastic,alben2008flapping,alben2008ffi,michelin2008vortex,shelley2011flapping}.
Another important
case that has received somewhat less attention is
extensible {\it membranes} of
zero bending modulus.
Membranes arise in various biological and technological applications including 
membrane aircraft and shape-morphing airfoils~\citep{lian2005numerical,hu2008flexible,stanford2008fixed,jaworski2012high,piquee2018aerodynamic,schomberg2018transition,tzezana2019thrust}, sails~\citep{colgate1996fundamentals,kimball2009physics}, parachutes~\citep{pepper1971aerodynamic,stein2000parachute}, membrane roofs~\citep{haruo1975flutter,knudson1991recent,sygulski1996dynamic,sygulski1997numerical,sygulski2007stability}, and the wings of flying animals such as bats~\citep{swartz1996mechanical,song2008aeromechanics,cheney2015wrinkle}.
The majority of previous studies of membranes showed that when they are held with their ends fixed in a uniform oncoming fluid flow, they tend to adopt steady shapes with a single hump (that is, when the flat state is unstable) \citep{song2008aeromechanics,mavroyiakoumou2020large}. In the current work, we show that periodic and chaotic oscillations can occur in a simple physical setup. In our investigation we consider a passive case, i.e., we do not impose heaving or pitching motions~\citep{jaworski2012thrust,tregidgo2013unsteady,gordnier2014impact,tzezana2019thrust}. We also do not have any forcing of oscillations from leading-edge vortex shedding (vortex induced vibrations), which can be important in membranes that are driven by heaving and pitching motions or held at nonzero angle of attack~\citep{rojratsirikul2011flow,jaworski2012high,sun2018bifurcations}. 

In~\citet{mavroyiakoumou2020large} we investigated how the membrane dynamics change when using different boundary conditions at the two ends of the membrane. In the first case, fixed-fixed, the membrane ends were held fixed, as in most previous studies of membrane flutter~\citep{le1999unsteady,sygulski2007stability,tiomkin2017stability,nardini2018reduced}. In the second case, fixed-free, we allowed the trailing edge of the membrane to move, but only in the direction perpendicular to the oncoming flow. This gives the free-end boundary condition for a string or membrane in classical mechanics~\citep{graff1975wave,farlow1993partial} and corresponds to a membrane end that can slide (without friction) perpendicularly to the membrane's flat equilibrium state. 
%Without friction, the force from the pole on the ring at the membrane end is horizontal. The tension force from the membrane on the ring must also be horizontal, or else the ring would have an infinite vertical acceleration. Therefore, $\mathbf{\hat{s}}=\mathbf{\hat{e}}_x$ at the trailing edge or, equivalently, $\partial_{\alpha}y(1,t)=0$.

Although well known in classical mechanics, free-end boundary conditions have not been studied much in membrane (as opposed to beam/plate) flutter problems. In~\citet{hu2008flexible}, the authors study membrane wings with partially free trailing edges and find that trailing edge fluttering may occur at relatively low angles of attack. Another recent experimental study found that membrane wing flutter can be enhanced by the vibrations of flexible leading and trailing edge supports~\citep{arbos2013leading}. Partially free edges occur also in sails: the shape of a sail membrane can be controlled by altering the tension in cables running along its free edges~\citep{kimball2009physics}. Flutter can occur when the tension in these edges is sufficiently low~\citep{colgate1996fundamentals}. A related application is to energy harvesting by membranes mounted on tensegrity structures (networks of rigid rods and elastic fibers) and placed in fluid flows~\citep{sunny2014optimal,yang2016modeling}. In such cases the membrane ends have some degrees of freedom akin to the free-end boundary conditions defined above.

Related work has studied the dynamics and flutter of membranes and cables under gravity with free ends~\citep{triantafyllou1994dynamic,manela2017hanging}. Here we neglect gravity to focus specifically on the basic flutter problem~\citep{shelley2011flapping}. Without gravity, some restriction on the motion of the free membrane ends is needed to avoid ill-posedness due to membrane compression~\citep{triantafyllou1994dynamic}. Such a restriction was realized experimentally by~\citet{kashy1997transverse}, with the membrane represented by an extensional spring that is tethered by steel wires to vertical supports. The membrane is thus free to move perpendicularly to its flat rest state, but remains stretched between the supports, allowing for stable dynamics. The current paper
uses this tethered boundary condition to study membrane dynamics in a fluid flow. We study both small- and large-amplitude dynamics when the membrane is attached to tethers---i.e., inextensible rods that rotate freely---or mounted on springs.

We will show that as the tether length is increased, the membrane dynamics change from static deflections with a single maximum, typical of the fixed-fixed case (similar to the shapes in~\citet{newman1984two,rojratsirikul2010effect,waldman2013shape,waldman2017camber,nardini2018reduced,tzezana2019thrust} to a wide range of oscillatory motions that have some commonalities with flapping plates and flags~\citep{shelley2011flapping}.
%Due to the many applications of fluid flow over flexible membranes we need a simplified model that can be analyzed faster than the existing models. 
We study the stability properties of tethered membranes via a nonlinear eigenvalue problem. The nonlinearity makes it difficult to solve in certain regions of parameter space. Therefore we consider an approximate problem---an infinite membrane mounted on a periodic array of Hookean springs---that is easier to solve and allows us to obtain asymptotic scaling laws for the eigenmodes' dependences on membrane pretension and mass density.

%%%%% table
\begin{table}[H]
\caption{Typical experimental values of parameter ranges relevant to our current model as used in previous membrane studies. Computational ($^c$), experimental ($^e$), or theoretical ($^t$) ranges of the dimensionless body mass density $R_1$, stretching modulus $R_3$,  and pretension~$T_0$.}\label{table:summary}
\centering
\vspace{.1cm}
\begin{tabular}{l|c|c|c|c}
\hline\hline
Reference & Material & $R_1=\displaystyle\frac{\rho_s h}{\rho_f L}$  & $R_3 = \displaystyle\frac{Eh}{\rho_f U^2L}$  & $T_0=\displaystyle\frac{\overline{T}}{\rho_f U^2L W}$\Tstrut\Bstrut\\
\hline
\citet{newman1991stability}$^t$& sail & 0--6  & --- & 0--2\Tstrutr\\
\citet{le1999unsteady}$^e$ & sail   & 0--0.8&  $10^1$,\,50,\,
$10^2$,\,500,\,$10^3$ & --- \\
\citet{sygulski2007stability}$^{e\,\&\,t}$&latex rubber &   0.1,\,1 &---  & 130.6,\,217 \\
\citet{jaworski2012high}$^{c \,\& \,e}$&  latex rubber & 2.4 &  100,\,200,\,400,\,614 & 4,\,10,\,20,\,30.7  \\
\citet{tiomkin2017stability}$^{c}$&---& 0--80&  --- & 0--6 \\
\citet{nardini2018reduced}$^{c}$& --- & 0--60 &---& 0--3\\
\citet{das2020nonlinear}$^e$ &silicone rubber
& 2.5--31.25  & $3.75\times 10^{-5}$--$0.04$ & 1--4\\
\citet{das2020deformation}$^e$&silicone rubber
& 2.5--31.25  & $3.75\times 10^{-5}$--$0.04$ & 1--4\\
\citet{mavroyiakoumou2020large}$^{c \,\& \, t}$ &--- 
& $10^{-3}$--$10^2$& $10^0$--$10^4$ & $10^{-3}$--$10^3$\\
\citet{mavroyiakoumou2021eigenmode}$^{c \,\& \, t}$ &--- 
& $10^{-3}$--$10^3$ & --- & $10^{-1.5}$--$10^2$\\
Current study$^{c \,\& \, t}$ &---
& $10^{-4}$--$10^4$ & $10^{0.5}$--$10^4$ & \,\,$10^{-3}$--$10^2$
 \Bstrutr\\
\hline\hline
\end{tabular}
\end{table}
% \christiana{Should I include the smallest $R_1$ and $T_0$ values used for the $k_s=0$ case in the infinite, periodic model?}
% \christiana{Add all the references in table from JFM paper and for material say either generic/non-specific/hyphen.}
% \christiana{Add our previous papers and current work.}

We show in Table~\ref{table:summary} typical ranges of membrane parameters---mass density ratios ($R_1$),  stretching rigidities ($R_3$), and pretensions ($T_0$)---from previous experimental, theoretical, and computational studies.
\citet{newman1991stability} used an infinite periodic membrane model with a low-mode approximation and found that stability is lost through divergence. \citet{le1999unsteady} used a vortex sheet model to study a more complex situation---the motions of a sail membrane under harmonic perturbations of the trailing edge and with randomly perturbed inflow velocities. \citet{sygulski2007stability} studied the membrane flutter threshold and divergence modes theoretically, with some experimental validation. Although most works omit specific values of the aspect ratio $h/L$ and the bending modulus $R_2 = Eh^3/(12\rho_f U^2L^3)=R_3(h/L)^2/12$, an example is given in~\citet{jaworski2012high} for a latex rubber, where the aspect ratio is $h/L=1/750$ and the bending modulus is therefore about a factor of $10^{-7}$ smaller
than $R_3$.
%$R_2=R_3/(6.75\times 10^{-6})=\{1.48,2.96,5.92,9.1\}\times 10^{-5}$ which is
%negligible compared to the stretching modulus $R_3=\{100,200,500,614\}$. 
In \citet{jaworski2012high}, they studied a heaving and pitching membrane airfoil in a fluid stream numerically at Reynolds number 2500, and found elastic modulus and prestress parameters that led to enhanced thrust and propulsive efficiency. \citet{tiomkin2017stability} presented a more detailed flutter threshold calculation using an
inviscid, small amplitude vortex sheet model. \citet{nardini2018reduced} compared a reduced-order model with direct numerical simulations to study the effect
of Reynolds number on the flutter stability threshold and small-amplitude membrane
deflection modes. \citet{das2020nonlinear} modeled the material properties of ultrasoft dielectric elastomers
over a wide range of elastic properties, prestretch, and thicknesses. They
measured the mechanical response of the silicone membranes and found that stiffer membranes harden at lower stretch ratios due to the increased fraction of polymer chains in them. \citet{das2020deformation} studied the deformations, forces, and flow fields associated with a highly compliant membrane disk
placed head-on in a uniform flow field. With increasing
flow velocity, the membrane deforms hyper-elastically into parachute-like shapes.  A resulting drag increase
correlates with the unsteady fluid-structure interactions between the membrane and the flow.

In the present study, we set $R_2$ to zero and study the dynamics of tethered membranes over wide ranges of the remaining parameters---$R_1$, $R_3$, and $T_0$---as well as the tether length or stiffness. We will show that the large-amplitude dynamics depend most strongly on the membrane mass density $(R_1)$ and less on the pretension $(T_0)$. The oscillation frequencies are the smallest at the largest $R_1$ and motions there are somewhat chaotic. Decreasing $R_1$ to about $10^{-0.75}$ makes the motions more periodic and symmetric. Further decreasing $R_1$ introduces very fine spatial features. The stretching modulus ($R_3$) mainly determines the overall magnitude of the membrane's deflection. 
These unsteady dynamics are possible because, unlike in previous studies, the membrane is attached to inextensible rod tethers whose lengths
set the 
transition between steady and unsteady motions.
%When the tether length is small then the membranes exhibit large---but physically reasonable---deflections that converge to a steady, fore-aft symmetric shape with a single maximum.

The paper is structured as follows. We begin in Sec.~\ref{sec:large} by presenting the membrane and vortex-sheet model and in Sec.~\ref{sec:rods} we present the boundary conditions when the  membranes are attached to inextensible-rod tethers. In Sec.~\ref{sec:largeAmplitudeResults} we present the results in the large-amplitude regime for this boundary condition. In Secs.~\ref{sec:springs} and~\ref{sec:verticalSprings} we study the related case of membranes mounted on Hookean springs. In
Sec.~\ref{sec:linearized} we present a linearized, small-amplitude version of our model and study the stability properties (Sec.~\ref{sec:eigenmodes}). We then study the stability behavior of an infinite periodic membrane mounted on a periodic array of springs and propose asymptotic scaling laws (Sec.~\ref{sec:array}). Finally, in Sec.~\ref{sec:conclusions}, we summarize our findings.

%%%%%%%%%%%%%%%%%%%%%%%%%%%%%%%%%%%%%%%%%%%%%%

\section{Membrane and vortex-sheet model}\label{sec:large}

\begin{figure}[H]
    \centering
    \includegraphics[width=.75\textwidth]{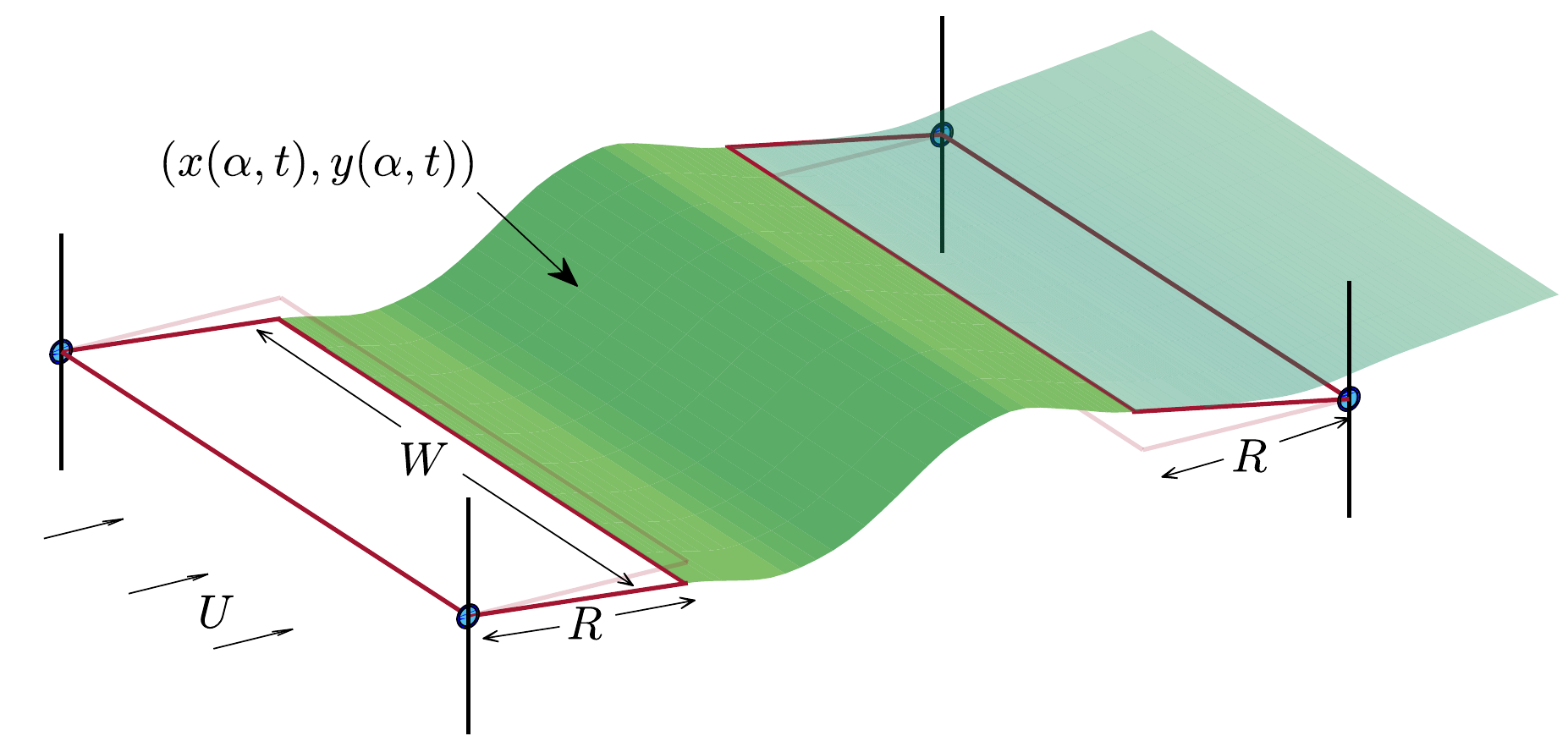}
    \caption{Schematic diagram of a flexible membrane (dark green surface) at an instant in time. $U$ is the oncoming flow velocity and~$W$ is the membrane's spanwise width. The leading edge of the membrane is attached to  inextensible rods (red rectangular frames) that rotate freely about their hinged ends (small black/blue circles). There is also a vortex wake (light green surface) emanating from the membrane's trailing edge.}\label{fig:schematicRods3d}
\end{figure}

We model the dynamics of an extensible membrane that is nearly aligned with a two-dimensional background fluid flow that has speed $U$ in the far field (see Fig. \ref{fig:schematicRods3d}). The membrane is shown as a dark green surface with the vortex wake (light green surface) emanating from its trailing edge.  
Each membrane end is attached to a rigid frame of inextensible rods (red solid lines) that pivots freely at the hinges shown by small black/blue circles in Fig.~\ref{fig:schematicRods3d} and therefore the membrane's ends are constrained to move along circles of radius $R$ centered at the hinges.
The motions of the membrane and rod frames are assumed to be invariant in the spanwise direction (along $W$), and the effect of gravity is neglected for simplicity.
The four clamping poles (black lines) at the end of the rod frame away from the membrane are sufficiently thin that their influence on the fluid flow is assumed to be negligible. 
%We assume that the membrane has a certain thickness that is sufficient to prevent it from sagging under gravity.

The membrane and flow models are the same as in~\citet{mavroyiakoumou2020large} but we repeat them briefly for completeness. 
The membrane dynamics are described by the unsteady extensible elastica equation with body inertia, stretching resistance, and fluid pressure loading, obtained by writing a force balance equation for a small section of membrane that lies between material coordinates $\alpha$ and $\alpha+\Delta \alpha$:
\begin{equation}\label{eq:dimensionalMembrane}
    \rho_s hW \partial_{tt}\zeta(\alpha,t)\Delta \alpha=T(\alpha+\Delta\alpha)\mathbf{\hat{s}}-T(\alpha,t)\mathbf{\hat{s}}-[p]_-^+(\alpha,t)\mathbf{\hat{n}}W(s(\alpha+\Delta\alpha,t)-s(\alpha,t)).
\end{equation}
Here $\rho_s$ is the mass per unit volume of the undeflected membrane, $h$ is the membrane's thickness, and~$W$ its spanwise width, all uniform along the length. In Eq.~\eqref{eq:dimensionalMembrane},
$\zeta(\alpha,t) = x(\alpha,t)+iy(\alpha,t)$ denotes the membrane position in the complex plane, parameterized by the material coordinate $\alpha$, $-L\leq \alpha\leq L$ ($L$ is half the initial length) and time $t$. $T$ is the tension in the membrane, $[p]_-^+$ is the pressure jump across it, $s(\alpha,t)$ is the local arc length coordinate, and the unit vectors tangent and normal to the membrane are $\mathbf{\hat{s}}=\partial_\alpha\zeta(\alpha,t)/\partial_\alpha s(\alpha,t)=e^{i\theta(\alpha,t)}$ and $\mathbf{\hat{n}}=i\mathbf{\hat{s}}=ie^{i\theta(\alpha,t)}$, respectively, with $\theta(\alpha,t)$ the local tangent angle and $\partial_\alpha s$ the local stretching factor. We use $+$ to denote the side towards which the membrane normal~$\mathbf{\hat{n}}$ is directed, and~$-$ for the other side. However, for the remainder of this paper, we drop the $+$ and $-$ for ease of notation.

Dividing Eq.~\eqref{eq:dimensionalMembrane} by $\Delta\alpha$ and taking the limit $\Delta\alpha\to 0$, we obtain
\begin{equation}\label{eq:dimensionalMembraneEq}
    \rho_shW\partial_{tt}\zeta(\alpha,t) = \partial_\alpha(T(\alpha,t)\mathbf{\hat{s}})-[p](\alpha,t)W\partial_\alpha s\mathbf{\hat{n}},
\end{equation}
where the membrane tension $T(\alpha,t)$ is given by linear elasticity~\citep{carrier1945non,narasimha1968non,nayfeh2008linear} as
\begin{equation}
    T(\alpha,t)=\overline{T}+EhW(\partial_{\alpha} s(\alpha,t)-1).
\end{equation}
Here $E$ is the Young's modulus and $\overline{T}$ is the tension in the initial, undeflected equilibrium state. After nondimensionalizing length by $L$, time by $L/U$, and pressure by $\rho_f U^2$, where $\rho_f$ is the density of the fluid and $U$ is the oncoming flow velocity, Eq.~\eqref{eq:dimensionalMembraneEq} becomes the nonlinear, extensible membrane equation
\begin{equation}\label{eq:membrane}
R_1\partial_{tt}\zeta -\partial_\alpha ((T_0+R_3(\partial_\alpha s-1))\mathbf{\hat{s}})=-[p]\partial_\alpha s\mathbf{\hat{n}}.
\end{equation}
In Eq.~\eqref{eq:membrane},  $R_1=\rho_sh/(\rho_f L)$ is the dimensionless membrane mass, $T_0=\overline{T}/(\rho_f U^2L W)$ is the dimensionless pretension, and finally, $R_3=Eh/(\rho_f U^2 L)$ is the dimensionless stretching rigidity.  We use Eq.~\eqref{eq:membrane} to study large-amplitude motions in Sec.~\ref{sec:largeAmplitudeResults}. We use a linearized, small-amplitude version to study membrane stability in Secs.~\ref{sec:linearized} and~\ref{sec:array}. 

We express the 2D flow past the membrane using $z=x+iy$, the complex representation of the $xy$ flow plane. The complex conjugate of the fluid velocity at any point $z$ not on the vortex sheets is a sum of the horizontal background flow with speed unity and the flow induced by the bound and free vortex sheets,
\begin{equation}
u_x(z)-iu_y(z)= 1 +\frac{1}{2\pi i}\int_{-1}^1\frac{\gamma(\alpha,t)}{z-\zeta(\alpha,t)}\partial_\alpha s \d\alpha+\frac{1}{2\pi i}\int_0^{s_{\max}} \frac{\gamma(s,t)}{z-\zeta(s,t)}\d s,
\end{equation}
where $s$ is the arc length along the free sheet starting at 0 at the membrane's trailing edge and extending to~$s_{\max}$ at the free sheet's far end. To determine the bound vortex sheet strength $\gamma$ we require that the fluid does not penetrate the membrane, which is known as the kinematic boundary condition. Here $\gamma$ represents the jump in the component of the flow velocity tangent to the membrane from the $-$ to the $+$ side, i.e.,\ $\gamma=-[(u_x,u_y)\cdot\mathbf{\hat{s}}]$. The normal components of the fluid and membrane velocities are equal along the membrane:
\begin{equation}\label{eq:kinematic}
    \mathrm{Re}(\mathbf{\hat{n}}\partial_t\overline{\zeta}(\alpha,t)) =\mathrm{Re}\left( \mathbf{\hat{n}}\left( 1 +\frac{1}{2\pi i}\int_{-1}^1\frac{\gamma(\alpha,t)}{z-\zeta(\alpha,t)} \partial_\alpha s\d\alpha+\frac{1}{2\pi i}\int_0^{s_{\max}} \frac{\gamma(s,t)}{z-\zeta(s,t)}\d s \right)\right),
\end{equation}
where $\mathbf{\hat{n}}$ is written as a complex scalar. Solving Eq.~\eqref{eq:kinematic} for $\gamma$ requires an additional constraint that the total circulation is zero for a flow started from rest. At each instant the part of the circulation in the free sheet, or alternatively, the strength of $\gamma$ where the free sheet meets the trailing edge of the membrane, is set by the Kutta condition which makes velocity finite at the trailing edge. At every other point on the free sheet, $\gamma$ is set by the criterion that circulation (the integral of $\gamma$) is conserved at fluid material points of the free sheet.
The vortex sheet strength $\gamma(\alpha,t)$ is coupled to the pressure jump $[p](\alpha,t)$ across the membrane using a version of the unsteady Bernoulli equation written at a fixed material point on the membrane:
\begin{equation}\label{eq:pressure}
    \partial_\alpha s\partial_t\gamma +\partial_\alpha\left(\gamma(\mu-\tau)\right)+\gamma(\partial_\alpha \tau-\nu\kappa\partial_\alpha s)=\partial_\alpha [p],
\end{equation}
% \begin{equation}\label{eq:pressure}
%     \partial_\alpha s\partial_t\gamma +(\mu-\tau)\partial_\alpha\gamma+\gamma(\partial_\alpha \mu-\partial_\alpha s\nu\kappa)=\partial_\alpha [p]_-^+,
% \end{equation}
where $\mu$ is the average flow velocity tangent to the membrane, $\tau$ and $\nu$ are the tangential and normal components of the membrane velocity, respectively, and $\kappa(\alpha,t)=\partial_\alpha\theta/\partial_\alpha s$ is the membrane's curvature. At the trailing edge, $[p]|_{\alpha=1}=0$. The derivation of Eq.~\eqref{eq:pressure} can be found in~\citet[Appendix A]{mavroyiakoumou2020large}.

%%%%%%%%%%%%%%%%%%%%%%%%%%%%%%%%%%%%%%%%%%%
%\silas{Boundary conditions should be a subsection of the previous Model section}
\subsection{Boundary conditions: inextensible-rod tethers \label{sec:rods}}

 \begin{figure}[H]
     \centering
     \includegraphics[width=.6\textwidth]{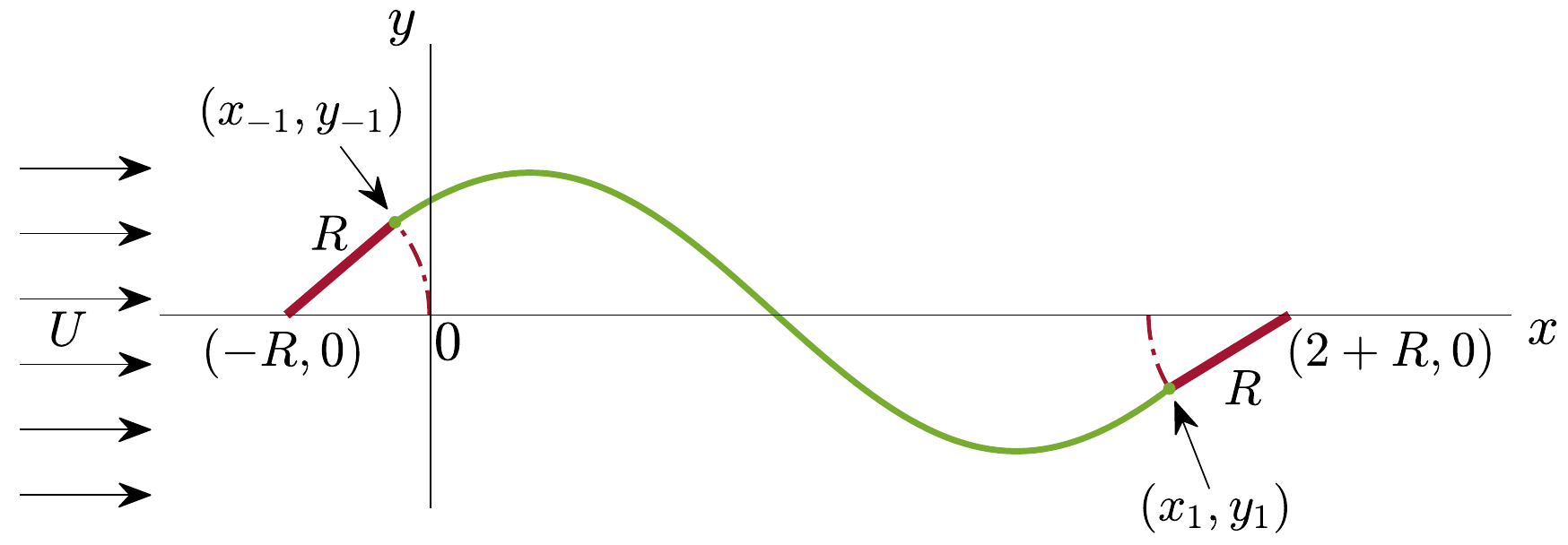}
 \caption{Slice through the membrane in Fig.~\ref{fig:schematicRods3d}. Schematic diagram of a flexible membrane (green line) at an instant in time. The leading edge of the membrane with position $(x(-1,t),y(-1,t))$ is attached to an inextensible rod frame (red line) whose motion is restricted to a circle of radius $R$ (length of rod frame) and whose other end is fixed at $(-R,0)$ for all time. The membrane's trailing edge with position $(x(1,t),y(1,t))$ is attached to another rod frame whose other end is fixed at $(2+R,0)$ for all time.}\label{fig:schematicRods}
\end{figure}
% could have used schematicRods with shifted y-axis
% \christiana{Include a 3D schematic diagram, with a rectangular frame on each side to avoid twisting effects.}
A slice through the membrane and rod frame in the 2D flow plane is shown schematically in Fig.~\ref{fig:schematicRods}. The rod frames pivot freely about the points $(-R,0)$ and $(2+R, 0)$, respectively. 
Because the frames are inextensible, the membrane ends are constrained to move along circular arcs of radius $R$. This is enforced by requiring
\begin{equation} \label{circle}
(x_{-1}-(-R))^2+y_{-1}^2=R^2\quad\text{and}\quad (x_{1}-(2+R))^2+y_{1}^2=R^2,
\end{equation}
for all time, where $x_{\pm 1}=x(\pm 1,t)$ and $y_{\pm 1}=y(\pm 1,t)$ are four unknowns that denote the 
$x$- and $y$-coordinates of the membrane ends, respectively. Two equations for the four unknowns are (\ref{circle}) and the remaining two equations
require the membrane and rod frames to be tangent where they meet:
\begin{equation}\label{eq:slopeRods}
\displaystyle\left.\frac{\partial_\alpha y}{\partial_\alpha x}\right|_{\alpha=-1}=\frac{y_{-1}-0}{x_{-1}-(-R)}\quad\text{and}\quad \displaystyle\left.\frac{\partial_\alpha y}{\partial_\alpha x}\right|_{\alpha=1}=\frac{0-y_{1}}{(2+R)-x_{1}},
\end{equation}
again for all time. Eqs.~\eqref{eq:slopeRods} follow from balancing the forces on an infinitessimal length of membrane near the membrane ends; because its mass is infinitessimal, the tension forces on it from the rods and from the adjacent portion of the membrane must be aligned. The rod tether length $R$ is an important parameter that influences the dynamics of the membrane. With short rods ($R\to 0$), we will show that the membrane dynamics are similar to fixed-fixed membranes, whereas with longer rods the dynamics resemble free-free membranes but without the large-scale translational motions seen
in~\citet{mavroyiakoumou2020large}.

\section{Large-amplitude results}\label{sec:largeAmplitudeResults}

We simulate the membrane starting from an initial condition in which the membrane is perturbed from the flat horizontal equilibrium state: it has a linear profile with a small nonzero
slope,
\begin{equation}\label{eq:nonzeroSlope}
    \zeta(\alpha,0) = (\alpha+1)(1+i\sigma),
\end{equation}
for $\sigma = 10^{-3}$.
We evolve the membrane and vortex sheet wake forward in time using a numerical method similar to those in~\citet{alben2009simulating,mavroyiakoumou2020large}.

\begin{figure}[H]
    \centering
    \includegraphics[width=\textwidth]{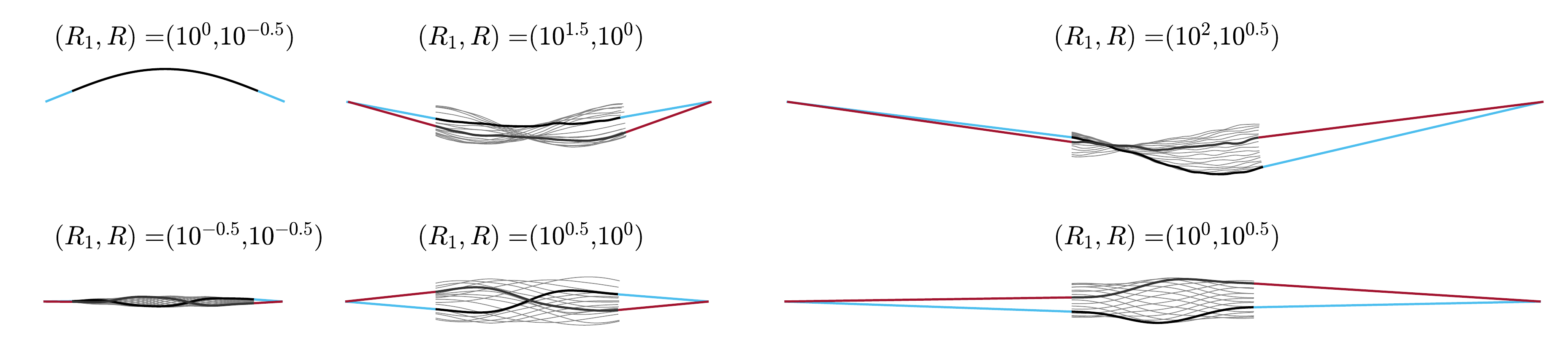}
    \caption{Examples of membrane (black) and rod (red and blue) snapshots at two different times, superposed on a larger set of membrane snapshots (gray) within a period. Each  column corresponds to a rod length $R$: $10^{-0.5}$ (left column), $10^0$ (middle column), and $10^{0.5}$ (right column). Here $R_3=10^{1.5}$ and $T_0=10^{-2}$.}
    \label{fig:rodscols}
\end{figure}

In Fig.~\ref{fig:rodscols} we show snapshots of membranes and rods for a fixed stretching rigidity ($R_3=10^{1.5}$) and pretension ($T_0=10^{-2}$), at six pairs of $(R_1,R)$ values that give typical dynamics. In each case, two of the snapshots show the rods (blue in one and red in the other) together with the membranes (black lines). The remaining 16 snapshots show only the membranes (gray lines), equally spaced in time within a period of motion. $R$ increases from left to right: $10^{-0.5}$ (left column), $10^0$ (middle column), and $10^{0.5}$ (right column). The membrane deflection may be very small, particularly at small $R_1$ (bottom left case), and may be steady, particularly at small $R$ (top left case). In the bottom row, middle column case (i.e.,\ $R_1=10^{0.5}$ and $R=10^0$) and in the bottom row, right column case (i.e.,\ $R_1=10^{0}$ and $R=10^{0.5}$) it is evident that the inextensible rods may deflect upwards or downwards.

We characterize the large-amplitude dynamics using three main quantities. One is the time-averaged deflection of the membrane, defined as
\begin{equation}\label{eq:avgDefl}
    \left<y_{\text{defl}}\right> \equiv \frac{1}{t_2}\int_{t_1}^{t_1+t_2}\left(\max\limits_{-1\leq \alpha \leq 1}y(\alpha,t)-\min\limits_{-1\leq \alpha \leq 1}y(\alpha,t)\right)\,\d t.
\end{equation}
Here, as in~\citet{mavroyiakoumou2020large}, $t_1$ and $t_2$ are sufficiently large (typically 50--100) that $\left<y_{\text{defl}} \right>$ changes by less than~1\% with further increases in these values. So, $\left<y_{\text{defl}} \right>$ is the maximum membrane deflection minus the minimum membrane deflection, averaged over time.
 
The second quantity used to characterize the large-amplitude dynamics is the time period. This is computed using the peak frequency in the power spectrum computed using the fast Fourier transform (\texttt{fft} function in \textsc{Matlab}). The power spectrum is obtained from a time series of the membrane's midpoint when the membrane has reached steady-state large-amplitude dynamics. The third quantity is the time-averaged number of zero crossings along the membrane, computed using the same temporal data as the power spectrum. Apart from the number of zero crossings, we also use the time-averaged number of local extrema as a different measure of the `waviness' of the membrane shape.

\begin{figure}[H]
  \centering
    \begin{minipage}{0.44\textwidth}
  \centering
\includegraphics[width=\textwidth]{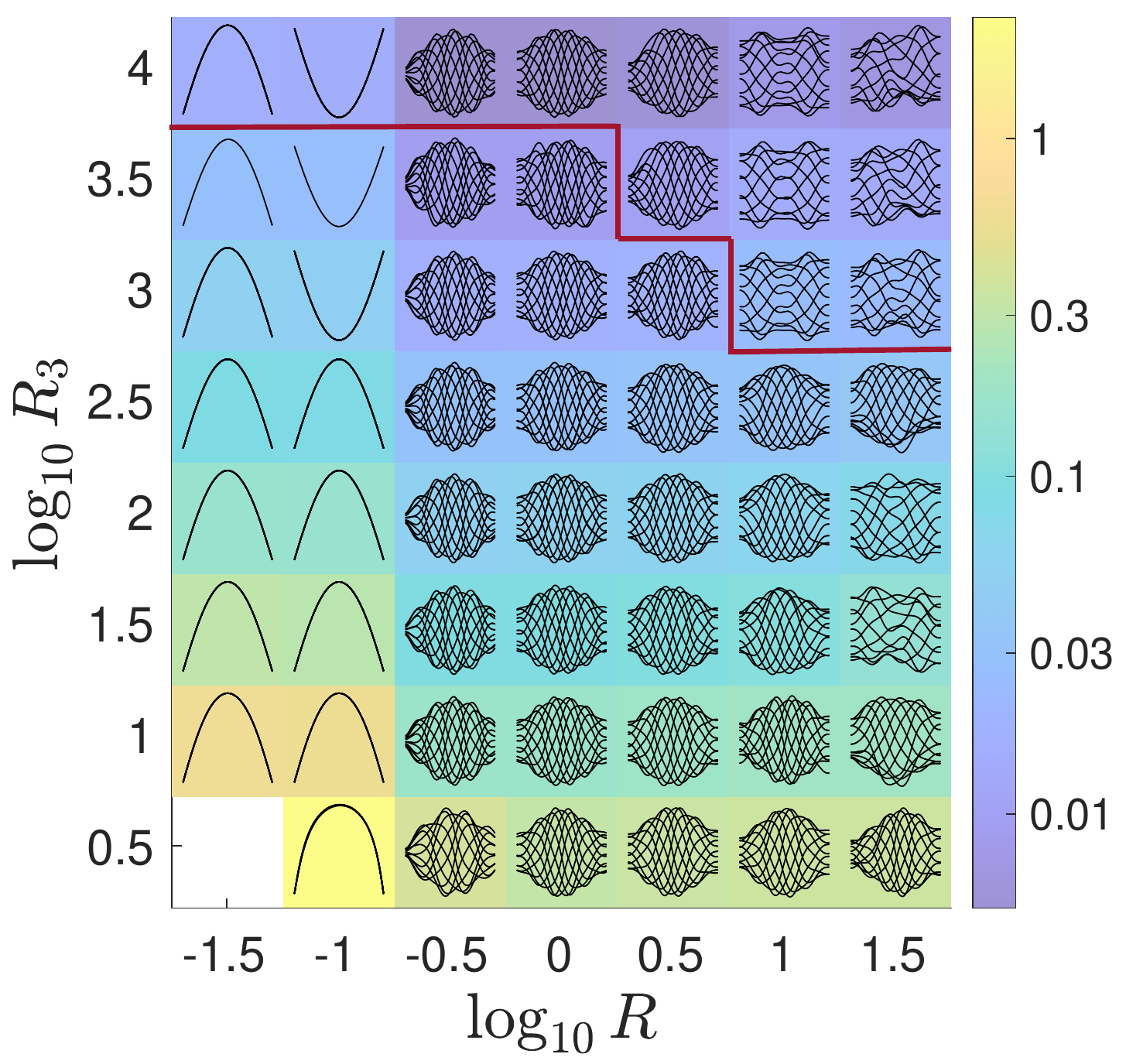}
\end{minipage}
\hspace{.08cm}
\begin{minipage}{0.5\textwidth}
    \centering
\fbox{\includegraphics[width=\textwidth]{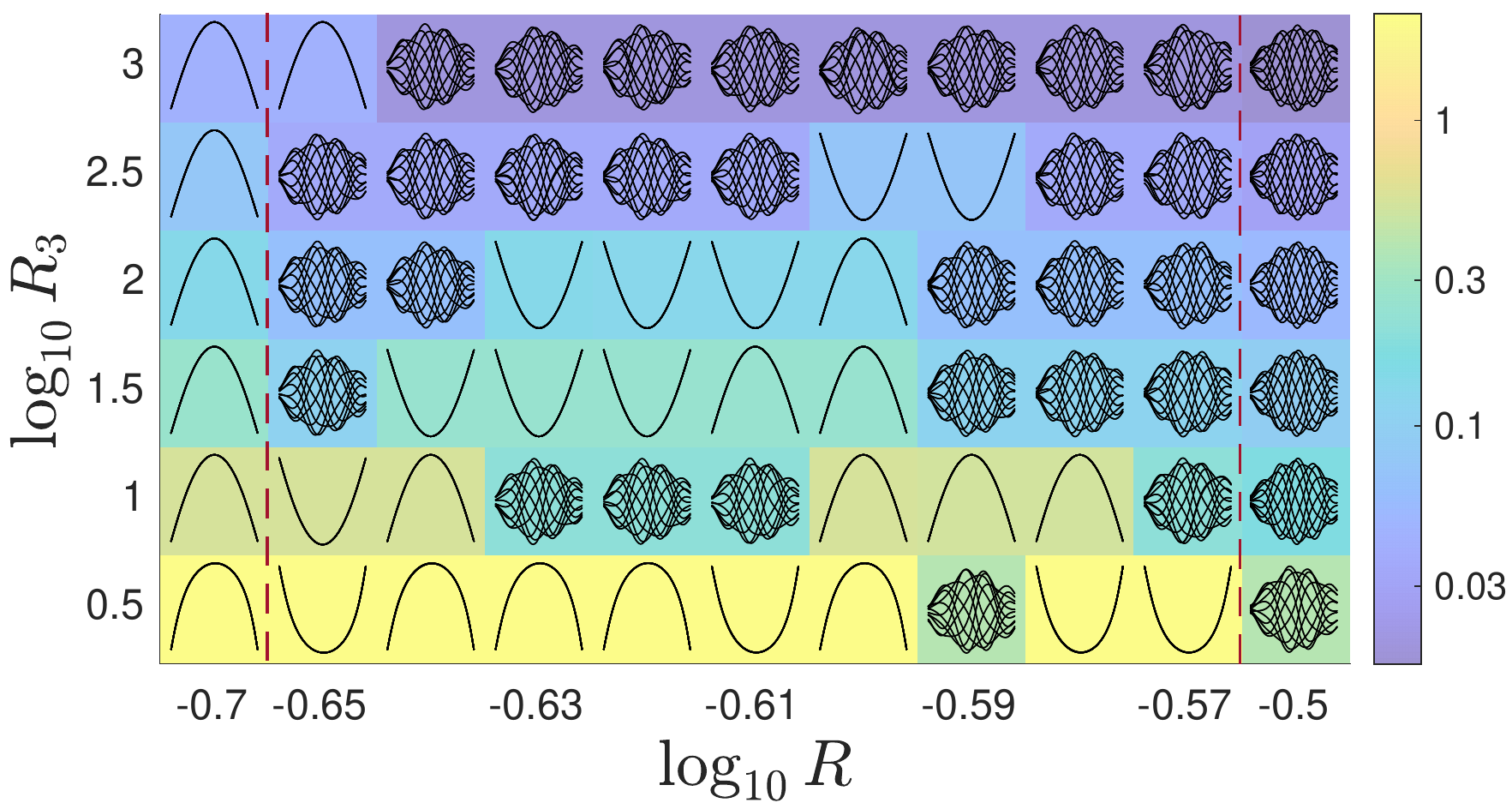}}
\end{minipage}
   \caption{(Inextensible rods) Snapshots of large-amplitude membrane motions in $R$--$R_3$ space for fixed $T_0=10^{-2}$ and $R_1=10^{-0.5}$. Colors denote the time-averaged deflection of the membranes defined by Eq.~\eqref{eq:avgDefl}. For rods with length $R\leq 10^{-1}$ the membranes behave similarly to those with fixed-fixed ends, yielding a single hump solution, whereas when $R\geq 10^{-0.5}$ the membranes oscillate as in some cases with free-free ends.
%   apart from $(R,R_3)=(10^{-0.5},10^{0.5})$ which is a single-hump solution as well. 
   At each $(R,R_3)$ value, the set of snapshots is scaled to fit within a colored rectangle centered at that value and normalized by the maximum deflection of the snapshots to show the motions more clearly. The red solid line separates membranes with $m=40$ points (above) and $m=120$ points (below). In the framed figure we look at a finer grid between $R=10^{-0.7}$ and $10^{-0.5}$, to investigate dynamics near the transition between the single-hump solution and the flapping state occurs. The red dashed lines indicate a jump in the increment of $R$ values.}
    \label{fig:deflRodsNeg05}
\end{figure}

In Fig.~\ref{fig:deflRodsNeg05} we show typical membrane snapshots in $R$--$R_3$ space, while fixing $T_0=10^{-2}$ and $R_1=10^{-0.5}$. At each $(R, R_3)$ value, the set of snapshots is normalized by
the maximum deflection of the snapshots to show the motions more clearly and
scaled to fit within a colored rectangle at the $(R, R_3)$ value. Each snapshot has the
corresponding $R$ value at its horizontal midpoint, and the~$R_3$ value at its average vertical position. 
Colors denote the time-averaged deflection defined by Eq.~\eqref{eq:avgDefl}. In the lower-left corner the snapshots are omitted because steady-state membrane motions were not obtained. Two main types of membrane behaviors are seen: at small~$R$, a steady single-hump shape that is fore-aft symmetric, similar to membranes that have both the leading and trailing edges fixed at zero deflection; at moderate-to-large $R$, an oscillatory motion.
The framed panel on the right-hand side of Fig.~\ref{fig:deflRodsNeg05} shows the transition between these states in finer detail,
between $R=10^{-0.65}$ and $10^{-0.57}$.
The red dashed lines show where larger increments of $R$ are taken, from $10^{-0.65}$ to $10^{-0.7}$ (where only single hump solutions are obtained for any~$R_3$) and from $R=10^{-0.57}$ to $10^{-0.5}$ (where only flapping membranes are observed, for any~$R_3$). In the framed panel we see that the initial condition of nonzero slope [Eq.~\eqref{eq:nonzeroSlope}] may evolve to single-hump shapes that are concave up, concave down, or to oscillatory motions when $R$ is changed slightly.
In the left panel, the oscillatory motions are mostly close to periodic and fore-aft symmetric, with some deviations particularly at $R = 10^{-0.5}$ and $10^{1.5}$, where a less wavy shape becomes more common. %and $R_3\geq 10^3$ the membranes are periodic. At larger~$R$ values, the membranes become less wavy, but are still mostly periodic.

For very large stretching rigidity $R_3\gtrsim 10^{3}$ the code reaches the steady-state regime only if we decrease the membrane discretization size to $m=40$ (from $m=120$ below the red dividing line). As we observed in~\citet{mavroyiakoumou2020large}, in many cases varying the stretching rigidity $R_3$ alters the overall deflection magnitudes but leaves the membranes' shapes nearly unchanged.

%The scaling of $\langle y_\mathrm{defl}\rangle\sim R_3^{-1/2}$ holds between shapes with a similar behavior, for instance at $R=10^{-0.6}$ and $R_3\leq 10^{2.5}$ or $R=10^{-0.57}$ and $R_3\geq 10^1$.

% The colors denote the time-averaged deflection of the membrane [Eq.~\eqref{eq:avgDefl}].

%and across several decades of $R_3$, the membranes converge to 

%From these late-time membrane shapes in this region, we see that the shapes are all steady and are, therefore, independent of $R_1$ here, since the acceleration term in Eq.~\eqref{eq:membrane} is zero.  

% \christiana{Check: Is there \textbf{bistability} here, meaning that the membrane could go to either the steady hump or the flapping state depending on the initial perturbation (i.e., it evolves a little differently at nearby parameter values).} 
% \christiana{Shall I decrease the number of snapshots here so that it's not as dark?}

\begin{figure}[H]
    \centering
    \includegraphics[width=.625\textwidth]{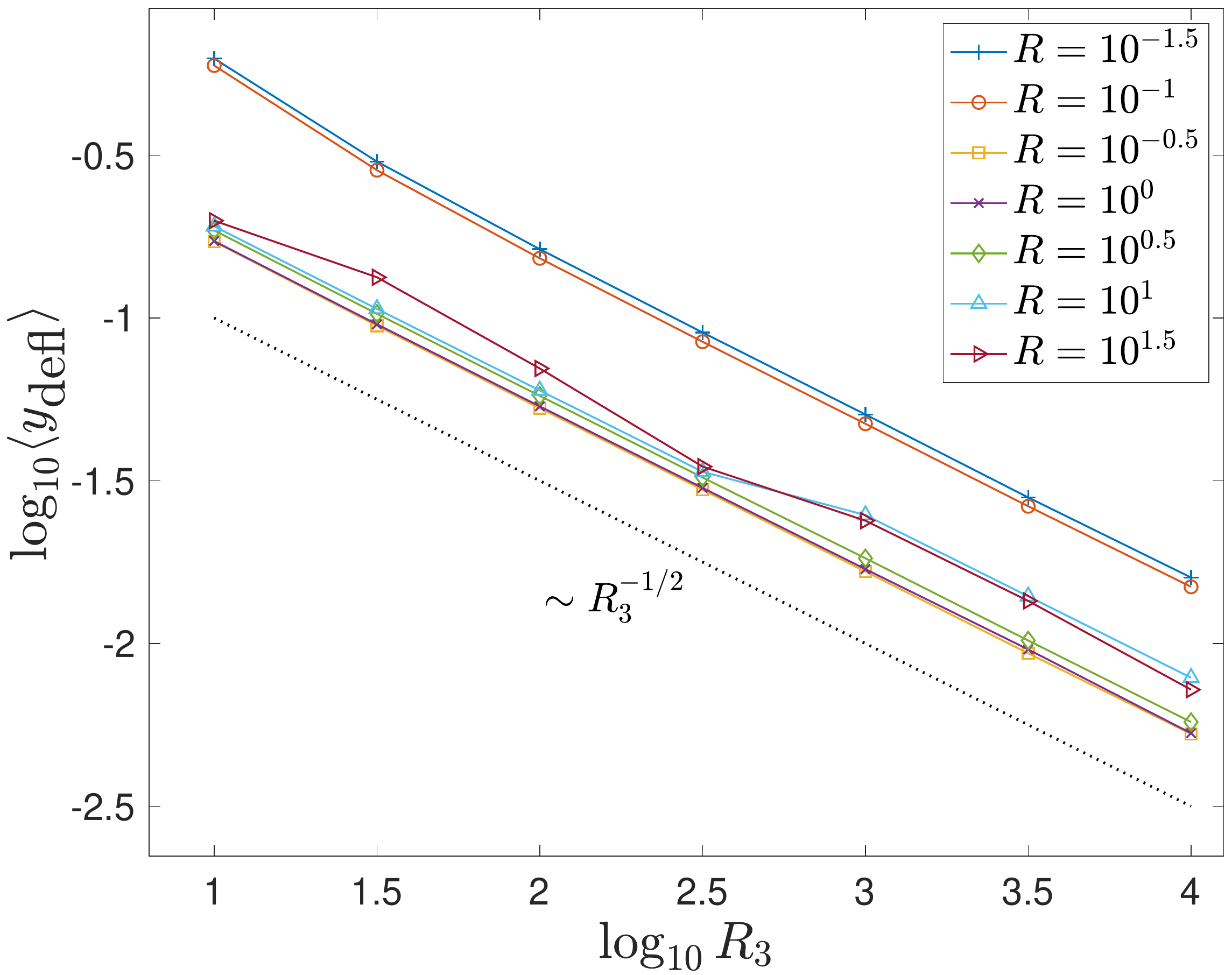}
    \caption{(Inextensible rods) Time-averaged deflections of the membranes (defined by Eq.~\eqref{eq:avgDefl}) versus $R_3$ for various $R$ and fixed $R_1=10^{0.5}$, $T_0=10^{-2}$. The dotted black line indicates the scaling $R_3^{-1/2}$.}
    \label{fig:linedeflR3}
\end{figure}

In Fig.~\ref{fig:linedeflR3} we show how the time-averaged deflection quantitatively depends on~$R_3$ at several fixed values of~$R\in\{10^{-1.5},10^{-1},\dots,10^{1.5}\}$, for $R_1=10^{0.5}$ and $T_0=10^{-2}$ here. The $\left<y_\mathrm{defl}\right>\sim R_3^{-1/2}$ dependence at large~$R_3$ is the same for other mass ratios from $R_1=10^{-0.5}$ to $R_1=10^2$, again with $T_0=10^{-2}$. This was observed also for fixed-fixed, fixed-free, and free-free membranes in~\citet{mavroyiakoumou2020large}. We include the explanation for how the scaling $\left<y_\mathrm{defl}\right>\sim R_3^{-1/2}$ arises from the $y$-component of the membrane equation [Eq.~\eqref{eq:membrane}] with small deflections. We assume that $\partial_\alpha y\ll 1$ and $\partial_\alpha x\approx 1$. Then $\partial_\alpha s-1=\sqrt{(\partial_\alpha x)^2+(\partial_\alpha y)^2}-1\approx \partial_\alpha y^2/2$ and $\hat{s}_y\approx \partial_\alpha y$. With these approximations, the $y$-components of the $T_0$ and $R_3$ terms in Eq.~\eqref{eq:membrane} are linear and cubic in deflection, respectively:
\begin{equation}
    \partial_\alpha(T_0\hat{s}_y)\approx T_0\partial_{\alpha\alpha}y;\quad \partial_\alpha(R_3(\partial_\alpha s-1)\hat{s}_y)\approx R_3\partial_\alpha((\partial_\alpha y)^3/2).
\end{equation}
The $R_1$ term that multiplies $\partial_{tt}y$ is also linear in deflection. The pressure jump is linear in the bound vortex sheet strength because the left-hand side of Eq.~\eqref{eq:pressure} $\approx \partial_t\gamma+\partial_\alpha\gamma$ with small deflections. The bound vortex sheet strength is linear in the deflection by the linearized version of Eq.~\eqref{eq:kinematic},
\begin{equation}
    \partial_t y(\alpha,t)\approx \frac{1}{2\pi}\Xint-_{-1}^1 \frac{\gamma(\alpha',t)}{x(\alpha,t)-x(\alpha',t)}\,\d \alpha'-\frac{1}{2\pi}\int_{0}^{\Gamma_+(t)}\frac{x(\alpha,t)-x(\Gamma',t)}{(x(\alpha,t)-x(\Gamma',t))^2+\delta(\Gamma',t)^2}\,\d \Gamma',
\end{equation}
in which the second integral consists of bound vorticity advected from the trailing edge, so it has the same dependence on deflection as the bound vorticity. Here, with small deflections, we have assumed that $\partial_\alpha x\approx 1$, and then the linearization is the same as in~\citet{alben2008flapping,mavroyiakoumou2020large}. Without viscous stresses, horizontal membrane deformations arise only through nonlinear terms in the elastic and pressure forces associated with large deflections, so it is reasonable to neglect them, and this is consistent with the simulation results. Balancing the terms that are linear in deflection with the product of $R_3$ and a term that scales with deflection cubed gives $\left<y_\mathrm{defl}\right>\sim R_3^{-1/2}$.
The slight increase in $\langle y_\mathrm{defl}\rangle$ between $R_3=10^{2.5}$ and $10^{3}$ when $R=10^1$ (light blue line with upward-pointing triangle) and $R=10^{1.5}$ (dark red line with right-pointing triangle) arises because for $R_3=10^{2.5}$ the discretization size of the membrane is $m=120$ whereas for $R_3=10^3$ it is $m=40$.

% \begin{figure}[H]
%     \centering
%     \includegraphics[width=.85\textwidth]{Figures/rodsFineR33.pdf}
%      \caption{(Inextensible rods) Snapshots of large-amplitude membrane motions in $R_1$--$R$ space for fixed $T_0=10^{-2}$ and $R_3=10^{3}$. Colors denote $\log_{10}$ of the average deflection defined by~\eqref{eq:avgDefl}.
%     %(Inextensible rods) Snapshots of large-amplitude membrane motions in $R$--$R_3$ space for fixed $T_0=10^{-2}$ and (a) $R_1=10^{0}$, (b) $R_1=10^{0.5}$. Colors denote $\log_{10}$ of the average deflection defined by~\eqref{eq:avgDefl}. For rods with length $R\leq 10^{-0.5}$ the membranes behave similar to fixed-fixed ends with a single hump solution, whereas when the rod's length is $R\geq 10^{0}$ the membranes behave similar to free-free ends. At each $(R,R_3)$ value, the set of snapshots is scaled to fit within a colored rectangle at the $(R,R_3)$ value and normalized by the maximum deflection of the snapshots to show the motions more clearly. In the lower-right corner of (b) snapshots are omitted because steady-state membrane motions were not obtained.
%     }
%     \label{fig:deflRodsR33}
% \end{figure}

\begin{table}[H]
\caption{Table of plots showing snapshots of large-amplitude membrane motions in $R_1$--$R$ space for two values of stretching rigidity $R_3$ ($10^{1.5}$ in left column, $10^{3}$ in right column) and two value of pretension $T_0$ ($10^{-1}$ in top row, $10^{-2}$ in bottom row).  Colors denote the time-averaged deflection defined by Eq.~\eqref{eq:avgDefl}.}\label{table:plots}
\centering
\vspace{.1cm}
\renewcommand{\arraystretch}{2}
\begin{tabular}{c||c|c}
% \cline{2-3}
\backslashbox{$T_0$}{$R_3$}& $10^{1.5}$ & $10^3$ \\ \hline\hline
$10^{-1}$& \begin{minipage}{0.41\textwidth}
    \centering
    \includegraphics[width=\textwidth]{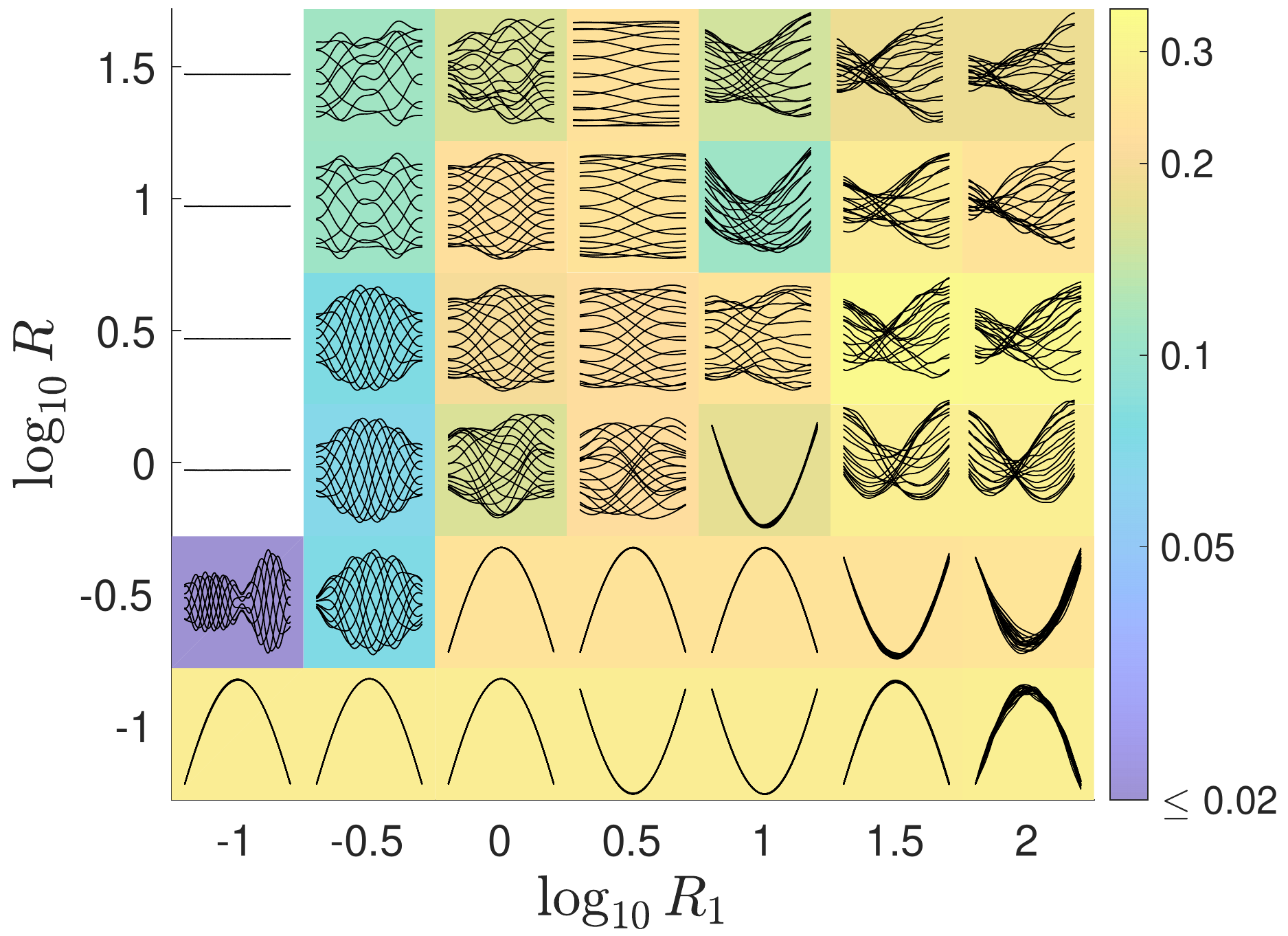}
\end{minipage} & \begin{minipage}{0.41\textwidth}
    \centering
    \includegraphics[width=\textwidth]{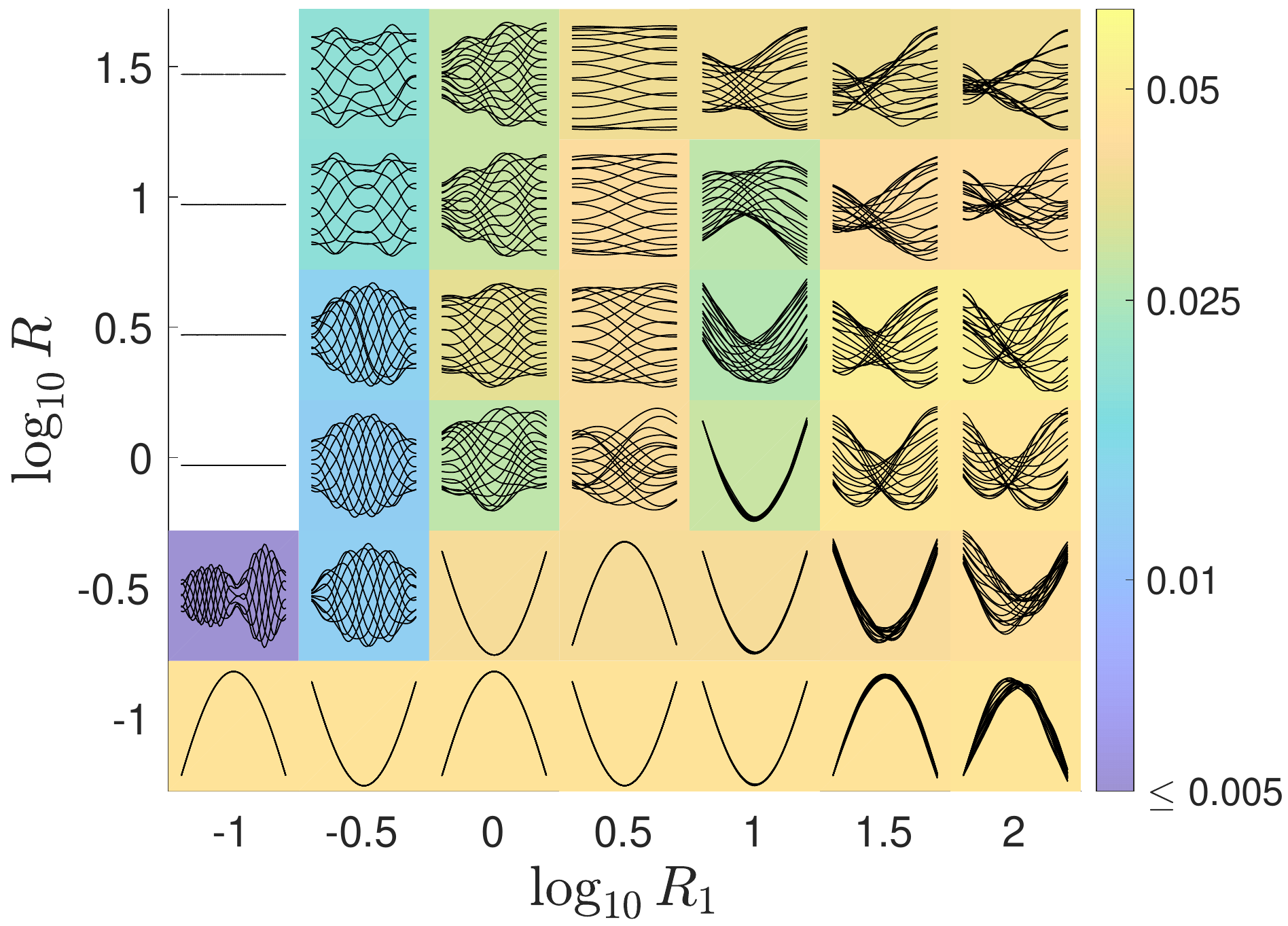}
\end{minipage}  \\ \hline
$10^{-2}$ & \begin{minipage}{0.41\textwidth}
    \centering
    \includegraphics[width=\textwidth]{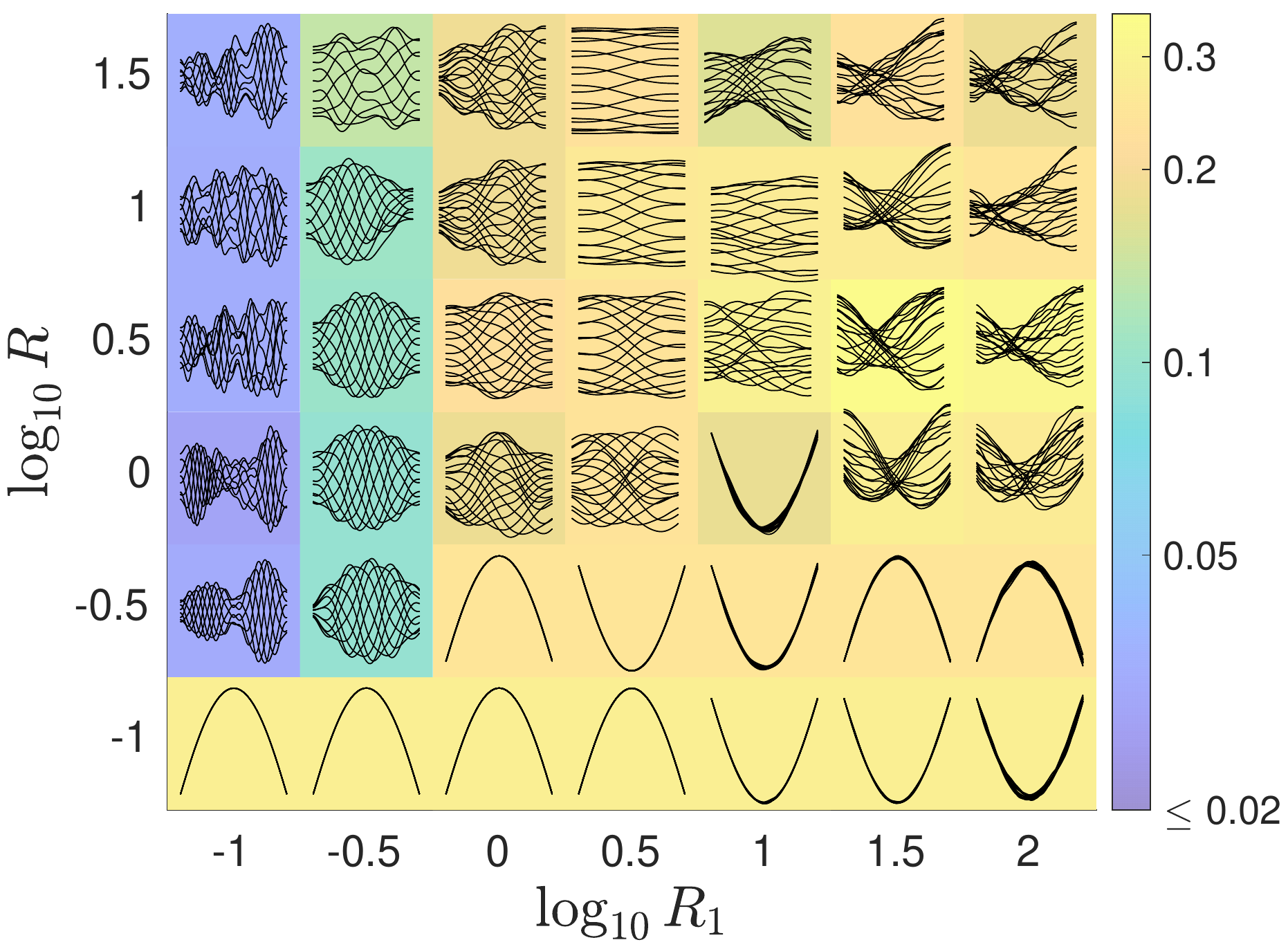}
\end{minipage}  & \begin{minipage}{0.41\textwidth}
    \centering
    \includegraphics[width=\textwidth]{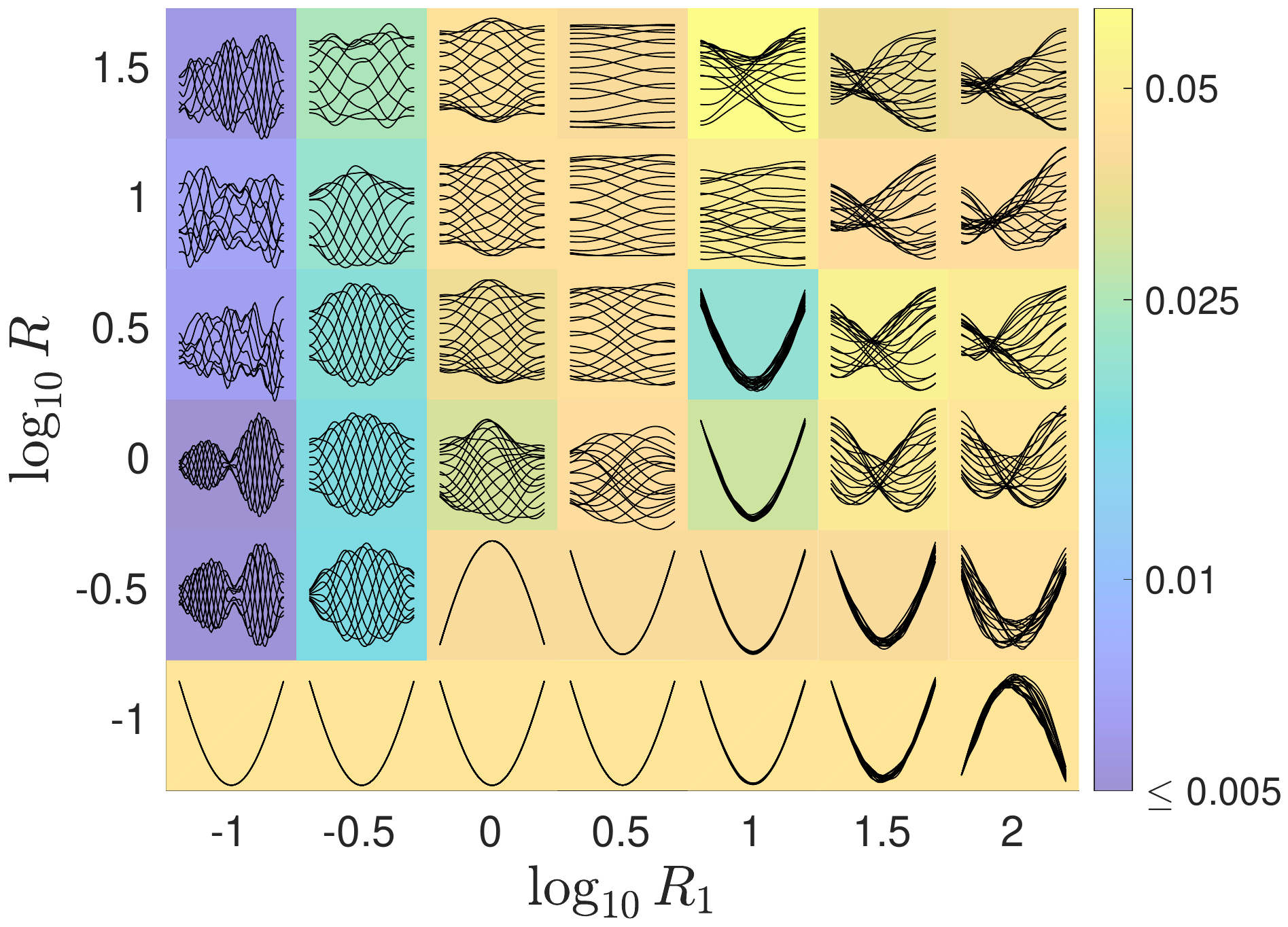}
\end{minipage} \\
\end{tabular}
\end{table}
In Fig.~\ref{fig:deflRodsNeg05} we saw that the motions do not change considerably with $R_3$ (apart from their amplitudes) except in the narrow transition region shown in the inset. We also find that the motions do not depend much on $T_0$ except close to the critical value of $T_0$ below which the flat state is unstable. In Table~\ref{table:plots}
we show membrane snapshots in the full four-dimensional parameter space $R_1$--$T_0$--$R_3$--$R$, collected into four
subpanels, each with a particular value of $T_0$ and $R_3$ (labeled at top and left, respectively), and
with a range of values of $R_1$ and $R$ within each subpanel. There is more variation within a given subpanel than between corresponding points in different subpanels, indicating that $R$ and $R_1$ have a stronger effect on the dynamics than
$T_0$ and $R_3$.
%, and show typical membrane snapshots in $R_1$–$R$
%space for a fixed stretching rigidity $R_3$ ($10^{1.5}$ in left column, $10^{3}$ in right column) and pretension $T_0$ ($10^{-1}$ in top row, $10^{-2}$ in bottom row). 
% \silas{please move/adapt the above lines to the earlier plot (R3 instead of R1)}
The white background and flat lines at $R_1=10^{-1}$ and $R\geq 10^0$ when $T_0=10^{-1}$ (top row) indicate stable membranes, so the deflection there is zero. 
From this comparison we see that, as in the previous figure, the deflections decrease with increasing $R_3$ (values in color bars at right) but often the snapshot shapes do not change much, at the same $(R_1,R)$ values. Some membranes with moderate values of $R_1$ ($10^0$ and $10^1$) have more prominent differences as $R_3$ is changed, sometimes by altering the location of a transition between different types of dynamics. Decreasing the value of $T_0$ can cause stable membranes to become unstable (e.g., at $R_1=10^{-1}$), but otherwise decreasing $T_0$ has a small effect, mainly to increase the deflection slightly at a given $R_3$. Below $10^{-2}$, the $T_0$ term in the membrane equation [Eq.~\eqref{eq:membrane}] becomes insignificant, as noted in \citet[]{mavroyiakoumou2020large}.

\begin{figure}[H]
    \centering
    \includegraphics[width=.95\textwidth]{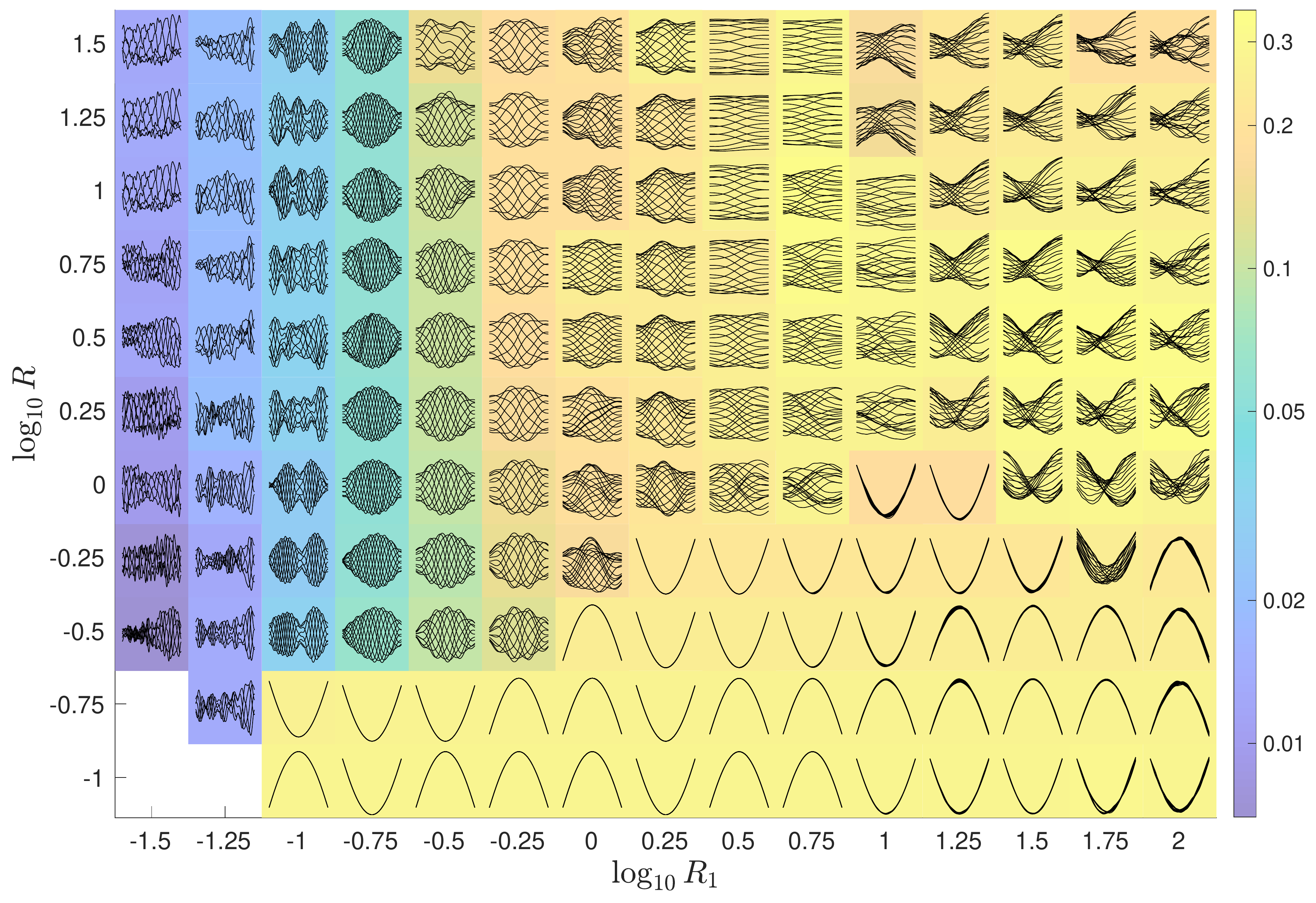}
     \caption{(Inextensible rods) Membrane profiles in the large-amplitude steady-state regime, in $R_1$--$R$ space for fixed $T_0=10^{-2}$ (dimensionless pretension) and $R_3=10^{1.5}$ (dimensionless stretching rigidity). The colored background denotes the time-averaged deflection of the membranes defined by Eq.~\eqref{eq:avgDefl}.} \label{fig:deflRodsR315}
\end{figure}

%and show typical membrane snapshots in the unstable region of %$R_1$–$R$
%space for fixed $R_3$ $(10^{1.5})$, a value giving moderately large deflections for the
%steady-state motion and $T_0=10^{-2}$. 
%The colors denote the average deflection of the membrane [Eq.~\eqref{eq:avgDefl}], now shown in a finer mesh of $R_1$ and $R$ compared to the plots in Table~\ref{table:plots}.
In Fig.~\ref{fig:deflRodsR315}, we focus on the lower left subpanel of Table \ref{table:plots}, but double the density of values of $R_1$ and $R$, and decrease the lower limit of $R_1$, to obtain a more comprehensive picture of the dynamics. 
The motions in Fig.~\ref{fig:deflRodsR315} have the largest deflection amplitudes at the largest $R_1=10^2$. As mentioned in~\citet{mavroyiakoumou2020large}, we hypothesize that at large~$R_1$ membrane inertia allows the membrane to maintain its momentum for longer
times against restoring fluid forces, and obtain larger deflections (with longer periods, as we will show) before reversing
direction. The same has been observed for flutter with bending rigidity~\citep{connell2007fdf,alben2008flapping}.
As $R_1$ decreases, the membrane deflection
amplitudes progressively decrease until the motions
become difficult to resolve numerically (for $R_1\lesssim 10^{-1}$). In this
region, we find chaotic membrane oscillations with very small amplitudes and high
spatial frequencies.  To obtain numerically-converged motions with respect to the spatial grid when $R_1\leq 10^{-1}$ we use more discretization points.
In the lower-left corner in Fig.~\ref{fig:deflRodsR315}, i.e.,\ $(R_1,R)=(10^{-1.5},10^{-1})$, $(10^{-1.5},10^{-0.75})$, and $(10^{-1.25},10^{-1})$, snapshots are omitted because steady-state membrane motions were not obtained.

Decreasing the membrane mass ratio ($R_1$) generally tends to introduce more oscillating states and fewer single-hump solutions for $R$ values in the range $(10^{-0.75},10^{0.25})$. For large $R_1$ (heavy membranes) the maximum deflection of the membrane occurs close to either the leading or trailing edge of the membrane. However, at $R_1\in[10^{-0.75},10^{0.25}]$ the maximum membrane deflection seems to occur close to the midpoint of the membrane, with the deflection at the endpoints decreasing with decreasing $R_1$ in this region.

%start here
\begin{figure}[H]
    \centering
    \includegraphics[width=.8\textwidth]{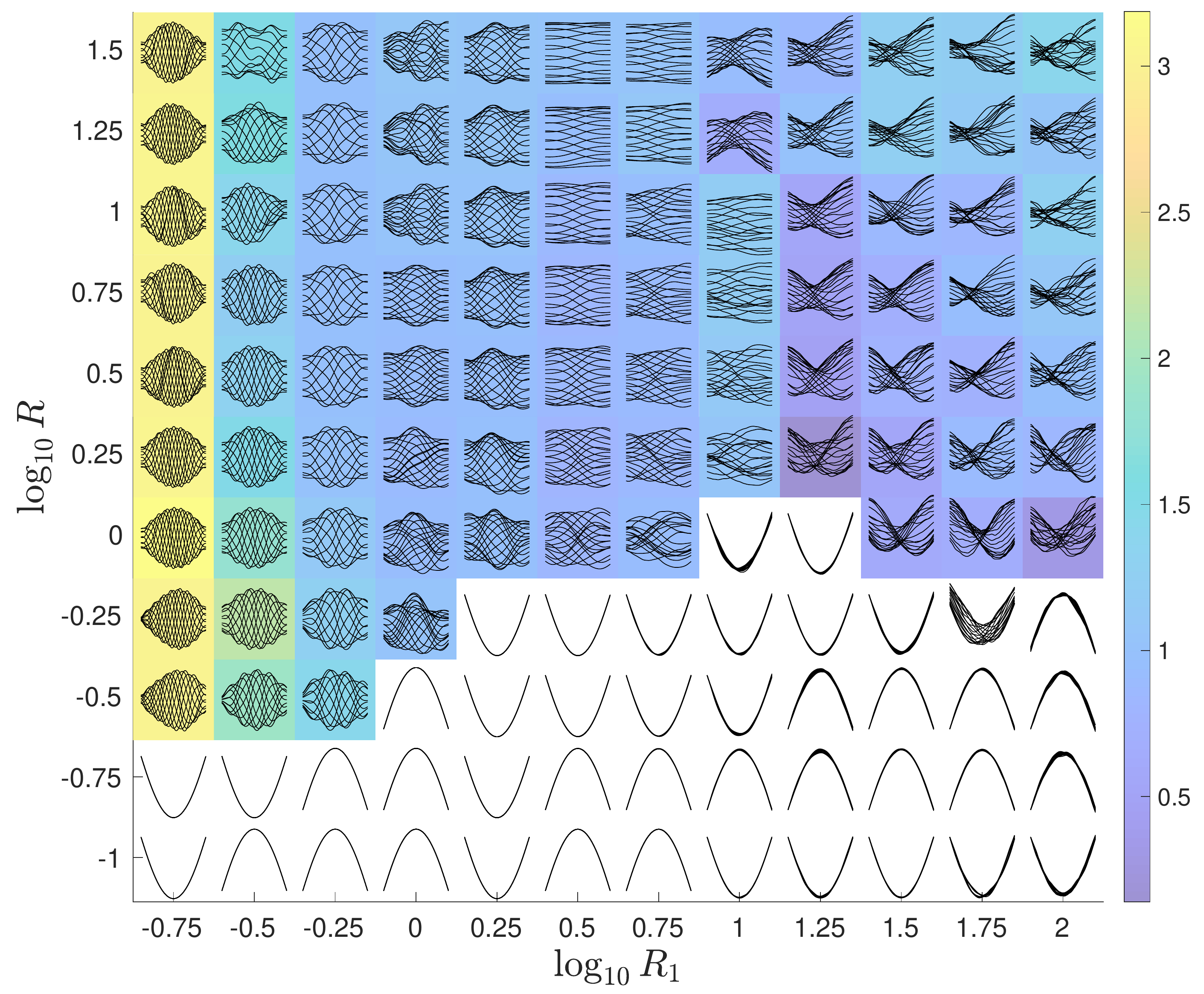}
     \caption{(Inextensible rods) Colors denote the time-averaged number of zero-crossings for membrane flutter in the $R_1$--$R$ parameter space for fixed $T_0=10^{-2}$ and $R_3=10^{1.5}$. Note that $R_1$ is the dimensionless membrane mass, $T_0$ is the dimensionless pretension, and $R_3$ is the dimensionless stretching modulus.  We also define $R$ to be the length of the inextensible rods at either end of the membrane. The white background corresponds to membranes with no zero-crossings. At each $(R_1,R)$ value the set of snapshots is normalized by the maximum deflection of the snapshots to show the motions more clearly.} \label{fig:zeroCrossRodsR315}
\end{figure}

We now quantify the membrane shapes in terms of the time-averaged number of `zero crossings'. Our definition is the number of crossings that a membrane
makes with the line connecting its two endpoints, averaged over time---excluding the endpoints.  This is one way to measure the `waviness' of a shape that is not sinusoidal and whose wavelength is thus not well defined~\citep{alben2015flag,mavroyiakoumou2020large}. 
We first focus on moderate-to-large values of $R_1$ where the membranes have fewer zero-crossings (Fig.~\ref{fig:zeroCrossRodsR315}). Decreasing~$R_1$ from the largest value ($10^2$), the average number of zero crossings changes non-monotonically. In most cases it decreases until about $R_1=10^{1.25}$ for $R\in[10^0,10^1]$. Further decreases in $R_1$ give rise to more periodic motions with slightly larger numbers of zero-crossings. Independent of $R_1$, when $T_0$ and $R_3$ are fixed at $10^{-2}$ and $10^{1.5}$, respectively, and when the rods have a length of $\leq 10^{-0.75}$ then the membrane behaves similarly to the fixed-fixed case, where a single-hump solution is obtained. We use a white background for membranes with no zero-crossings (single hump solutions).

\begin{figure}[H]
\begin{minipage}{.45\textwidth}
    \flushright
    \includegraphics[width=.65\textwidth]{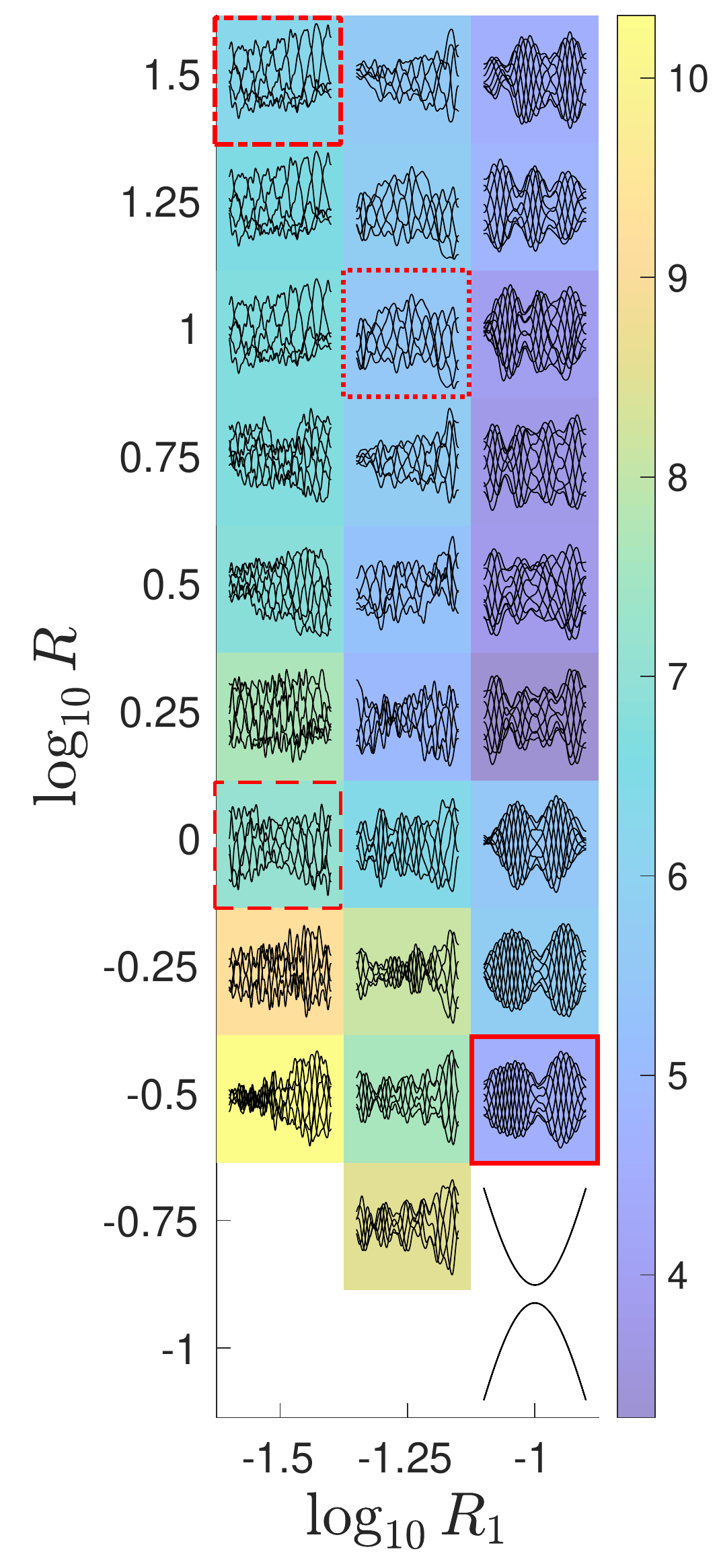}
\end{minipage}\hspace{.4cm}
\begin{minipage}{.37\textwidth}
    \includegraphics[width=\textwidth]{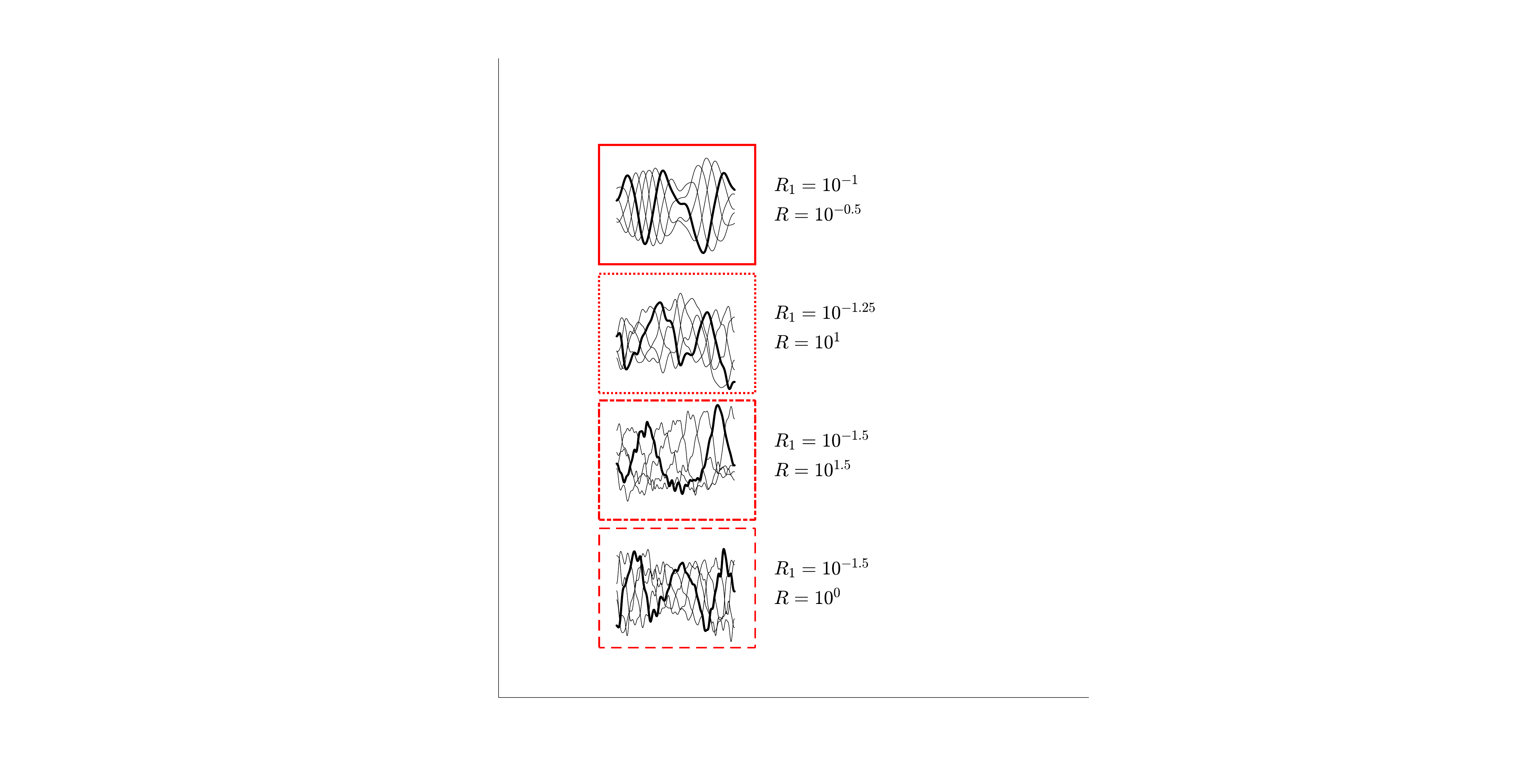}
\end{minipage}
\caption{(Inextensible rods) Colors denote the time-averaged number of zero-crossings for membrane flutter in the $R_1$--$R$ parameter space for fixed $T_0=10^{-2}$ and $R_3=10^{1.5}$ for light membranes ($R_1\leq 10^{-1}$). Snapshots of these large-amplitude membrane motions are superposed to show the motions clearly in this region.}
\label{fig:zeroCrossRodssmallR315}
\end{figure}
In Fig.~\ref{fig:zeroCrossRodssmallR315} we present the zero-crossings in the small $R_1$ ($\leq 10^{-1}$) region, where higher spatial frequency components occur with decreasing $R_1$. The motions also become more irregular at the smallest $R_1$ values, where we increase the spatial grid density to resolve the fine undulations that appear on the membranes. On the right-hand side of Fig.~\ref{fig:zeroCrossRodssmallR315} we show four panels with examples of sequences of membrane snapshots, equally spaced in time (with the thicker black line representing the membrane at the last time), to emphasize that even though the number of zero-crossings is a good measure of waviness it also misses some features of the shapes. For example, we see that the shape at $R_1=10^{-1.5}$ and $R=10^0$ (bottom row of right-most column) has small undulatory features that are not reflected in the number of zero-crossings. In the small-$R_1$ region, the numbers of zero-crossings (shown by the colors) vary more rapidly compared to Fig.~\ref{fig:zeroCrossRodsR315}. In the lower-left corner, snapshots are omitted because steady-state membrane motions were not obtained. %A white background is again used for membranes with no zero-crossings (single hump solutions), i.e., at $R_1=10^{-1}$ when $R=10^{-1}$ and $10^{-0.75}$.
% \begin{figure}[H]
%     \centering
%     \includegraphics[width=.95\textwidth]{Figures/rodsR315T0Minus2zeroCrossFine.pdf}
%      \caption{(Inextensible rods) Snapshots of large-amplitude membrane motions in $R_1$--$R$ space for fixed $T_0=10^{-2}$ and $R_3=10^{1.5}$. Colors denote the number of zero-crossings.} \label{fig:zeroCrossRodsR315}
% \end{figure}

%The definition based on the number of times the membrane crosses a straight line, ignores the fact that these membranes have small undulations about a line with non-zero net slope. This leads us to another way of measuring the waviness of a shape which is not sinusoidal;
\begin{figure}[H]
    \centering
    \includegraphics[width=.58\textwidth]{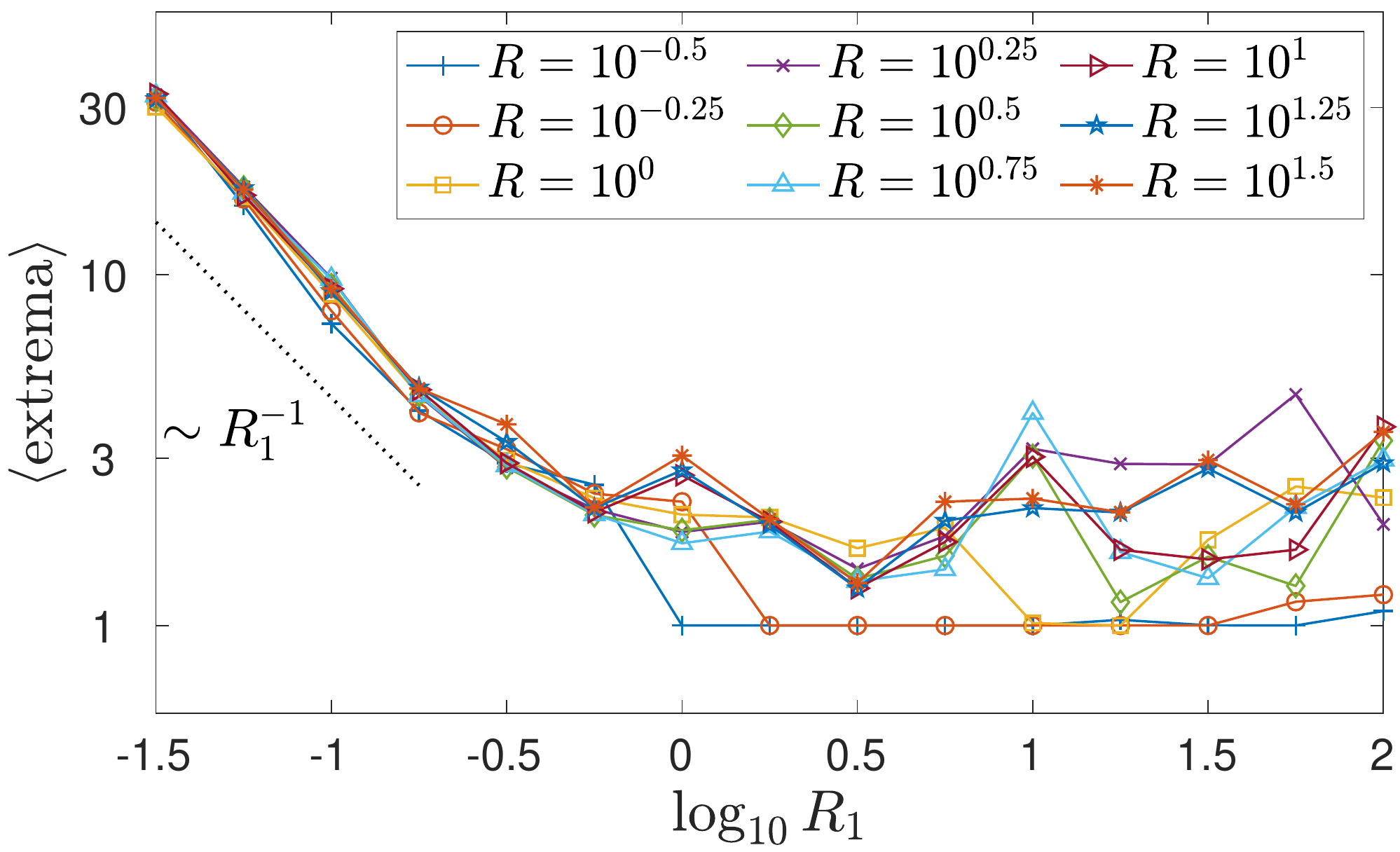}
    \caption{(Inextensible rods) Time-averaged number of local extrema of the membranes versus the dimensionless mass density $R_1$ for various $R$ and fixed $R_3=10^{1.5}$ and $T_0=10^{-2}$. The  dotted black line at small $R_1$  indicates the scaling~$R_1^{-1}$.}
    \label{fig:lineExtremaR1}
\end{figure}
% \christiana{Use \texttt{langle, rangle} around extrema for latex.}

% \christiana{We classify periodicity in the following way:
% For different values of $R$ (length of rod) we present in Fig.~ (left column) power spectra (using the Welch method~\citep{welch1967use}). In the middle column we present the limit cycles of $\mathrm{d}y/\mathrm{d}t$ versus $y$ at the membrane's midpoint, and in the right column plots of the membrane's midpoint against time (i.e., $y(1/2,t)$). The results for a fixed $R$ value do not seem to change much with varying $R_3$.}
% \christiana{Change the period plots to use the mean period from pwelch. I'll have to adjust which data I'm using in each case.}
% \christiana{Mention in the text that if the dominant peak is very close to zero then the long period corresponds to almost the whole time series and thus we look at the second most dominant peak.}
To quantify the small undulatory features on the membranes, we calculate the time-averaged number of local extrema of deflection. In Fig.~\ref{fig:lineExtremaR1} we show that for fixed $R_3=10^{1.5}$, $T_0=10^{-2}$, and various fixed values of~$R$,  the time-averaged number of local extrema for small $R_1$ scales as $R_1^{-1}$ approximately. At moderate-to-large values of~$R_1$ (i.e.,\ $[10^0,10^2]$) and $R$ small, the membranes tend to fore-aft symmetric, single-hump solutions and therefore the average number of extrema is one. For the oscillatory shapes that occur at larger values of~$R$ in the same region of~$R_1$,  the average number of local extrema is
not large (i.e., between 1 and 5).

\begin{figure}[H]
    \centering
    \includegraphics[width=.95\textwidth]{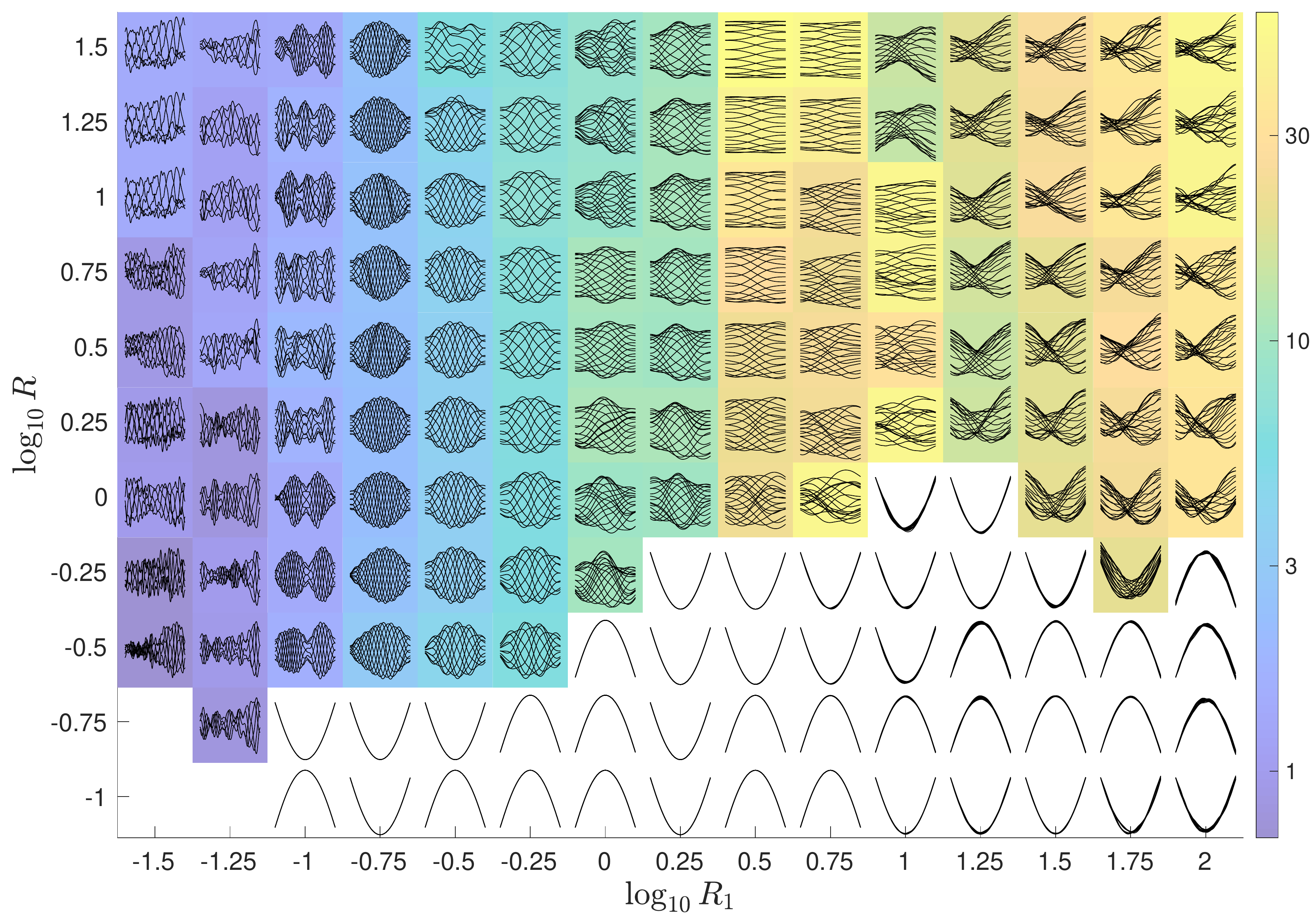}
     \caption{(Inextensible rods) Colors denote the dominant periods of large-amplitude motions for various $R_1$ and $R$, and fixed $T_0=10^{-2}$ and $R_3=10^{1.5}$. The data in the bottom-left corner are obtained for a shorter time and so we neglect the computational results for those values of $R_1$ and $R$.} \label{fig:periodRodsR315}
\end{figure}
We have considered the amplitude of membrane deflection and its spatial frequency (in terms of zero crossings and numbers of extrema). The third main quantity we consider is the temporal period. 
%Along with the 
%We have already mentioned the trend to higher spatial frequency components with decreasing $R_1$. 
%In Fig.~\ref{fig:periodRodsR315} we look at the temporal dynamics corresponding to these motions and quantify them by 
We compute the power spectra of the time series of the membrane's midpoint, $y(1/2,t)$, using the fast Fourier transform.  
We identify the dominant frequency as that corresponding to the largest local maximum in the power spectrum (in a few cases excluding the peak closest to zero, which represents the time scale of the entire time series, and occurs because of the discontinuity in $y(1/2,t)$ at the beginning and end of the time series). The background color in Fig.~\ref{fig:periodRodsR315} denotes the dominant period, defined as the reciprocal of the dominant frequency, and is white for the steady single-hump solutions. 
%(at smaller $R$) are steady, so the dominant period is undefined and thus we use a white background. 

We find different types of power spectra in different regions of $R_1$--$R$ space, corresponding to the different motions illustrated in Fig.~\ref{fig:periodRodsR315}.
At small $R_1$ ($\lesssim 10^{-1}$) the motions are more chaotic and there, the power spectra have a broad band of frequencies.  At small-to-moderate values of $R_1$---between $10^{-0.75}$ and $10^{0.25}$---the motions are periodic and thus the power spectra have a discrete set of peaks. At moderate values of $R_1$---between $10^{0.5}$ and $10^{1}$---the peak frequencies are decreased. Finally, for large values of $R_1$ ($\geq 10^{1.25}$) the motions become somewhat chaotic again (as at the smallest $R_1$), and with little dependence on $R$ except at values greater than $10^1$, where there is a slight increase in the dominant period. 
%For these heavy membranes ($R_1\geq 10^{1.25}$)  the dominant frequency is mostly zero. Therefore, we again pick the second most dominant frequency in most cases. 
\begin{figure}[H]
    \centering
    \includegraphics[width=.635\textwidth]{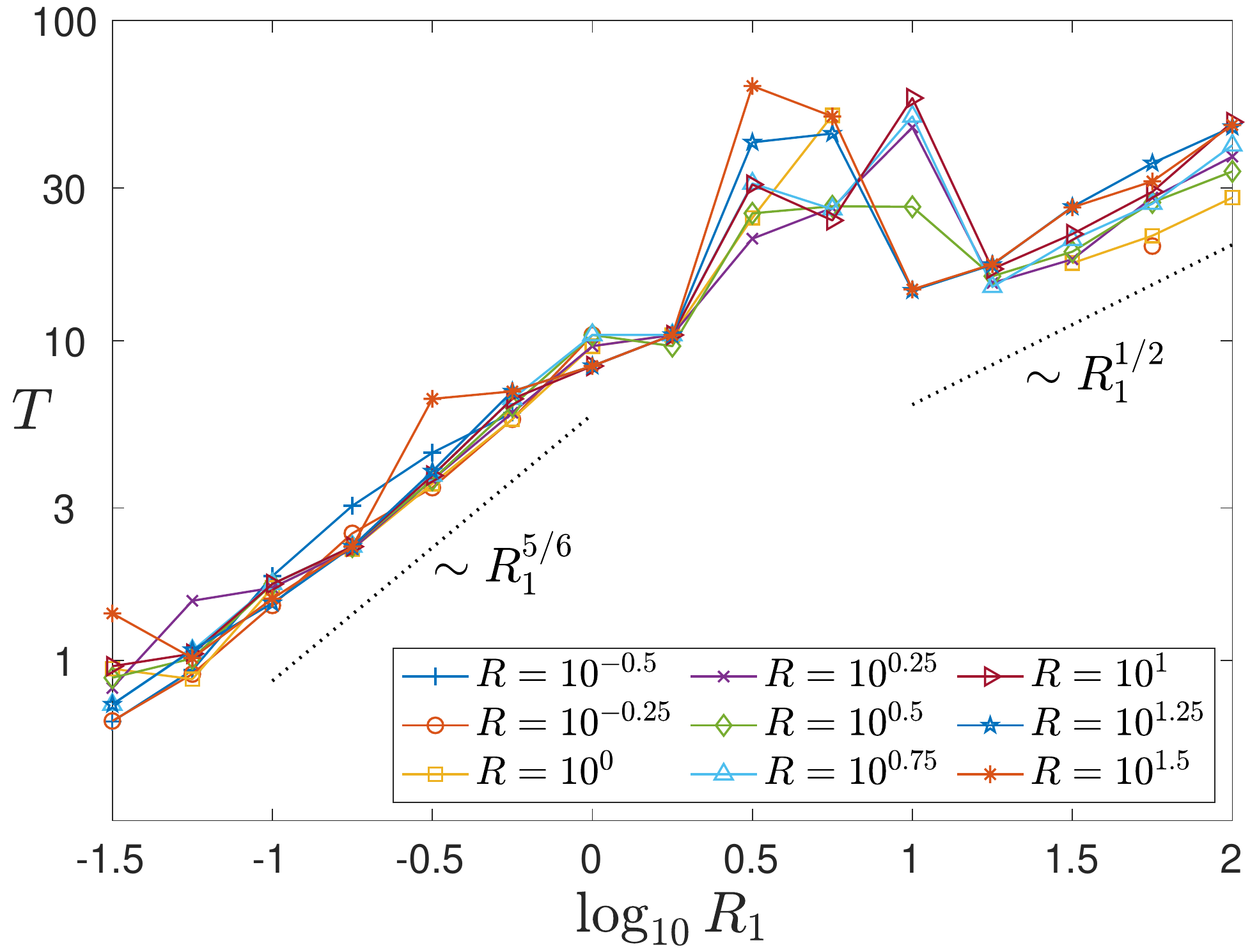}
    \caption{Plots of the dominant period ($T$) versus mass density $R_1$ for various $R$ and fixed $R_3=10^{1.5}$ and $T_0=10^{-2}$. The  dotted black line at large $R_1$ shows the scaling $R_1^{1/2}$ and the dotted black line at small $R_1$ shows the scaling $R_1^{5/6}$.}
    \label{fig:lineperiodR1}
\end{figure}
In Fig.~\ref{fig:lineperiodR1} we show how the dominant period varies with
$R_1$ for various fixed values of $R$.
The trend at the largest~$R_1$ is approximately $T\sim R_1^{1/2}$ (admittedly over a short range of~$R_1$). This
scaling arises when one approximates the normal component of the membrane equation [Eq.~\eqref{eq:membrane}] by its $y$-component, and chooses a characteristic time scale $t_0$ so that
$R_1\partial_{tt}y$ balances other terms that depend on $y$ but not its time derivatives (i.e.,\ the $R_3$ and $T_0$ terms and some of the fluid pressure terms). At large $R_1$, $R_1\partial_{tt}y$ is comparable to the other terms when $R_1/T^2\sim 1$ giving a typical period $T\sim R_1^{1/2}$. For some values of $R$, when $10^{0.25}< R_1< 10^{1.25}$ and $R_1>10^{1.5}$, the period increases to $>30$ as can be seen in Fig.~\ref{fig:lineperiodR1}. This range of moderate $R_1$ is a transition region, and at smaller $R_1$, (here, \ $10^{-1.5}\leq R_1<10^{0.25}$), another power law behavior is observed: $T\sim R_1^{5/6}$.

\subsection{Hookean springs}\label{sec:springs}

\begin{figure}[H]
    \centering
    \includegraphics[width=.75\textwidth]{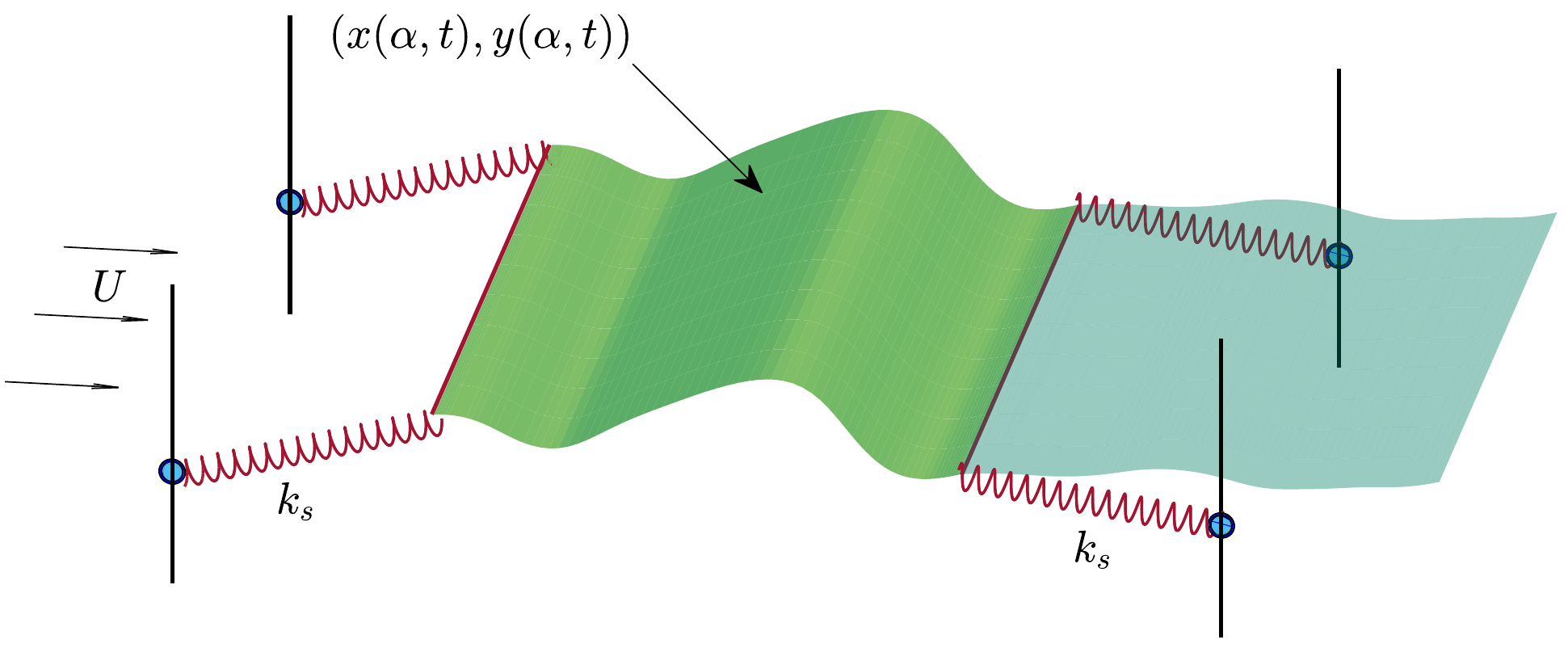}
    \caption{Schematic diagram of a flexible membrane (green surface) at an instant in time. $U$ is the oncoming flow velocity.  There is also a vortex wake (light green surface) emanating from the membrane's trailing edge. The leading edge of the membrane with position $(x(-1,t),y(-1,t))$ is attached to springs (red coils) of spring constant $k_s$ whose other ends are fixed at $(0,0)$ for all time. The membrane's trailing edge with position $(x(1,t),y(1,t))$ is attached to another spring whose other end is fixed at $(2,0)$.}\label{fig:schemSprings3d}
\end{figure}

The inextensible rods are a particular choice of tether motivated by the experiment of~\citet{kashy1997transverse}. In this section we briefly explore some alternative tethers involving Hookean springs.
In the first case, we replace the inextensible rods at the ends of the membrane with springs of rest length zero that obey Hooke's law~\citep{hooke1678potentia}.
We illustrate schematically this alternative configuration in Fig.~\ref{fig:schemSprings3d}. The four prescribed dimensionless parameters are: membrane mass~$R_1$, stretching rigidity~$R_3$, pretension~$T_0$, and spring stiffness~$k_s$. 
% \begin{figure}[H]
%     \centering
%     \includegraphics[width=.6\textwidth]{Figures/schematicSprings.pdf}
%     \caption{Schematic diagram of a flexible membrane (black line) at an instant in time. The leading edge of the membrane with position $(x(-1,t),y(-1,t))$ is attached to a spring (red zig-zag line) of spring constant $k_s$ whose other end is fixed at the origin for all time. The membrane's trailing edge with position $(x(1,t),y(1,t))$ is attached to another spring whose other end is fixed at $(2,0)$.}\label{fig:schematicSprings}
% \end{figure}
We solve for the four endpoint unknowns ($x_{\pm 1}$, $y_{\pm 1}$) with four boundary conditions. At the membrane-spring contact, the tension forces must be equal in magnitude and direction to avoid infinite acceleration at the membrane ends, as for the rod tethers. Here the forces are equal in magnitude when:
\begin{equation}\label{eq:hookes}
    k_s\sqrt{x_{-1}^2+y_{-1}^2}=T_{-1}\quad\text{and}\quad k_s\sqrt{(x_{1}-2)^2+y_{1}^2}=T_{1}.
\end{equation}
Here $T_{\pm 1}$ is the tension force at $\alpha=\pm 1$ and $\sqrt{x_{-1}^2+y_{-1}^2}$ is the stretch of the spring (change in length from its rest length, zero).
The directions of the tensions in the membrane and springs are equal if the slopes of the membrane and springs are equal:
\begin{equation}
    \displaystyle\left.\frac{\partial_\alpha y}{\partial_\alpha x}\right|_{\alpha=-1}=\frac{y_{-1}-0}{x_{-1}-0}\quad\text{and}\quad\displaystyle\left.\frac{\partial_\alpha y}{\partial_\alpha x}\right|_{\alpha=1}=\frac{0-y_{1}}{2-x_{1}}.
\end{equation}
%The tension term in the membrane equation [Eq.~\eqref{eq:membrane}] is $T(\alpha,t)=T_0+R_3(\partial_\alpha s-1)$. Using this term and the boundary conditions [Eqs.~\eqref{eq:hookes}], we can derive the initial conditions as: $x_{-1}=T_0/(R_3+k_s)$ and $x_1=2-T_0/(R_3+k_s)$.\silas{I'm not seeing this--let's discuss}

When we simulate the spring-tethered membrane for various $k_s$, we find that for sufficiently large $k_s$, the membrane behaves like the fixed-fixed case, converging to a steady single-hump shape when the flat state is unstable. As we decrease $k_s$, the single hump solution continues until a threshold value of $k_s$ (near unity) where the membrane develops a sharp spike at the trailing edge at early times and the simulations fail to converge beyond a short time. Unlike the inextensible-rod tethers, here the springs are too soft to ensure that the membrane remains under tension during the dynamics, and the membrane equation is ill-posed under compression~\citep{triantafyllou1994dynamic}.

%Focusing on the limit of small and large spring constants, we seek to observe the response of the membrane dynamics. Since a larger value of $k_s$ implies that the spring is stiffer, we expect that the dynamics obtained for larger $k_s$ will resemble the behavior exhibited by fixed-fixed membranes. Indeed that is true. The code, however, does not converge for smaller values of $k_s$. Therefore, in the next section, we restrict the motion of the Hookean springs to be only in the vertical direction.

%%%%%%%%%%%%%%%%%%%%
\subsection{Vertical Hookean springs} \label{sec:verticalSprings}

\begin{figure}[H]
    \centering
    \includegraphics[width=.72\textwidth]{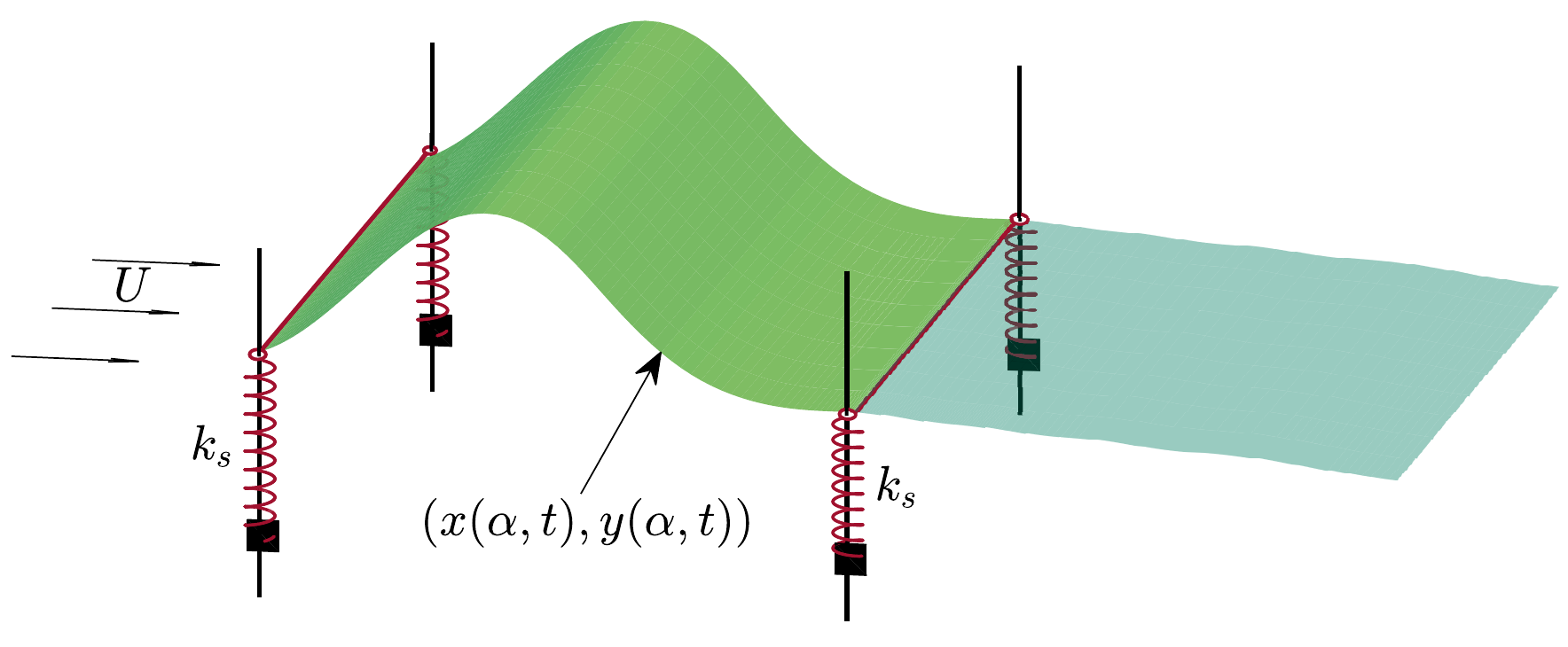}
    \caption{Schematic diagram of a flexible membrane (green surface) at an instant in time. $U$ is the oncoming flow velocity.  There is also a vortex wake (light green surface) emanating from the membrane's trailing edge. The leading edge of the membrane with position $(0,y(-1,t))$ is attached to vertical springs (red coils) of spring constant $k_s$  whose other end is fixed at (0,0) for all time. The membrane's trailing edge with position $(2,y(1,t))$ is attached to another vertical spring whose other end is fixed at $(2,0)$.}\label{fig:schemVerticalSprings3d}
\end{figure}

More interesting dynamics occur with springs in an alternative configuration, in which the springs are attached to massless rings that slide along vertical poles, shown in Fig.~\ref{fig:schemVerticalSprings3d}. This is the same as the free-free boundary condition except that the vertical motion is not free but instead resisted by springs.
As in the free-free case, the vertical poles ensure that the membrane does not experience significant compression, and thus stable long-time oscillatory dynamics can occur. We will show that this boundary condition is equivalent to that of the inextensible-rod tethers in the limit of small deflections, so it provides an alternative way to understand the effect of the rods. Both the rods and vertical springs allow for a difference in resistance to transverse and in-plane motions, and hence allow for stable oscillatory large-amplitude flutter.

%In this section, we consider a membrane attached to a massless spring at its leading and trailing edge and assume that only transversal oscillations are allowed (Fig.~\ref{fig:schemVerticalSprings3d}). This boundary condition is closer in nature to the free-free membranes we studied in~\citet{mavroyiakoumou2020large}, where we assumed that the membrane can only deflect in the vertical direction
%with the membrane end fixed to a massless ring that slides without friction along a vertical pole.
%~\citep{figotin2015lagrangian}

Here, by balancing the vertical forces on the rings, we obtain the mixed boundary conditions:
\begin{equation}\label{eq:verticalBC}
T_{-1}\left.\frac{\partial_\alpha y}{\partial_\alpha s}\right|_{\alpha=-1}-k_sy_{-1} = 0\quad\text{and}\quad
-T_1\left.\frac{\partial_\alpha y}{\partial_\alpha s}\right|_{\alpha=1}-k_sy_1=0.
\end{equation}
%where $T(\alpha,t)$ is the tension term and $k_s$ is the spring constant.
The free-free case corresponds to $k_s=0$~\citep{mavroyiakoumou2020large}. The fixed-fixed case ($y(-1,t)=y(1,t)=0$) occurs when $k_s\to \infty$. 
The remaining boundary conditions are $x_{\pm 1} = 2$, due to the poles.

% \begin{figure}[H]
%     \centering
%     \includegraphics[width=.6\textwidth]{Figures/schemVerticalSprings.pdf}
%     \caption{Schematic diagram of a flexible membrane (black line) at an instant in time. The leading edge of the membrane with position $(0,y(-1,t))$ is attached to a vertical spring (red zig-zag line) of spring constant $k_s$  whose other end is fixed at the origin for all time. The membrane's trailing edge with position $(2,y(1,t))$ is attached to another vertical spring whose other end is fixed at $(2,0)$.}\label{fig:schemVerticalSprings}
% \end{figure}

\begin{figure}[H]
    \centering
    \begin{minipage}{0.48\textwidth}
        \centering
\includegraphics[width=\textwidth]{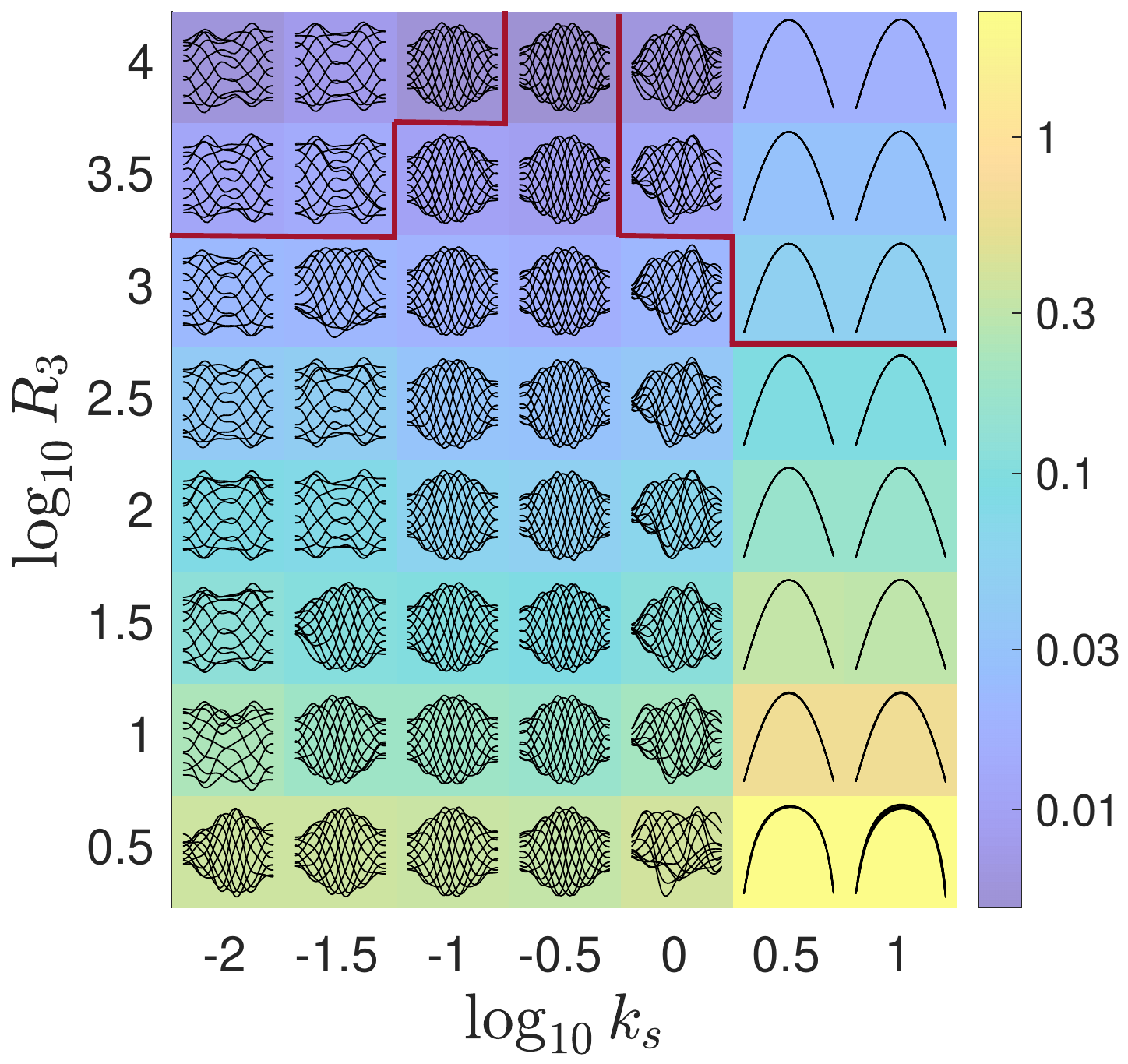}
\end{minipage}
\hspace{.2cm}
\begin{minipage}{0.32\textwidth}
    \centering
\fbox{\includegraphics[width=\textwidth]{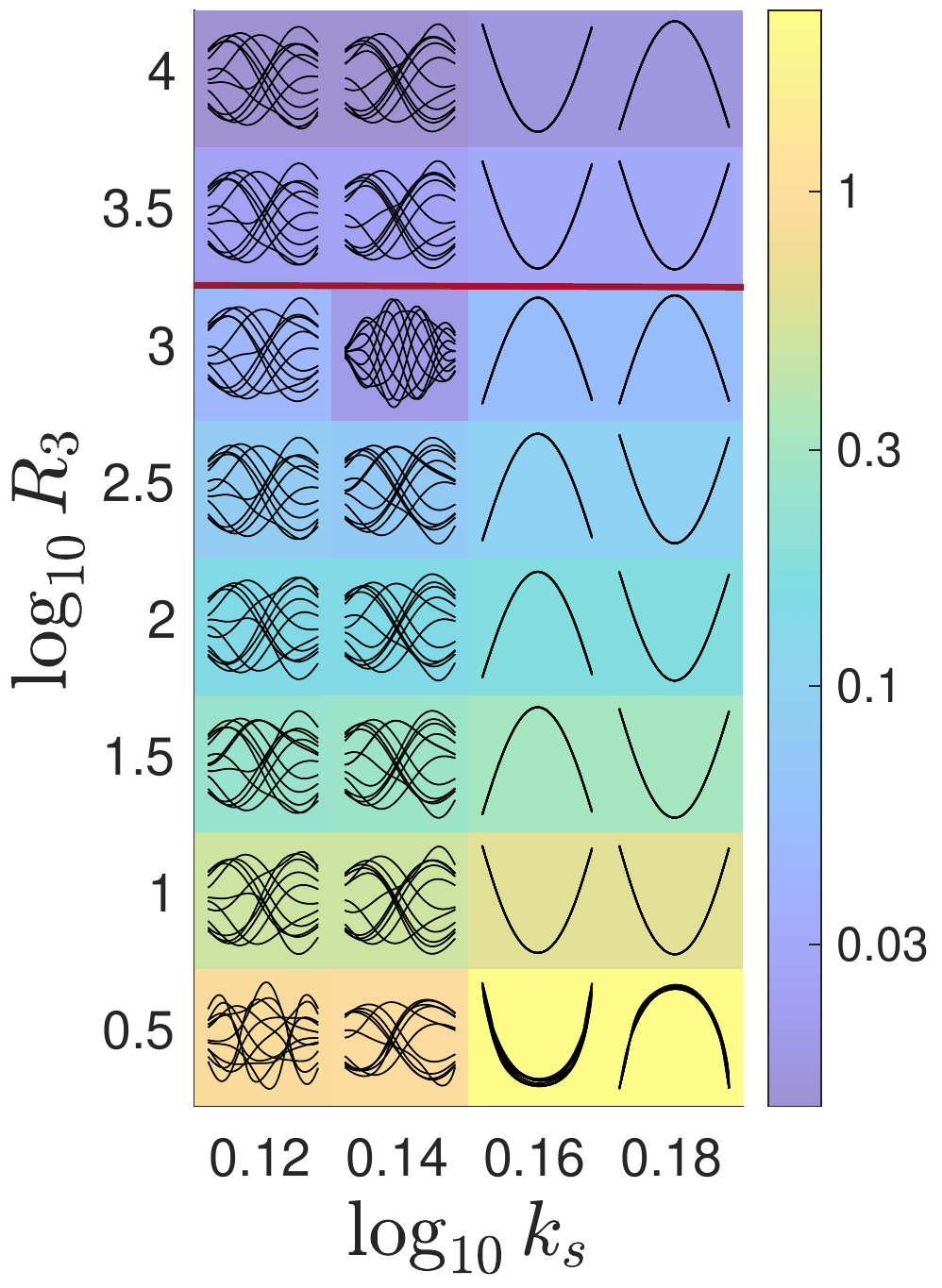}}
\end{minipage}
\caption{(Vertical springs) Snapshots of large-amplitude membrane motions in $k_s$--$R_3$ space for fixed $T_0=10^{-2}$ and $R_1=10^{-0.5}$. Colors denote the time-averaged deflection of membranes defined by Eq.~\eqref{eq:avgDefl}. Oscillatory ($k_s\leq 10^0$) and steady single-hump solutions ($k_s\geq 10^{0.5}$) are obtained. At each $(k_s, R_3)$ value, the set of snapshots is scaled to fit within a colored rectangle at the $(k_s, R_3)$ value and normalized by the maximum deflection of the snapshots to show the motions more clearly. The framed panel at right shows a finer grid between $k_s=10^{0.12}$ and $10^{0.18}$, near the transitional $k_s$ value. The red line separates membranes with $m=40$ points (above) and $m=80$ points (below).}
    \label{fig:deflVertical}
\end{figure}

In Fig.~\ref{fig:deflVertical} we show membrane snapshots in the~$k_s$--$R_3$ parameter space for fixed $T_0=10^{-2}$ and $R_1=10^{-0.5}$. The shapes are superposed on colors that denote the time-averaged deflections of the membranes [Eq.~\eqref{eq:avgDefl}]. As for the rod tethers, the stretching rigidity $R_3$ mainly affects the deflection of the membrane, not its shape. For $R_3=10^{0.5}$ and $k_s\in[10^{0.5},10^{1}]$ the deflections are so large that vortex shedding might not be confined to the trailing edge in reality, but we include these results to illustrate the model's behavior. The red line separates simulations with $m=80$ points (below) and $m=40$ (above); the smaller value is needed when $R_3\geq 10^3$ to reach the steady-state regime. When $k_s\geq 10^{0.5}$ the membranes reach the single hump state, as in the fixed-fixed case, and for the rods with $R\leq 10^{-1}$ in Fig.~\ref{fig:deflRodsNeg05}. 

 There is a critical value of $k_s$ at which the membrane transitions from the steady single-hump solutions to oscillatory motions. In the framed panel on the right-hand side of Fig.~\ref{fig:deflVertical}, we show the dynamics close to the transition.
 %locate the critical  spring stiffness constant for which we transition from more wavy membranes to less wavy ones, and finally to the single-hump solutions. 
 From $k_s = 10^0$ to $10^{0.12}$ the membrane shapes become less wavy. At $k_s = 10^{0.12}$ and $10^{0.14}$ they have only one ``neck'' in their deflection envelopes, apart from
 $(k_s,R_3)=(10^{0.12},10^{0.5})$ and $(10^{0.14},10^3)$.
 %the membranes attached to springs with stiffness values at   appear to
%follow an up-down symmetric, periodic (as expected) oscillation with only .

%%%%%%%%%%%%%%%%%%%%%%%%%%%%%%%%%%%%%%%%%%%%%%%

\section{Linearized membrane model}\label{sec:linearized}

In this section we analyze the small-amplitude behavior of the system described in Sec.~\ref{sec:large}. We are able to present the small-amplitude motions of the membranes at a wide range of parameter values (membrane mass and pretension) by computing the eigenvalues and eigenmodes in detail, and after further simplifications, obtain asymptotic scaling laws. The modes resemble the large-amplitude motions qualitatively, and quantitatively in some cases.
We consider small deflections $y(x,t)$ from the straight configuration, aligned with the flow. Since the membrane stretching factor is $\partial_\alpha s\approx 1+\partial_xy^2/2$, to linear order $\alpha\approx s\approx x$, all $\alpha$-derivatives in Eq.~\eqref{eq:membrane} are $x$-derivatives, and $\zeta(\alpha,t)\approx \zeta(x,t)=x+iy(x,t)$. 
At linear order, the tangent and normal vectors are:
\begin{equation}\label{eq:linearNormal}
    \mathbf{\hat{s}}\approx(1,\partial_xy)^\top,\quad \mathbf{\hat{n}}\approx(-\partial_xy,1)^\top.
\end{equation}
The linearized version of the membrane equation [Eq.~\eqref{eq:membrane}] is 
\begin{equation}\label{eq:linearmembrane}
R_1\partial_{tt}y-T_0\partial_{xx} y = -[p].
\end{equation}
When considering the linearized problem the term in the tension force $T(\alpha,t)=T_0+R_3(\partial_\alpha s-1)$ involving~$R_3$ (dimensionless stretching rigidity) is neglected since it is of quadratic order, and so the linear dynamics are governed by the dimensionless membrane mass $R_1$ and the dimensionless pretension~$T_0$. 

The linearized conditions from Secs.~\ref{sec:rods} and~\ref{sec:verticalSprings} are:
\begin{align}
    \text{Inextensible rods:} &\quad  x(-1,t)=0,\,\, x(1,t) =2,\,\, \partial_x y(-1,t)=\frac{1}{R}y(-1,t), \,\,\partial_x y(1,t) =-\frac{1}{R}y(1,t),\label{eq:bcrods}\\  
    \text{Vertical Hookean springs:} &\quad  T_0\partial_xy(-1,t)-k_sy(-1,t)=0, \,\, -T_0\partial_x y(1,t)-k_sy(1,t)=0.\label{eq:bcsprings} 
\end{align}
We note that the boundary conditions in Eqs.~\eqref{eq:bcrods} are equivalent to Eqs.~\eqref{eq:bcsprings} with $1/R=k_s/T_0$. In~\citet{mavroyiakoumou2020large} the boundary conditions were (i) fixed-fixed: $y(\pm 1,t)=0$, (ii) fixed-free: $ y(-1,t)=0$, $\partial_x y( 1,t)=0$, and (iii) free-free: $\partial_x y(\pm 1,t)=0$.

The dynamics of the membrane are coupled to the fluid flow through the pressure jump term $[p](x,t)$. The linearized version of the pressure jump equation [Eq.~\eqref{eq:pressure}] is
\begin{equation}\label{eq:linearpressure}
\partial_t\gamma+\partial_x \gamma = \partial_x [p].
\end{equation}
The set of equations is closed by relating the vortex sheet strength $\gamma(x,t)$ back to the membrane position $y(x,t)$, through the kinematic condition [Eq.~\eqref{eq:kinematic}], which in linearized form is:
 \begin{equation}\label{eq:linearKinematic}
 \partial_t y(x,t) =-\partial_x y(x,t) + \frac{1}{2\pi}\Xint- _{-1}^1 \frac{v(x',t)}{\sqrt{1-x'^2}(x-x')}\,\d x'+\frac{1}{2\pi} \int_{1}^{\ell_w+1}\frac{\gamma(x',t)}{x-x'}\,\d x',\quad -1<x<1.
 \end{equation} 
Here, we use that $\partial_t\overline{\zeta}(x,t) \approx -i\partial_t y$ and from Eq.~\eqref{eq:linearNormal}, the normal velocity component is $\mathrm{Re}(\mathbf{\hat{n}} \partial_t\overline{\zeta})\approx \partial_t y$.
The general solution $\gamma(x,t)$ has inverse square-root singularities at $x=\pm 1$ and so we define $v(x,t)$, the bounded part of $\gamma(x,t)$, by $\gamma = v/\sqrt{1-x^2}$. The second integral in Eq.~\eqref{eq:linearKinematic} represents the velocity induced by the vortex sheet wake, which extends downstream from the membrane on the interval $1<x<\ell_w+1$, $y=0$. Therefore, the eigenvalue problem assumes a free vortex wake of a given fixed length~$\ell_w$, which we take to be large, 39 here, as in~\citet{mavroyiakoumou2021eigenmode}. In that work, we found that the modes are essentially unchanged at larger values of $\ell_w$. This long flat wake corresponds to starting with a deflection that is sufficiently small that we remain in the small-amplitude regime for large times.

The circulation in the wake,
\begin{equation}
\Gamma(x,t)=-\int_x^{\ell_w+1} \gamma (x',t)\,\d x',
\end{equation}
is conserved along material points of the wake by Kelvin's circulation theorem. At linear order, the wake moves at the constant speed (unity) of the free stream; self-interaction is negligible. 

At each time $t$, the total circulation in the wake, $\Gamma(1,t)$, is set by the Kutta condition, i.e., \begin{equation}\label{eq:kutta}
    v(1,t)=0.
\end{equation}
%Note that in linearized form it is unchanged. \silas{this was never stated in the nonlinear form, was it? If not, perhaps delete this sentence, but instead say very briefly what the equation means physically.}
%\christiana{It was not stated in the large amplitude case. So maybe we can change it to: Note that the vortex sheet strength at the trailing edge, $\gamma(1,t)$, must be finite and therefore the velocity must be zero there.}
Using the system of Eqs.~\eqref{eq:linearmembrane},~\eqref{eq:linearpressure},~\eqref{eq:linearKinematic}, and~\eqref{eq:kutta} we solve for the following unknowns: the motion of the membrane and the strength of the vortex sheets along the membrane and in the wake.

For the linearized system, we may write solutions in the following form:
% \begin{align}
% y(x,t)&=Y(x)e^{i\sigma t},\label{eq:yexp}\\
% \gamma(x,t)&=g(x)e^{i\sigma t},\label{eq:gammaaexp}\\
% v(x,t)&=V(x)e^{i\sigma t},\label{eq:vexp}\\
% \Gamma(1,t)&=\Gamma_0 e^{i\sigma t},\label{eq:Gammaexp}
% \end{align}
\begin{equation}\label{eq:eigenmodesYGammaV}
y(x,t)=Y(x)e^{i\sigma t},\quad
\gamma(x,t)=g(x)e^{i\sigma t},\quad
v(x,t)=V(x)e^{i\sigma t},\quad
\Gamma(1,t)=\Gamma_0 e^{i\sigma t},
\end{equation}
where $Y$, $g$, $V$, and $\Gamma_0$ are components of eigenmodes with complex eigenvalues $\sigma=\sigma_\Re
+ i \sigma_\Im\in \mathbb{C}$. The real parts of the eigenvalues are the angular frequencies and the imaginary parts are the temporal growth rates. If $\sigma_\Im>0$, small perturbations decay exponentially and the mode is stable, while if $\sigma_\Im<0$, small perturbations grow exponentially and the mode is unstable. If $\sigma_\Im=0$ the mode is neutrally stable.
We wish to identify the region of $R_1$--$T_0$ space in which unstable eigenmodes exist, and when there are multiple unstable modes, identify the fastest growing mode.

Since $\Gamma$ is conserved at material points of the free vortex sheet as they move downstream (at speed~1), and the material point at location $x\geq 1$ at time $t$ was at location $x=1$ at time $t-(x-1)$ we can write
\begin{align}
    \Gamma(x,t) &= \Gamma_0e^{i\sigma(t-(x-1))}=\Gamma_0e^{-i\sigma(x-1)}e^{i\sigma t},\quad 1<x<\ell_w+1,\\
    \gamma(x,t) &= \partial_x\Gamma(x,t) = -i\sigma\Gamma_0e^{-i\sigma(x-1)}e^{i\sigma t},\quad\,\,\, 1<x<\ell_w+1,
\end{align}
using $\Gamma(1,t)$ from Eq.~\eqref{eq:eigenmodesYGammaV}.
% Eq.~\eqref{eq:Gammaexp}.
Inserting the eigenmodes~\eqref{eq:eigenmodesYGammaV}
% \eqref{eq:yexp}--\eqref{eq:Gammaexp} 
into the governing Eqs.~\eqref{eq:linearmembrane} and~\eqref{eq:linearKinematic}, yields
\begin{equation}\label{eq:Ymembrane}
-\sigma^2R_1 Y=T_0\partial_{xx} Y-i\sigma \int_{-1}^1 g\,\d x-g,
\end{equation}
and
\begin{equation}\label{eq:2intB}
i\sigma Y=-\partial_x Y + \frac{1}{2\pi}\Xint- _{-1}^1
\frac{V(x')}{\sqrt{1-x'^2}(x-x')}\,\d x'-\frac{1}{2\pi}i\sigma \Gamma_0 \int_1^{\ell_w+1} \frac{e^{-i\sigma (x'-1)}}{x-x'}\,\d x', \quad -1<x<1,
\end{equation}
respectively. Because $\sigma$ appears in the exponential in the second integral in Eq.~\eqref{eq:2intB}, this is a nonlinear eigenvalue problem. 
We solve the nonlinear eigenvalue problem iteratively by the method shown in~\ref{app:numericalEigenmodes}, the same as in~\citet{mavroyiakoumou2021eigenmode}.
%\silas{move rest of paragraph to appendix} 

% Similar to~\cite{tiomkin2017stability,nardini2018reduced}, we  classify the stability of the membrane by focusing on the eigenvalues $\sigma=\sigma_\Re+i\sigma_\Im$. Since the membrane solutions consist of $e^{i\sigma t}=e^{i\sigma_\Re t}e^{ -\sigma_\Im t}$, we have:
% \begin{enumerate}
%   \setlength\itemsep{-.2em}
%     \item $\sigma_\Im> 0$: stable,
%     \item $\sigma_\Im=0$ and $\sigma_\Re\neq 0$: neutrally-stable flutter,
%     \item $\sigma_\Im<0$ and $\sigma_\Re=0$: divergence,
%     \item $\sigma_\Im<0$ and $\sigma_\Re\neq 0$: flutter and divergence.
% \end{enumerate}

%%%%%%%%%%%%%%%%%%
\subsection{Eigenmode analysis of membranes attached to vertical Hookean springs}\label{sec:eigenmodes}

For the small-amplitude analysis we focus on membranes attached to vertical Hookean springs, equivalent to rods (shown by Eqs.~\eqref{eq:bcrods} and~\eqref{eq:bcsprings} with $1/R=k_s/T_0$). 

\begin{figure}[H]
    \centering
    \includegraphics[width=.49\textwidth]{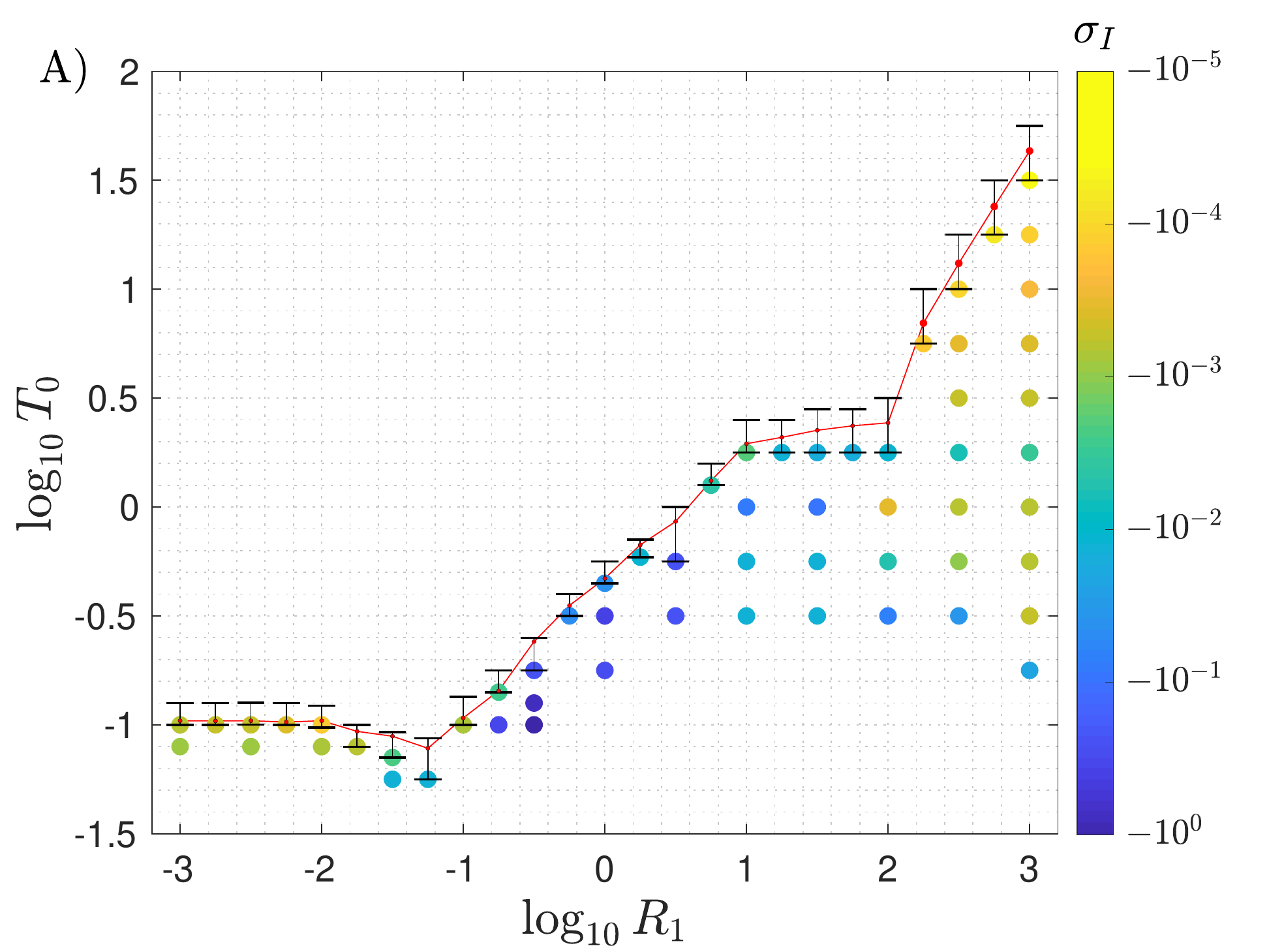}
    \includegraphics[width=.49\textwidth]{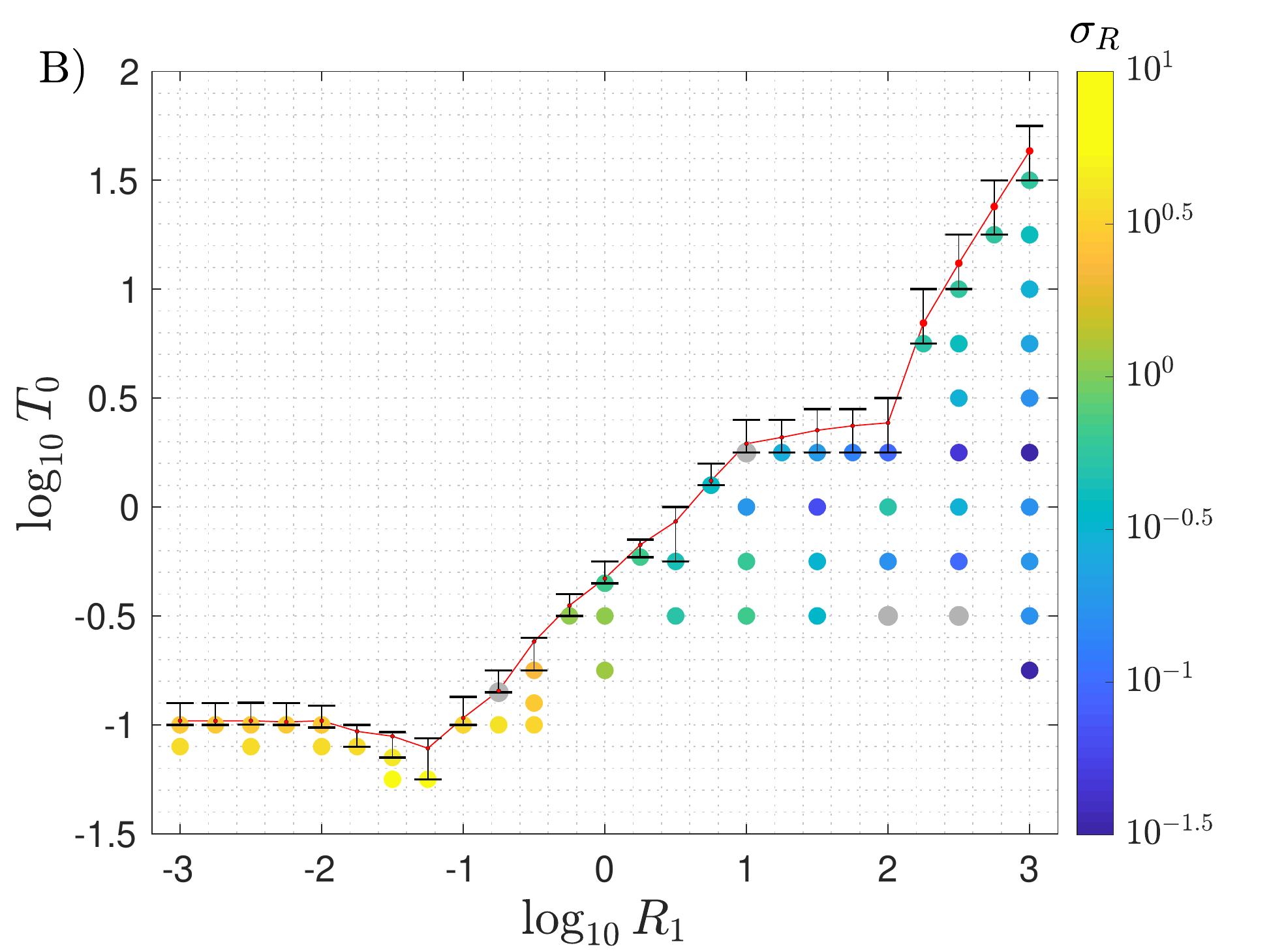}
    \caption{(Vertical springs) The region in $R_1$--$T_0$ space in which membranes are unstable. The springs attached at the leading and trailing edges of the membrane have spring constant $k_s=10^{-1}$. The red line and red dots indicate the position of the stability boundary computed using linear interpolation between $\sigma_\Im$ of the smallest $T_0$ that gives a stable membrane and the $\sigma_\Im$ of the largest~$T_0$ that gives an unstable membrane (shown in the error bars). The color of the dots below the stability boundary labels: A) The imaginary part of the eigenvalue ($\sigma_\Im$) corresponding to the most unstable modes. It represents the temporal growth rate. B) The real part of the eigenvalues ($\sigma_\Re$) for the most unstable mode, representing the angular frequency. The gray dots correspond to modes that lose stability by divergence and have $\sigma_\Re\leq 10^{-9}$.}
    \label{fig:scatterksMinus1}
\end{figure}
In Fig.~\ref{fig:scatterksMinus1} we plot the imaginary (Fig.~\ref{fig:scatterksMinus1}A) and real parts (Fig.~\ref{fig:scatterksMinus1}B) of the most unstable eigenvalues in the
region of instability for membranes attached to springs with spring constant $k_s=10^{-1}$ in $R_1$--$T_0$ space. The red line marks the boundary where the
eigenvalues change from all $\sigma_\Im>0$ (stable membranes) to at least one  $\sigma_\Im<0$ (unstable membranes). The stability boundary moves to larger pretension ($T_0$) values with
increasing membrane mass ($R_1$), starting at $R_1 = 10^{-1.25}$. As $R_1$ decreases below $10^{-1.75}$
the critical pretension reaches a lower plateau.
%As in~\citet{mavroyiakoumou2021eigenmode}, we plot the stability boundary as the red dots connected by red lines in Figs.~\ref{fig:scatterksMinus1}A and~\ref{fig:scatterksMinus1}B. 
Below and to the right of the red line is the unstable region. The red dots that mark the stability boundary are computed by linear interpolation of $\sigma_{\Im}$ between neighboring $T_0$ values (shown by the horizontal black bars) that bracket the boundary: all $\sigma_\Im$ are positive at the larger of the $T_0$ values and above, but one $\sigma_\Im$ is negative at the smaller of the $T_0$ values.
The four gray dots in Fig.~\ref{fig:scatterksMinus1}B indicate
negative $\sigma_\Im$ and nearly zero $\sigma_\Re$ ($\sigma_\Re\leq 10^{-9}$) for the most unstable eigenmode, which corresponds to divergence without flutter; they occur at $(R_1,T_0)=(10^{-0.75},10^{-0.85})$, $(10^1,10^{0.25})$, $(10^{2},10^{-0.5})$, and $(10^{2.5},10^{-0.5})$.
The colored dots in Fig.~\ref{fig:scatterksMinus1}B indicate a nonzero real part (value in color bar at right) for the most unstable
eigenmode, corresponding to 
flutter and divergence.

\begin{figure}[H]
    \centering
    \includegraphics[width=.8\textwidth]{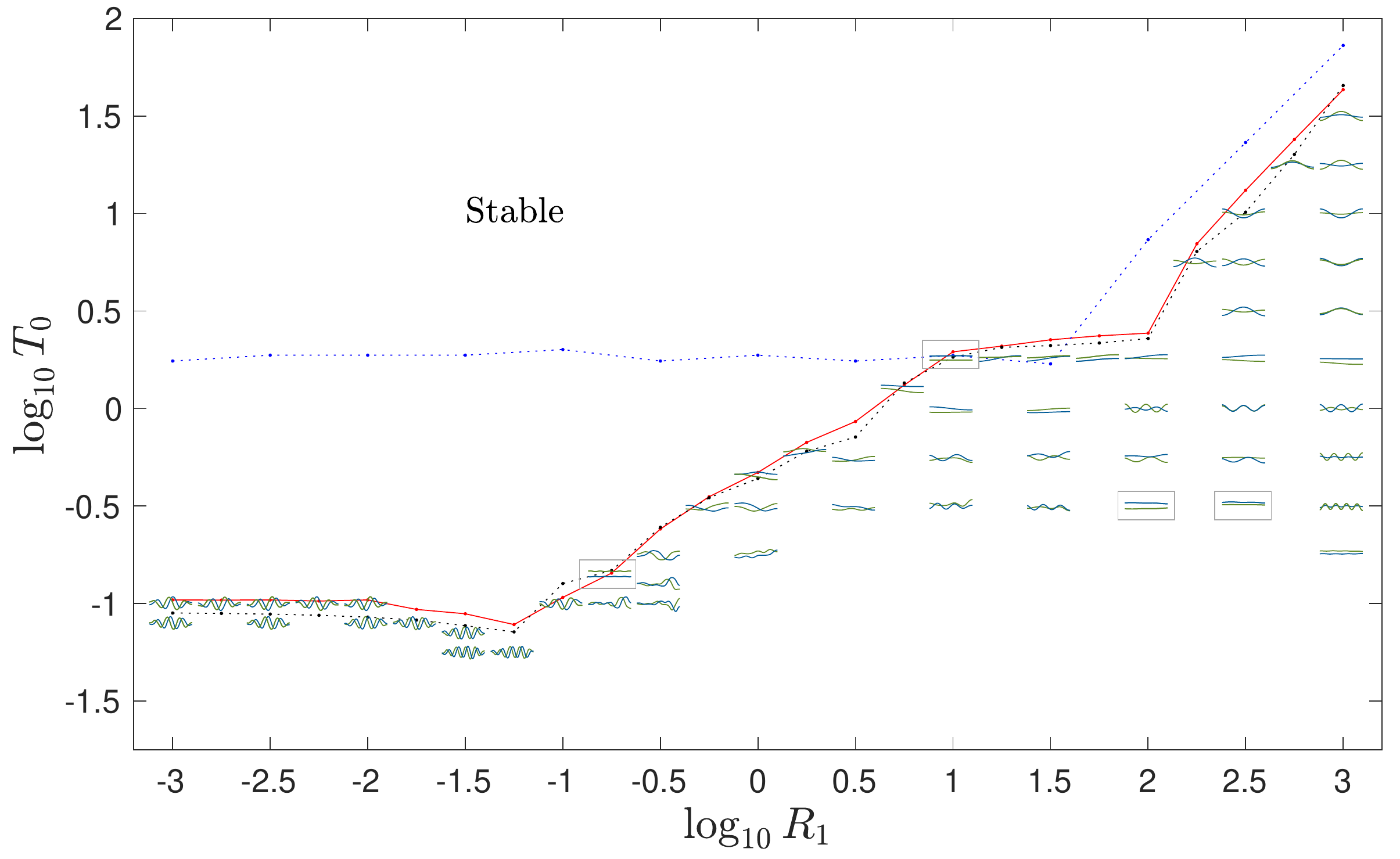}
    \caption{(Vertical springs) The shapes $Y(x)$ of the most unstable eigenmode as a function of $R_1$ and $T_0$ with springs that have a spring stiffness of $k_s=10^{-1}$. The real part of $Y(x)$ is shown in green and the imaginary part of $Y(x)$ is shown in blue. Each shape is scaled, both vertically and horizontally, to fit within the plot. Modes exhibiting a divergence instability have a gray rectangle outline. The shapes are superposed on the same stability boundary (red line) as in Fig.~\ref{fig:scatterksMinus1}. The blue dotted line represents the stability boundary for fixed-fixed membranes and the black dotted line represents the stability boundary for free-free membranes from~\citet{mavroyiakoumou2021eigenmode}. We include them here for comparison.}
    \label{fig:stabksMinus1}
\end{figure}

In Fig.~\ref{fig:stabksMinus1} we examine the variations in the most unstable eigenmodes in the same $(R_1,T_0)$ space as Fig.~\ref{fig:scatterksMinus1}, corresponding to the eigenvalues shown there. We also include our results from~\citet[Figs.~5 and~13]{mavroyiakoumou2021eigenmode} for the stability boundary when both ends of the membrane are fixed (dotted blue line) and when both ends of the membrane are free (dotted black line). The real part of the eigenmode $Y(x)$ is shown in green and the imaginary part of $Y(x)$ is shown in blue. We place gray rectangles around the modes that lose stability by divergence.
The shapes do not change noticeably for the wavier motions at $R_1\in[10^{-3},10^{-2}]$. At these small $R_1$ values the deflection at the trailing edge is nearly zero. With $R_1$ increased to $(10^{-1},10^{-0.25})$, however, the maximum deflection occurs at the trailing edge of the membrane in most cases. Here and at some larger values of~$R_1$, the mean slope of the membrane is nonzero.
 When $R_1\in[10^{1.25},10^3]$ and $T_0=10^{0.25}$ the modes are nearly alike and their growth rates $(\sigma_{\Im}$, Fig.~\ref{fig:scatterksMinus1}A) and angular frequencies  $(\sigma_{\Re}$, Fig.~\ref{fig:scatterksMinus1}B) are almost equal. 
 %At moderate values of $R_1$ [i.e.,\ $(10^{-1},10^{-0.25})$] the maximum deflection occurs in most cases at the trailing edge of the membrane. 
 
 In the limit $R_1,T_0\gg 1$, the fluid pressure is negligible and the linearized membrane equation reduces to the homogeneous wave equation 
\begin{equation}
    R_1\partial_{tt}y-T_0\partial_{xx}y=0,
\end{equation}
which after substituting 
the form of $y(x,t)$ from Eq.~\eqref{eq:eigenmodesYGammaV}
% Eq.~\eqref{eq:yexp}
becomes
\begin{equation}
    -\sigma^2R_1Y-T_0\partial_{xx}Y=0.
\end{equation}
The eigenmodes are combinations of $\cos(kx)$ and $\sin(kx)$, with $k=\pm \sigma\sqrt{R_1  /T_0}$, satisfying the two boundary conditions in Eqs.~\eqref{eq:bcsprings}. We find $k$ by determining where the determinant of the matrix
\begin{equation}\label{eq:determinant}
\begin{pmatrix}
-kT_0\sin(-k)-k_s\cos(-k) & kT_0\cos(-k)-k_s\sin(-k)\\
kT_0\sin(k)-k_s\cos(k) & -kT_0\cos(k)-k_s\sin(k)
\end{pmatrix}
\end{equation}
vanishes, which occurs if $k\sin(k)-(k_s/T_0)\cos(k)=0$ or $k\cos(k)+(k_s/T_0)\sin(k)=0$. 
%To solve for $k$ ($k\neq 0$), we have to fix the spring constant $k_s$ and the pretension parameter $T_0$. 
The numerical solutions of these two nonlinear equations for $k_s=10^{-1}$ and $T_0=10^1$ are:
\begin{equation}\label{eq:kVaccum}
    k=0.0998,\,\,
    1.5771,\,\,
    3.1448,\,\,
    4.7145,\,\,
    6.2848,\,\,
    7.8553,\,\,
    9.4258,\,\,
   10.9965,\,\,
   12.5672.
\end{equation}
The eigenmodes are given by
\begin{equation}\label{eq:eigenmodeks}
    Y(x)=\cos(k(x+1))+\left(\frac{k_s}{T_0}\right)\frac{1}{k}\sin(k(x+1)),
\end{equation}
with $k$ from Eq.~\eqref{eq:kVaccum}, for $-1\leq x\leq 1$.
%, and an arbitrary amplitude.
Heavy membranes $(R_1>10^2)$ with $T_0$ between $10^{0.25}$ and $T_{0C}(R_1)$ (i.e., the stability boundary) all lose stability with the third mode, $k=3.1448$ in Eq.~\eqref{eq:eigenmodeks}.

\begin{figure}[H]
    \centering
    \includegraphics[width=.49\textwidth]{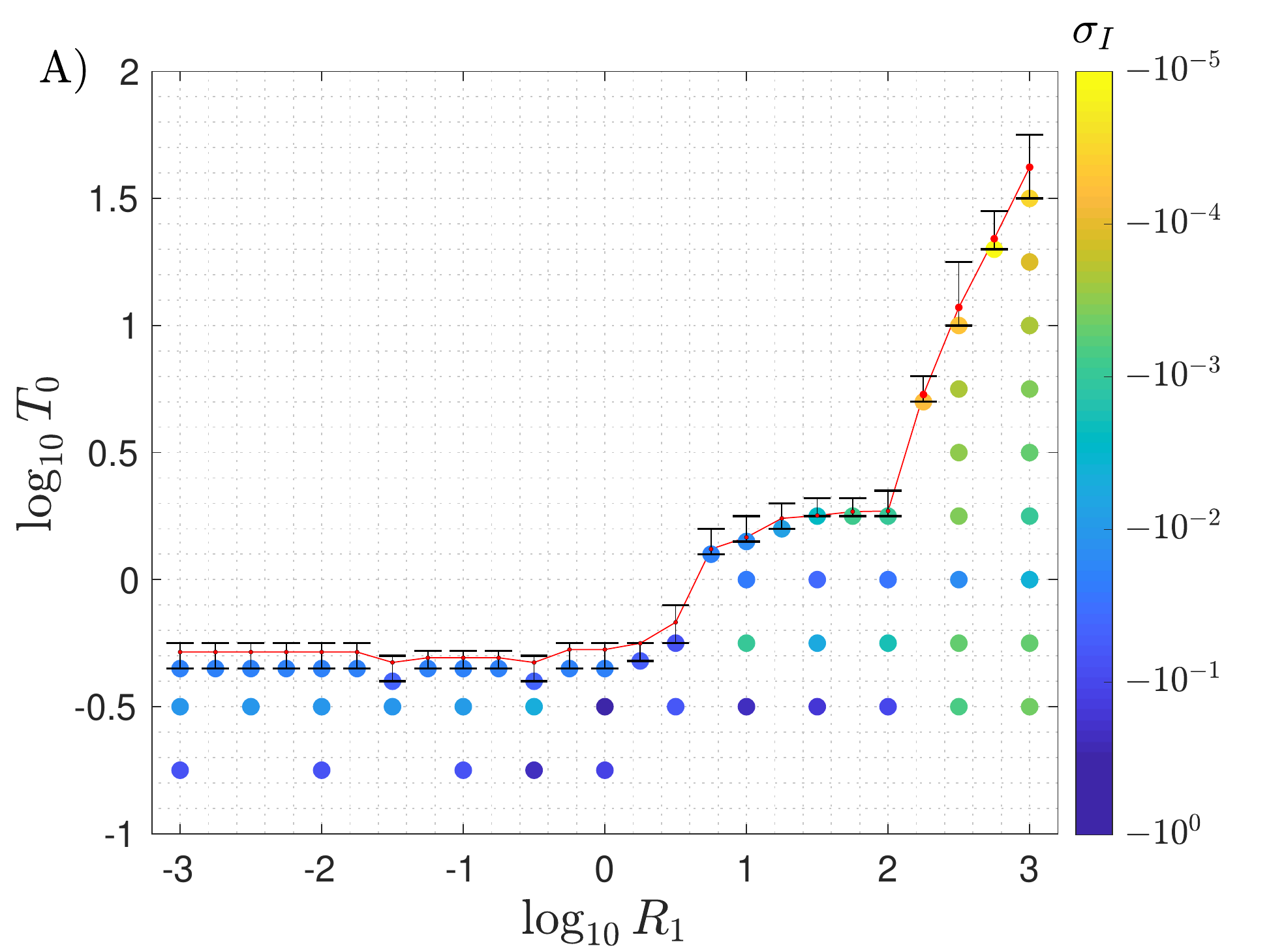}
    \includegraphics[width=.49\textwidth]{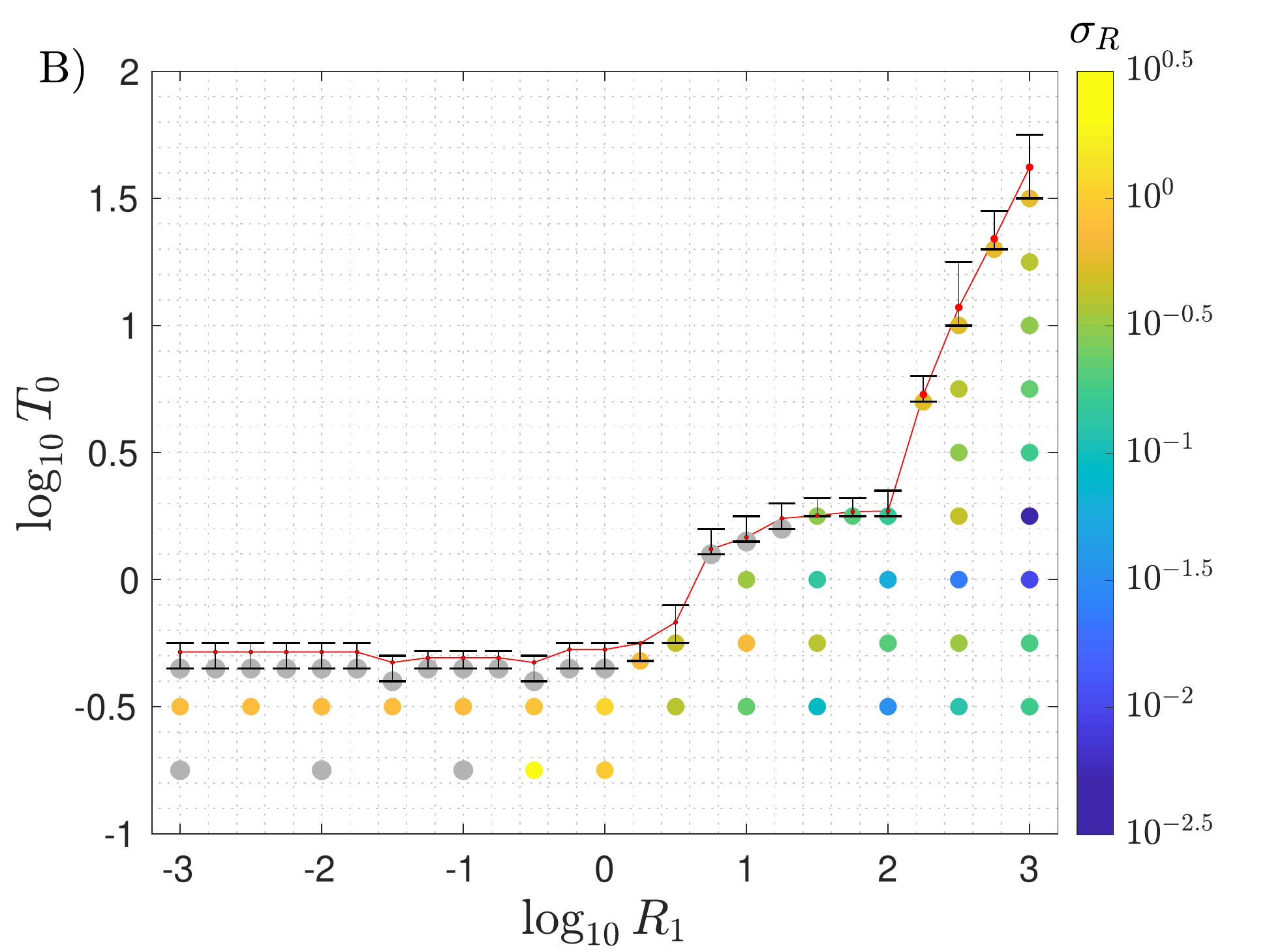}
    \caption{(Vertical springs) Same as Fig.~\ref{fig:scatterksMinus1} but with $k_s=10^{0}$.}
    \label{fig:scatterks0}
\end{figure}

We now consider the analogous results when the Hookean spring constant is increased to $k_s=10^0$. The stability boundary is shown as the red dots connected by red lines in Figs.~\ref{fig:scatterks0}A and~\ref{fig:scatterks0}B. 
%Below and to the right are the unstable membranes.
%In Fig.~\ref{fig:scatterks0} we plot the imaginary (Fig.~\ref{fig:scatterks0}A) and real parts (Fig.~\ref{fig:scatterks0}B) of the most unstable eigenvalues in the
%region of instability for membranes attached to springs with spring constant $k_s=10^{0}$ in $R_1$--$T_0$ space. 
As with $k_s=10^{-1}$, the stability boundary moves to larger pretension~($T_0$) values with
increasing membrane mass~($R_1$), starting at $R_1=10^2$. Now the critical pretension reaches a lower plateau at $R_1$ = $10^0$ and below.
The gray dots in Fig.~\ref{fig:scatterks0}B again indicate divergence without flutter 
(negative $\sigma_\Im$ and nearly zero $\sigma_\Re$ ($\leq 10^{-9}$) for the most unstable eigenmode). We observe this for all $R_1\leq 10^0$ and $R_1\in[10^{0.75},10^{1.25}]$ close to the stability boundary, as well as for $R_1\in[10^{-3},10^{-1}]$ with $T_0=10^{-0.75}$.
%The colored dots in Fig.~\ref{fig:scatterks0}B indicate a nonzero real part (value in color bar at right) for the most unstable
%eigenmode, corresponding to 
%flutter and divergence. 
Therefore, an increase in the spring stiffness not only changes the location and shape of the stability boundary but also leads to more instances of the divergence instability.
\begin{figure}[H]
    \centering
    \includegraphics[width=.75\textwidth]{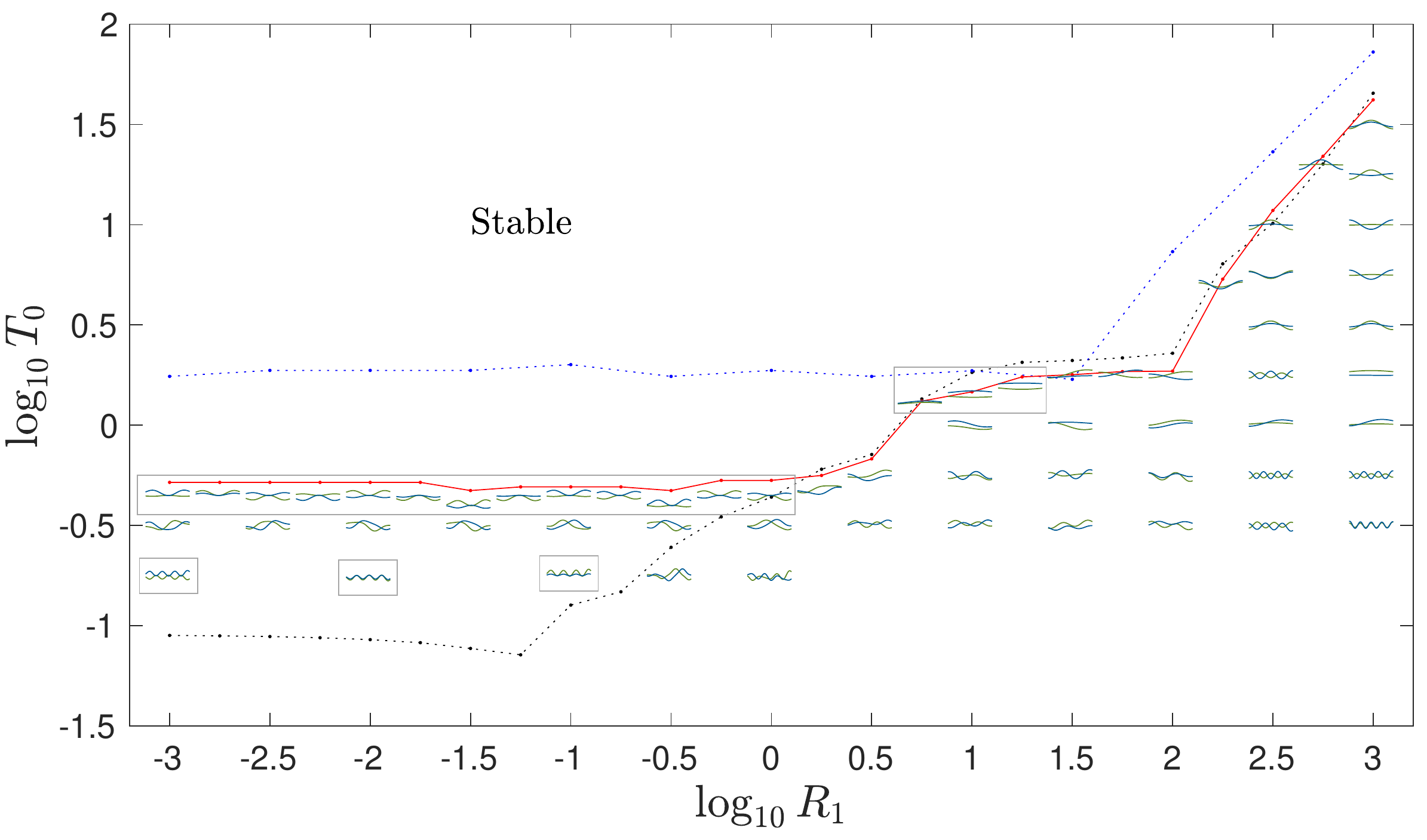}
    \caption{(Vertical springs) Same as Fig.~\ref{fig:stabksMinus1} but with $k_s=10^0$.}
    \label{fig:stabks0}
\end{figure}

The corresponding eigenmodes are shown in Fig.~\ref{fig:stabks0}. The critical pretension for $R_1<10^1$ is larger for $k_s=10^0$ than for $k_s=10^{-1}$ and lies almost midway between the stability boundary for fixed-fixed membranes ($k_s\to \infty$, blue dotted line) and for free-free membranes ($k_s=0$, black dotted line).
The mode shapes of light membranes $R_1\leq 10^0$ close to the stability boundary have three extrema and are mostly symmetric.
%nearly zero deflection at the ends
%\silas{doesn't look like zero to me--the real and imag values are different, so they can't both be zero}   \silas{what are turning points? inflection points? I don't see three}. 
The shapes do not vary noticeably with $R_1$ at these $R_1$ values. The eigenvalues in Fig.~\ref{fig:scatterks0} were also nearly constant in this region for fixed $T_0$. In general, as $T_0$ decreases the most unstable mode changes to a ``wavier'' profile. However, there are exceptions: the membrane modes at $R_1\geq 10^1$ and $T_0=10^0$ all have a similar shape (small but nonzero mean slope) but the associated eigenvalues vary more significantly there, as can be seen from Fig.~\ref{fig:scatterks0}.

Using $k_s=10^0$ and $T_0=10^{1}$ we have that the determinant of Eq.~\eqref{eq:determinant} vanishes when 
\begin{equation}\label{eq:kVaccumks0}
    k= 0.3111,\,\,
    1.6320,\,\,
    3.1731,\,\,
    4.7335,\,\,
    6.2991,\,\,
    7.8667,\,\,
    9.4354,\,\,
   11.0047,\,\,
   12.5743.
\end{equation}
 When the mass density is between $10^{0.75}$ and $10^2$ (especially close to the boundary), the membranes are similar in shape to those with $k_s=10^{-1}$. The modes for heavy membranes $(R_1>10^2)$, with~$T_0$ between $10^{0.5}$ and $T_{0C}(R_1)$, all lose stability again with the third mode, $k=3.1731$ in Eq.~\eqref{eq:eigenmodeks}.

\begin{figure}[H]
    \centering
    \includegraphics[width=.49\textwidth]{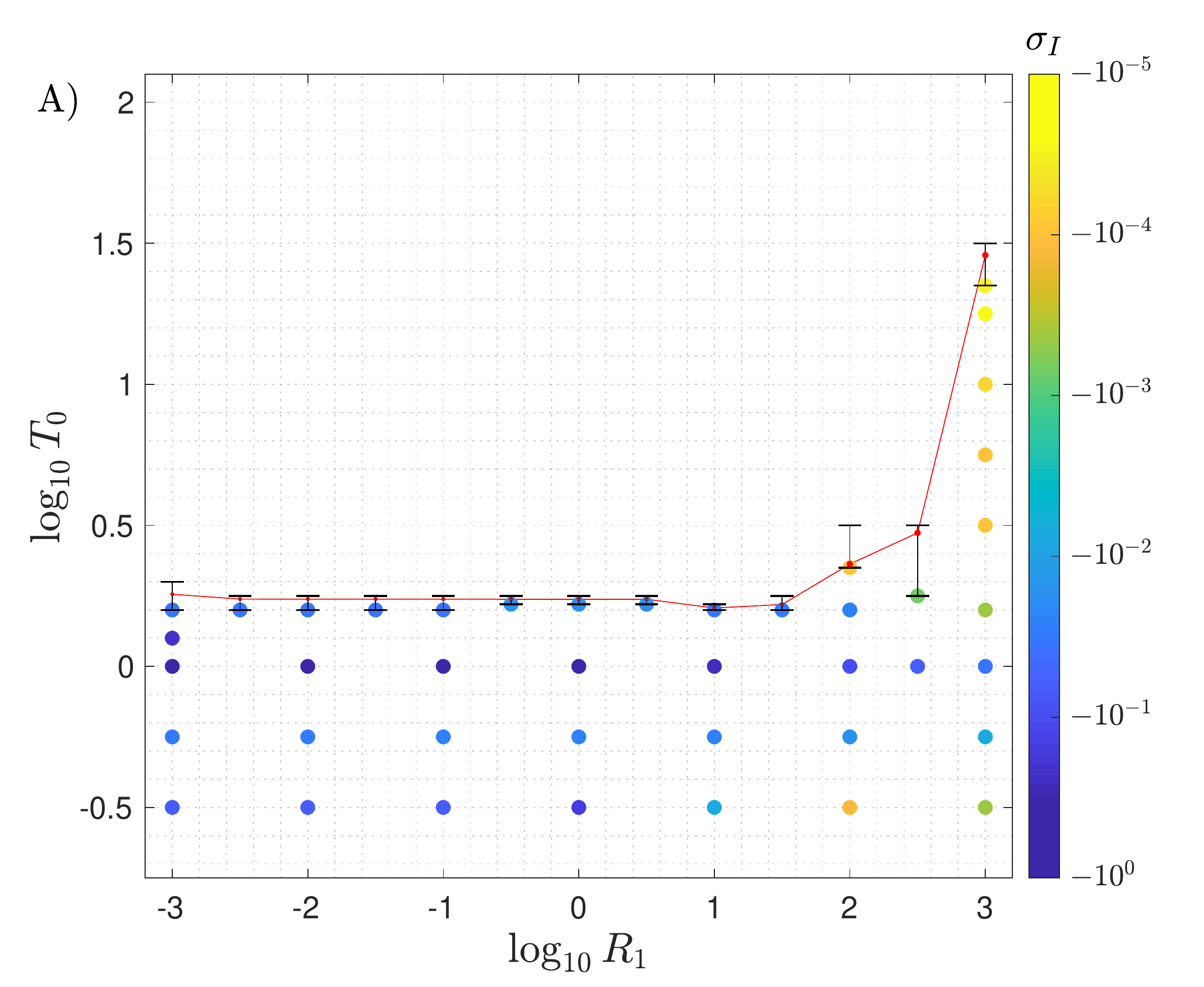}
    \includegraphics[width=.5\textwidth]{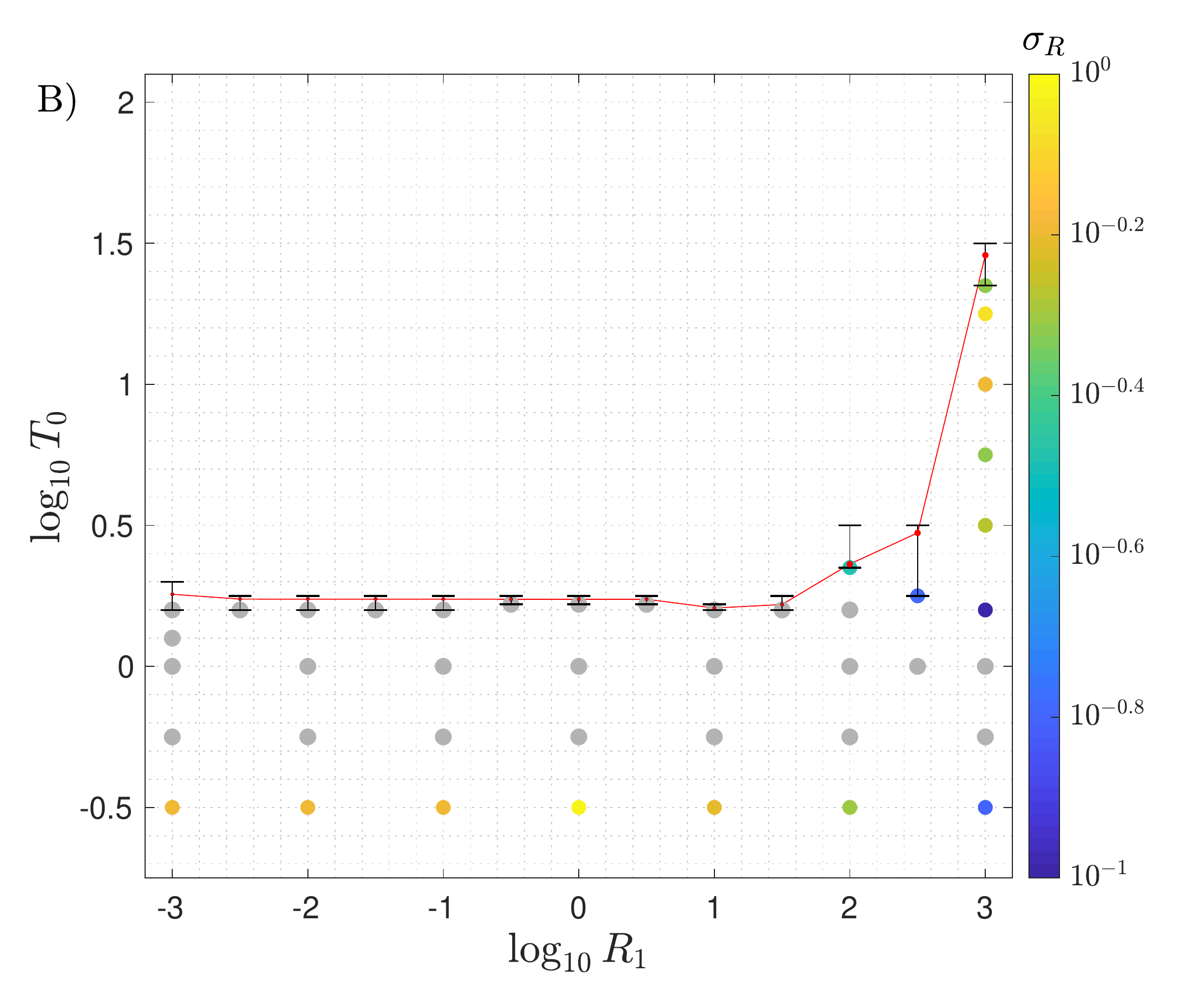}
    \caption{(Vertical springs) Same as Fig.~\ref{fig:scatterksMinus1} but with $k_s=10^{1}$.}
    \label{fig:scatterks10}
\end{figure}
\begin{figure}[H]
    \centering
    \includegraphics[width=.8\textwidth]{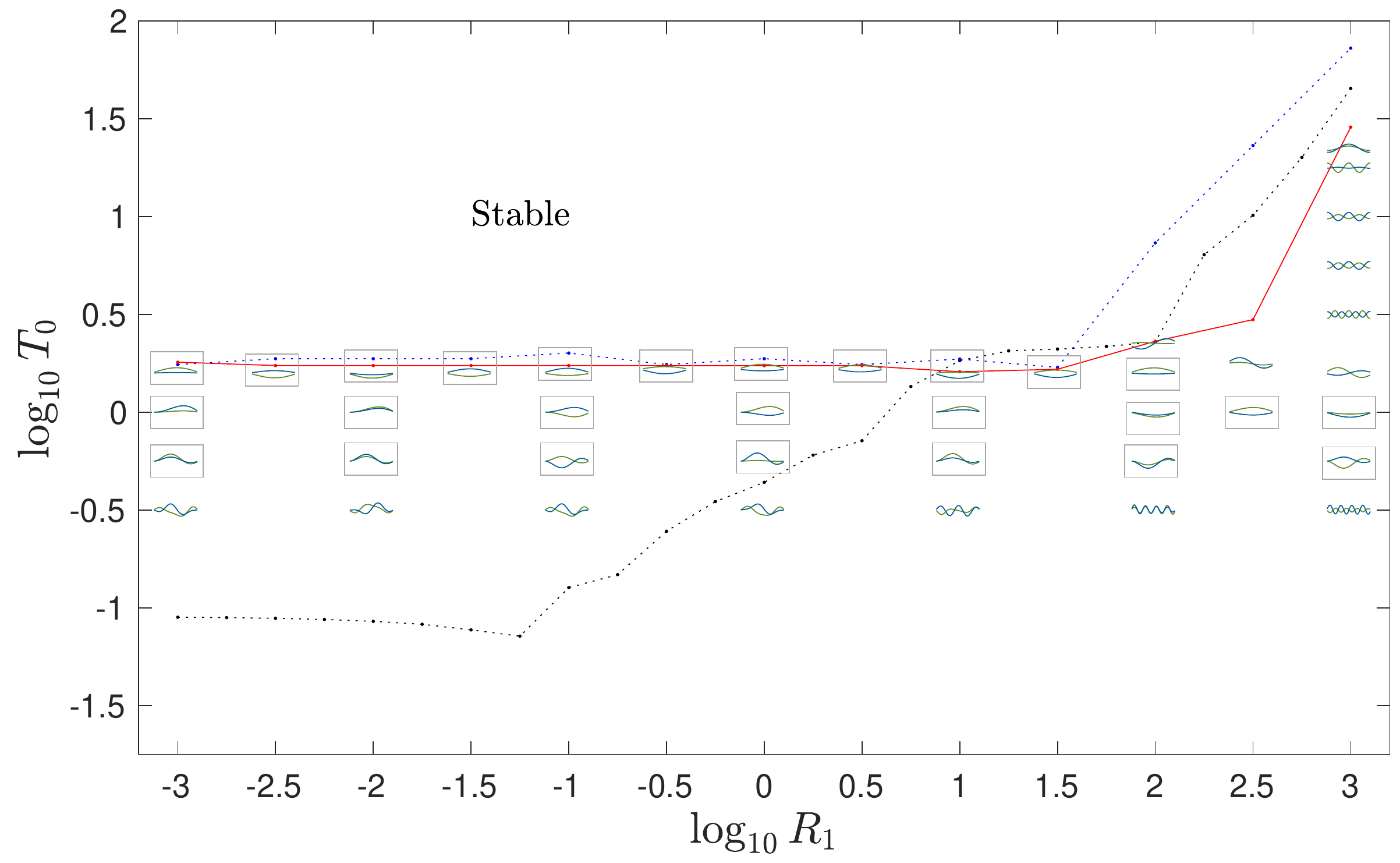}
    \caption{(Vertical springs) Same as Fig.~\ref{fig:stabksMinus1} but with $k_s=10^1$.}
    \label{fig:stabks10}
\end{figure}

Increasing $k_s$ further to $10^1$ we approach the small-amplitude dynamics of a membrane whose edges are both fixed at zero deflection. In Fig.~\ref{fig:scatterks10} the colored dots give the imaginary (Fig.~\ref{fig:scatterks10}A) and real parts (Fig.~\ref{fig:scatterks10}B) of the most unstable eigenvalues (with corresponding eigenmodes shown later, in Fig.~\ref{fig:stabks10}). There are now many more cases of divergence without flutter (gray dots in Fig.~\ref{fig:scatterks10}B).
%indicate membranes with nearly zero $\sigma_{\Re}$ ($\leq 10^{-9}$) and so those membranes lose stability by divergence. 
At $T_0=10^{-0.5}$, divergence with flutter occurs (colored dots in Fig.~\ref{fig:scatterks10}B).
%the membranes lose stability by flutter and divergence ($\sigma_{\Im}<0$ in Fig.~\ref{fig:scatterks10}A, $\sigma_{\Re}\neq 0$ in Fig.~\ref{fig:scatterks10}B).

In Fig.~\ref{fig:stabks10} we see that the critical pretension for $R_1\leq 10^{1.5}$ is almost the same as the stability boundary for fixed-fixed membranes ($k_s\to \infty$, blue dotted line in Fig.~\ref{fig:stabks0}). We place gray rectangles around the modes that lose stability by divergence. The shapes are also similar to the ones seen for a fixed-fixed membrane: For $R_1<10^2$ and $T_0$ just below $T_{0C}$, the unstable eigenmode is a single-hump shape that is nearly fore-aft symmetric. As the pretension is decreased further below $T_{0C}$ (at $T_0=10^0$ and $R_1<10^{2.5}$) the divergent eigenmode becomes asymmetric, its maximum deflection point shifting towards the trailing edge. As the membrane mass $(R_1)$ is increased to $10^{2.5}$, the maximum camber point moves towards the midchord and the membrane shape becomes almost fore-aft symmetric. At a smaller $T_0$ ($10^{-0.25}$) the membranes still lose stability by divergence but there is now an inflection point approximately at the membrane's midchord, with the maximum point on the membrane being closer to the fore part. Below $T_0=10^{-0.25}$ membranes lose stability by divergence with flutter, with the shapes varying more with $R_1$ now, especially when $R_1\geq 10^1$. Even though the membrane mode shapes generally look very similar to the fixed-fixed membranes in~\citet[Fig.~5]{mavroyiakoumou2021eigenmode} when $k_s = 10^1$, this is not the case for $R_1>10^{2.5}$. The critical pretension in Fig.~\ref{fig:stabks10} starts to increase when $R_1 \gtrsim 10^{2.5}$ as opposed to a smaller mass, i.e., $R_1\gtrsim 10^{1.5}$ for fixed-fixed membranes, and the mode shapes there are also very different.
As for the other $k_s$ values, here we use $k_s=10^1$ and for a fixed value of $T_0$ determine the value of $k$ such that the determinant of Eq.~\eqref{eq:determinant} is equal to zero. In Fig.~\ref{fig:stabks10} the first membrane that becomes unstable just below the stability boundary (at $T_0=10^{1.35}$) is approximately the third sinusoidal mode ($k=3.28$). At $T_0=10^1$ and $10^{0.5}$ the most unstable modes are approximately the fifth and seventh sinusoidal modes ($k=6.44$ and 9.74, respectively). The trend of odd-numbered modes does not continue when $T_0<10^{0.5}$.
% \begin{equation}
%     k=  0, 0.8603, 2.0288,3.4256,4.9132,6.4373,7.9787,9.5293,11.0855,12.6453.
% \end{equation}
%Finally, further decreasing $T_0$ to  the most unstable mode corresponds to $k=9.7387$ which is the seventh mode.
% \begin{equation}
%     k=0,1.2064,2.4771,3.8316,5.2542,6.7229,8.2212,9.7387,11.2692,12.8084.
% \end{equation}

To summarize, we have found that the stability boundary has an upward slope
for large $R_1$, whereas for small-to-moderate~$R_1$ values, the critical $T_0$ is smaller. At small $R_1$ the critical pretension for instability reaches a plateau value that depends on the spring stiffness. When~$R_1$ and~$T_0$ are dominant over fluid pressure forces, the membrane eigenmodes tend to neutrally-stable
sinusoidal functions. Increasing the spring stiffness $k_s$ introduces more divergence instabilities, in agreement with the fixed-fixed case studied in~\citet{mavroyiakoumou2021eigenmode}. In general, the most unstable modes become more wavy at smaller~$T_0$. The nonlinear eigenvalue problem for the linearized membrane model has allowed us to
extend results from the large-amplitude model in Sec.~\ref{sec:largeAmplitudeResults} to a wider range of $R_1$--$T_0$ space. Next, we study a more analytically tractable model---that of an infinite, periodic array of springs attached to an infinite membrane. This model allows us to compute solutions for a much wider range of parameters and obtain asymptotic scaling laws.
%%%%%%%%%%%%%%%%%%%%%%%%%%%%%%%%%%%%

\section{Periodic array of springs on an infinite membrane}\label{sec:array}

\begin{figure}[H]
    \centering
    \includegraphics[width=.85\textwidth]{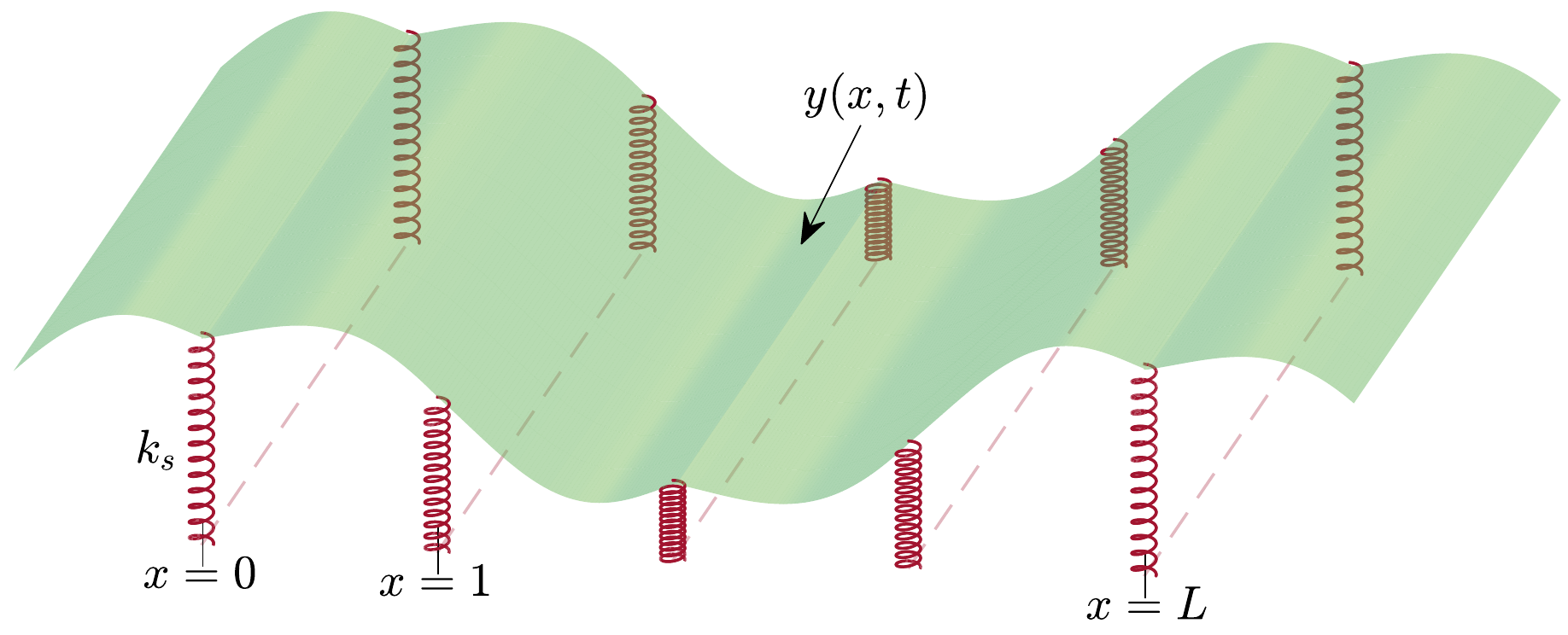}
    \caption{Schematic diagram of a section of an infinite, flexible membrane (green surface) at an instant in time. Here~$L$ is the $x$-period of the membrane, $y(x,t)$ is the membrane deflection and the red springs of stiffness $k_s$ are spaced one unit apart. The distance between springs is smaller than the membrane's period ($L>1$, $L\in \mathbb{N}$).}
    \label{fig:schemPeriodic3dView}
\end{figure}

We have seen that the eigenvalue problem for a membrane tethered with springs (or rods) interpolates between the fixed-fixed and free-free cases. The vortex sheet wake results in a nonlinear eigenvalue problem, requiring an iterative solver that is time-consuming, particularly at small $T_0$. We now consider a simplified model with spatially periodic solutions that will allow us to derive asymptotic scaling laws. We assume the membrane extends to infinity upstream and downstream, and is tethered by  
an infinite, periodic array of Hookean springs (with stiffness $k_s$). The horizontal spacing between the springs (unity) is analogous to the length of the finite membrane in the previous section.  
%springs attached to the lower surface of the membrane are spaced apart by one in dimensionless terms. 
This problem is shown schematically in Fig.~\ref{fig:schemPeriodic3dView}, where the green surface represents a section of the infinite membrane at an instant in time and the pairs of red coils on either side of the membrane span represent the springs. The membrane has period $L$. By taking $L$ larger than the distance between the springs, the infinite periodic membrane may have different deflections at streamwise-adjacent spring locations, as occurs for the tethered finite membrane. As $L$ increases, the membrane can assume a wider range of shapes, but the eigenvalue problem becomes more costly to solve. We choose $L = 4$ as a compromise between these competing considerations.
The flow velocity is again uniform at infinity (far above and below the membrane). With an infinite membrane there is no free vortex wake, and the nonlinear eigenvalue problem is reduced to a quadratic eigenvalue problem, which has analytic solutions for the eigenvalues when $k_s = 0$. 
%By replacing the vortex wake with more , does not include a vortex wake which is a significant approximation. However, in cases where the membrane has small deflection at the trailing edge, we think this is sufficient.  
In~\citet{newman1991stability} a related approximate model was considered---an infinite membrane with two- and three-harmonic truncations that were used to approximate fixed-fixed boundary conditions.
%for standing sine waves since the authors consider a membrane that has fixed ends. %.8
% \begin{figure}[H]
%     \centering
%     \includegraphics[width=.8\textwidth]{Figures/schemArray.pdf}
%     \caption{Schematic representation of an extensible membrane attached to an infinite array of springs, of constant spring stiffness $k_s$, that are spaced apart by one unit.}
%     \label{fig:schematicArray}
% \end{figure}
% Since in this model there is a vertical point forcing from the periodic array of springs, we have a periodic delta function in the membrane equation.
The system of governing equations is:
\begin{align}
R_1\partial_{tt}y-T_0\partial_{xx}y&=-[p]-k_sy(x,t)\delta_1(x),\label{eq:infiniteMembraneDisL}\\
    \partial_t y+\partial_x y&=\frac{1}{2\pi}\int_{-\infty}^{\infty}\frac{\gamma(x',t)}{x-x'}\,\text{d}x',\label{eq:infiniteKinematic}\\
    \partial_{t}\gamma+\partial_x\gamma&=\partial_x[p].\label{eq:infinitePressure}
\end{align}
In Eq.~\eqref{eq:infiniteMembraneDisL}, $\delta_1(x)$ is a periodic Dirac delta function with period one, resulting in a spring force at each integer~$x$, and proportional to $y(x,t)$, the vertical deflection there. 
We next write the membrane position, vortex sheet strength, and pressure jump across the membrane, each as a Fourier series with period~$L$, and the periodic Dirac delta function as a Fourier series with period one:
\begin{align}
    y(x,t)=\sum_{k=-\infty}^\infty \hat{y}_ke^{i(2\pi k/L) x}e^{i\sigma t},\quad
    \gamma(x,t)=\sum_{k=-\infty}^\infty \hat{\gamma}_ke^{i(2\pi k/L) x}e^{i\sigma t},\label{eq:YandGammaFourierL}\\
    [p](x,t)=\sum_{k=-\infty}^\infty \widehat{[p]}_ke^{i(2\pi k/L) x}e^{i\sigma t},\quad\delta_1(x) = \sum_{k=-\infty}^\infty e^{i(2\pi k) x},\label{eq:pandDeltaFourierL}
\end{align}
respectively, where $\hat{y}_k$, $\hat{\gamma}_k$, $\widehat{[p]}_k$ are complex Fourier coefficients to be found. 
%To ensure that the period of the membrane  does not coincide with the period of the springs (one unit apart) we choose $L>1$, with $L\in\mathbb{N}$.

Using Eqs.~\eqref{eq:YandGammaFourierL} and~\eqref{eq:pandDeltaFourierL}, the membrane equation [Eq.~\eqref{eq:infiniteMembraneDisL}] can be written as
\begin{equation}\label{eq:membraneFourierL}
    \sum_{k=-\infty}^{\infty}\left(-\sigma^2R_1\hat{y}_k+T_0\left(\frac{2\pi k}{L}\right)^2\hat{y}_k\right)e^{i(2\pi k/L)x}=-\sum_{k=-\infty}^{\infty}\widehat{[p]}_ke^{i(2\pi k/L) x}-k_s\sum_{k'=-\infty}^{\infty}\hat{y}_{k'}e^{i(2\pi k'/L)x}\sum_{k''=-\infty}^{\infty}e^{i(2\pi k'')x},
\end{equation}
having divided throughout by the common factor $e^{i\sigma t}$. 
% Here $Y(x)e^{i\sigma t}\equiv y(x,t)$. 
Substituting Eqs.~\eqref{eq:YandGammaFourierL} into Eq.~\eqref{eq:infiniteKinematic}, we obtain
\begin{equation}
    \sum_{k=-\infty}^{\infty}\left(i\sigma+i\frac{2\pi k}{L}\right)\hat{y}_ke^{i(2\pi k/L) x}e^{i\sigma t}=\sum_{k=-\infty}^{\infty}-\frac{i}{2}\mathrm{sgn}\left(\frac{2\pi k}{L}\right)\hat{\gamma}_ke^{i(2\pi k/L) x}e^{i\sigma t},
\end{equation}
which implies that 
\begin{equation}\label{eq:fourierGammaandYL}
    i\left(\sigma+\frac{2\pi k}{L}\right)\hat{y}_k=-\frac{i}{2}\mathrm{sgn}\left(\frac{2\pi k}{L}\right)\hat{\gamma}_k.
\end{equation}
Similarly, if we substitute Eqs.~\eqref{eq:YandGammaFourierL} and~\eqref{eq:pandDeltaFourierL} into Eq.~\eqref{eq:infinitePressure}, we get
\begin{equation}\label{eq:fourierPandYL}
    i\left(\sigma+\frac{2\pi k}{L}\right)\hat{\gamma}_k=i\frac{2\pi k}{L}\widehat{[p]}_k.
\end{equation}
Using Eqs.~\eqref{eq:fourierGammaandYL} and \eqref{eq:fourierPandYL} in Eq.~\eqref{eq:infiniteKinematic} and in Eq.~\eqref{eq:infinitePressure}, we obtain
\begin{align}
    % \hat{\gamma}_k&=-2\mathrm{sgn}\left(\frac{2\pi k}{L}\right)\left(\sigma+\frac{2\pi k}{L}\right)\hat{y}_k,\\
     \hat{\gamma}_k&=-2\mathrm{sgn}\left(k\right)\left(\sigma+\frac{2\pi k}{L}\right)\hat{y}_k,\\
    \widehat{[p]}_k&=-\frac{L}{\pi|k|}\left(\sigma+\frac{2\pi k}{L}\right)^2\hat{y}_k,
\end{align}
respectively, where we use that $\mathrm{sgn}\left(2\pi k/L\right)=\mathrm{sgn}(k)$ and thus write Eq.~\eqref{eq:membraneFourierL}, in terms of  $\hat{y}_k$ only, as
\begin{align}
    \sum_{k=-\infty}^{\infty}\left(-\sigma^2R_1+T_0\left(\frac{2\pi k}{L}\right)^2\right)\hat{y}_ke^{i(2\pi k/L)x}=\sum_{k=-\infty}^{\infty}\frac{L}{\pi|k|}\left(\sigma+\frac{2\pi k}{L}\right)^2\hat{y}_ke^{i(2\pi k/L) x}\nonumber\\ -k_s\sum_{k'=-\infty}^{\infty}\hat{y}_{k'}\left(\sum_{k''=-\infty}^{\infty}  e^{i(2\pi (k''L + k')/L)x} \right).\label{eq:membraneFourierLnew}
\end{align}
% -k_sY(0)\sum_{k=-\infty}^{\infty}\hat{\delta}_{1,k}e^{i(2\pi k/L)x}
% Using the Galerkin method, we multiply Eq.~\eqref{eq:membraneFourierLnew} by $e^{-i(2\pi \bar{k}/L)x}$ and integrate from $x=0$ to $x=L$. This integration step picks out terms related to $k=\bar{k}$, as follows:
% This picks out terms related to $k=\bar{k}$, as follows:
% \begin{equation}\label{eq:membraneykL}
%     -\sigma^2R_1\hat{y}_{\bar{k}}+T_0\left(\frac{2\pi \bar{k}}{L}\right)^2\hat{y}_{\bar{k}}=\frac{L}{\pi|\bar{k}|}\left(\sigma+\frac{2\pi \bar{k}}{L}\right)^2
%     \hat{y}_{\bar{k}}-k_s\sum_{k=-\infty}^{\infty}\hat{y}_ke^{i(2\pi k/L) 0}, \quad \bar{k}=-\infty,\dots,\infty.
% \end{equation}
% The case where $k=0$ is an exception and so should be treated separately. It can be shown that $\hat{y}_0=0$ using that the Hilbert transform of a constant is equal to zero.
We match coefficients of $e^{i(2\pi k/L)x}$ in Eq.~\eqref{eq:membraneFourierLnew} and obtain
\begin{equation}\label{eq:quadraticEigL}
    \left(-R_1-\frac{L}{\pi|k|}\right)\sigma^2\hat{y}_{k}-\frac{4 k}{|k|}\sigma \hat{y}_{k}+\left(\frac{2\pi k}{L}\right)^2\left(T_0-\frac{L}{\pi |k|}\right)\hat{y}_{k}+k_s\sum_{k'\equiv k \!\!\!\!\mod L} \hat{y}_{k'}=0,
\end{equation}
for ${k}=-N,\dots,-1,1,\dots,N$. 
%Here, $\sum_{k'\equiv k\!\!\!\mod L} \hat{y}_{k'}$ for $k'=-N,\dots,N$ is a rank-$L$ matrix whose entries are either one if $k'\equiv k\mod L$ or zero, otherwise.
The last sum in Eq.~\eqref{eq:quadraticEigL} includes
those $k'$ that are equal to $k$ plus a multiple of $L$.
% \begin{equation}
%     \sum_{k=-N}^N\hat{y}_k\hat{\delta}_{1,\bar{k}-k}=\sum_{k=-N}^N\hat{y}_k\left(\sum_{k=-N}^N e^{i 2\pi (k-\bar{k})}\right)=\sum_{k=-N}^N\hat{
%     y}_{k-\bar{k}}.
% \end{equation}
If we make the truncation approximation that $\hat{y}_{k}=0$ for $|k|>N$ then Eq.~\eqref{eq:quadraticEigL} is a system of $2N+1$ equations in $2N+1$ unknowns $\hat{y}_k$. In the derivation we assumed ${k}\neq 0$. From Eq.~\eqref{eq:fourierGammaandYL} we see that $\hat{y}_0=0$ (Hilbert transform of a constant is equal to zero). Therefore, we insert 0 for $\hat{y}_0$ in the system of equations and remove $\hat{y}_0$ from the unknowns, resulting
in $2N$ equations in $2N$ unknowns.
%component corresponding to $k=0$ from the matrices $A_2$, $A_1$, and $A_0$, so instead of having size $(2N+1)\times (2N+1)$ they now have size $2N\times 2N$. When we are constructing the eigenmodes
%\begin{equation}\label{eq:trunctatedyfourierL}
%    y(x)=\sum_{k=-N}^N\hat{y}_ke^{i(2\pi k/L)x},
%\end{equation}
%we have to include the case %$\hat{y}_0=0$.

% \begin{equation}
%     \sum_{k=-\infty}^\infty \hat{y}_k\hat{\delta}_{L,\bar{k}}=\begin{bmatrix}
%     1/L & \cdots  & 1/L & 0 & 0 & 1/L & \cdots  & 1/L\\
%     \vertbar&&\vertbar&\vertbar&\vertbar&\vertbar& &\vertbar\\
%     1/L  & \cdots  & 1/L & 0 & 0 & 1/L & \cdots  & 1/L\\
%     \end{bmatrix}.
% \end{equation}
% \christiana{It's actually a rank-4 matrix so it's not looking exactly as above.}
Eq.~\eqref{eq:quadraticEigL} is a quadratic eigenvalue problem of the form 
\begin{equation}\label{eq:quadraticFourier}
    (A_2\sigma^2+A_1\sigma+A_0)\hat{\mathbf{y}}=0,
\end{equation}
where $A_2$ and $A_1$ are diagonal matrices, $A_0$ is a rank-$L$ matrix, and $\hat{\mathbf{y}}$ is the eigenvector of Fourier coefficients $\{\hat{y}_{k}, k=-N,\dots,-1,1,\dots,N\}$. Using \texttt{polyeig} in \textsc{Matlab} we solve for the eigenvalues $\sigma$ and determine the fastest growing eigenmode, i.e., corresponding to the most negative $\sigma_\Im$. 
%.444,.45,.442,.45
\begin{figure}[H]
    \centering
    \includegraphics[width=.444\textwidth]{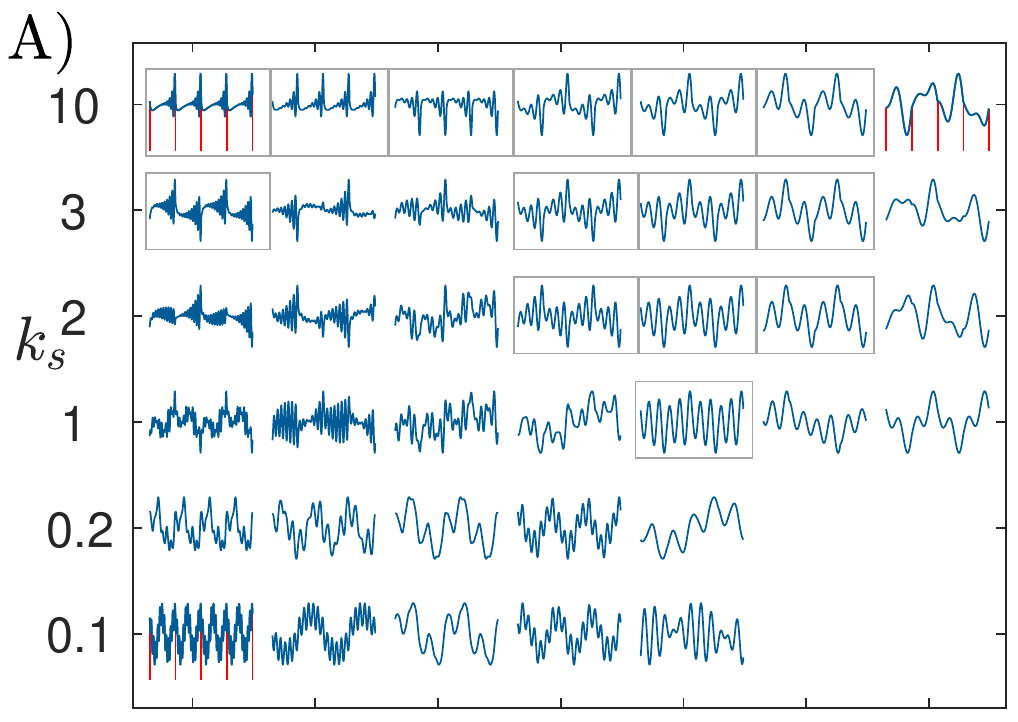}
    \includegraphics[width=.45\textwidth]{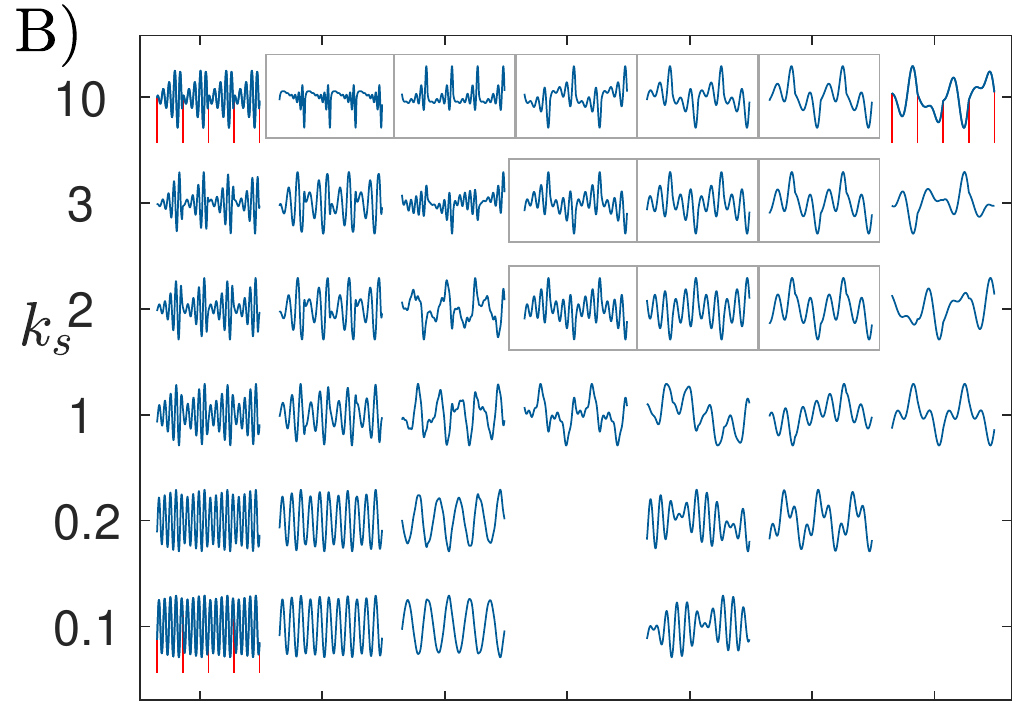}
    \includegraphics[width=.442\textwidth]{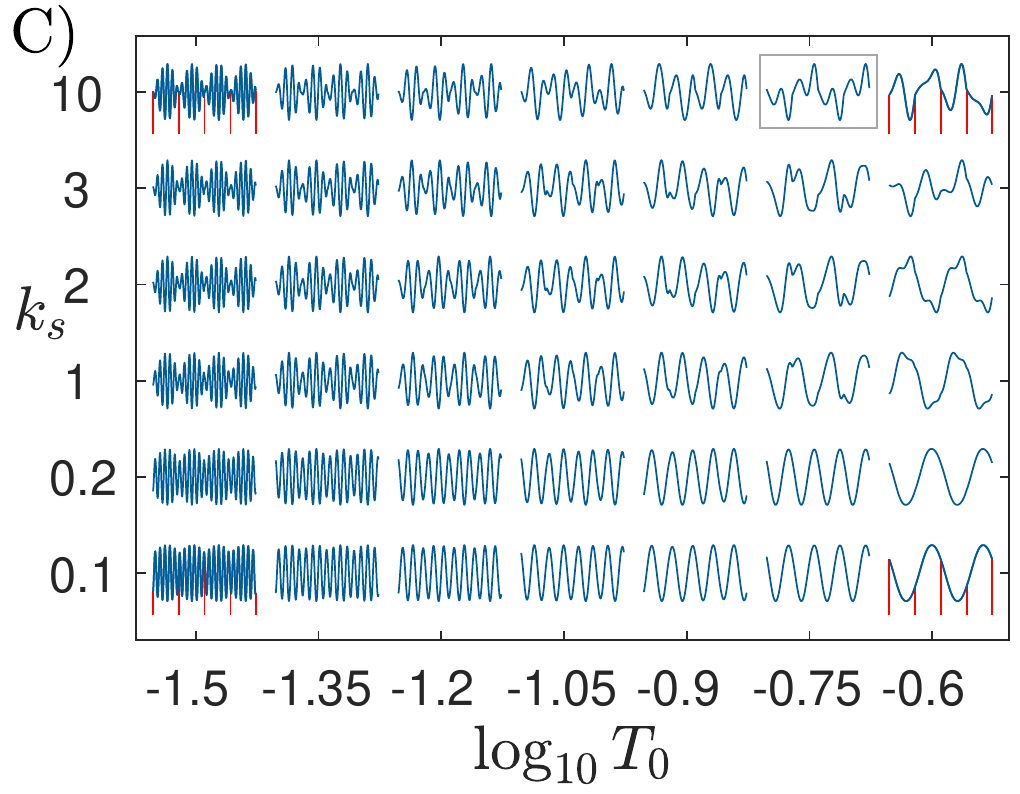}
    \includegraphics[width=.45\textwidth]{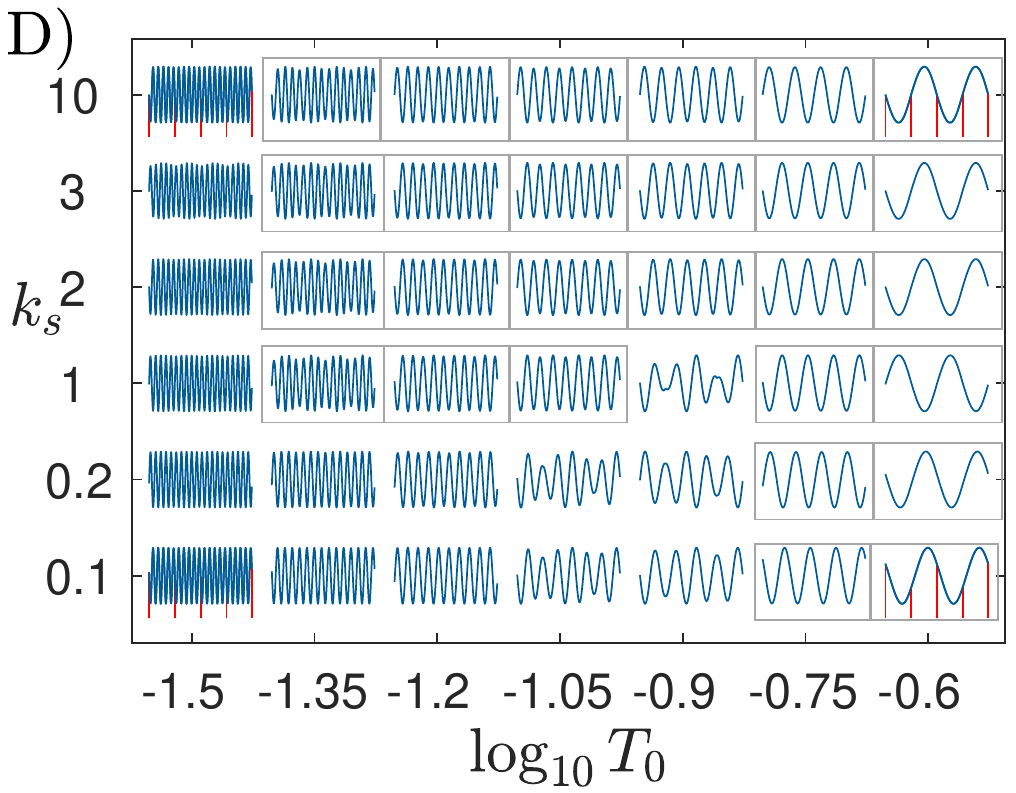}
    \caption{Imaginary part of the most unstable eigenmode [$\mathrm{Im}(y(x))$] in $T_0$--$k_s$ parameter space for A) $R_1=10^{-4}$, B) $R_1=10^{-1}$, C) $R_1=10^{0}$, and D) $R_1=10^{4}$. Modes exhibiting a divergence instability with $\sigma_\Re\leq 10^{-9}$ have a gray rectangle outline. In all the panels, we use $N=2^9$.}
    \label{fig:imagModesFixedR1}
\end{figure}

In Fig.~\ref{fig:imagModesFixedR1} we show the imaginary parts of the most unstable modes for the periodic membrane problem, over one period $0 \leq x \leq 4$, and thus with 4 subintervals between springs shown. In a few examples (at the corners) in panels A--D, we show the locations of the springs by small red lines. In many (but not all cases), the shapes seem to repeat 4 times. This is particularly true at larger $k_s$, where the springs are stronger and impose a period-1 component more strongly in the eigenmode. The real parts are similar and are omitted. Membranes that lose stability by divergence without flutter are again outlined with gray rectangles. We compute the relative error in the eigenvalues when $N=2^8$ and $2^9$:
\begin{equation}\label{eq:relativeError}
    \text{relative error}=\left|\frac{\sigma_{2^8}-\sigma_{2^9}}{\sigma_{2^9}}\right|.
\end{equation}
The maximum relative error is small for the cases in Fig.~\ref{fig:imagModesFixedR1}: $0.0437$ when $R_1=10^{-4}$ (Fig.~\ref{fig:imagModesFixedR1}A), $0.0269$ when $R_1=10^{-1}$ (Fig.~\ref{fig:imagModesFixedR1}B), $0.00267$ when $R_1=10^0$ (Fig.~\ref{fig:imagModesFixedR1}C), and $1.31\times 10^{-5}$ when $R_1=10^4$ (Fig.~\ref{fig:imagModesFixedR1}D).

%\silas{maybe it would help to add small vertical red lines (like hash marks) at the spring locations for some of the cases in Fig. 21--say at the ones in the upper left, upper right, and lower left corners. Otherwise it might be hard for the readers to see easily what is being shown here.}

The periodic membrane modes do not align precisely with those in the membrane-vortex-wake model due to the different membrane boundary conditions (periodic versus finite with a trailing vortex wake). However, there are many qualitative similarities.
In both cases, the modes become sharper (or wavier) as we decrease~$T_0$. At large~$R_1$ the membranes are more sinusoidal with single bumps between the springs at large values of~$T_0$ (in Fig.~\ref{fig:imagModesFixedR1}D for the periodic membrane). At small~$R_1$ (Figs.~\ref{fig:imagModesFixedR1}A and~\ref{fig:imagModesFixedR1}B), the membranes are less sinusoidal and less symmetric. Another similarity at small $R_1$ is that increasing the spring stiffness $k_s$ causes the maximum deflection point of the membrane to move downstream (to the right) with sharp peaks close to the spring locations (Figs.~\ref{fig:imagModesFixedR1}A and~\ref{fig:imagModesFixedR1}B as well as
Fig.~\ref{fig:stabksMinus1}).
Also true for both models is that the stability boundary shifts to lower~$T_0$ at small $R_1$ and small $k_s$.
As a result, at some locations
in the lower right of panels
A and B, membranes are omitted because all modes are stable, unlike at the corresponding locations in panels C and D (where $R_1$ is larger).

%The white background without a  superposed eigenmode in Figs.~\ref{fig:imagModesFixedR1}A and~\ref{fig:imagModesFixedR1}B indicates that the membrane at that combination of $(R_1,T_0,k_s)$ is stable. 

The membrane deflections at the springs increase when $R_1$ and $T_0$ increase relative to $k_s$. This can be seen by moving from left to right in some of the rows of Figs.~\ref{fig:imagModesFixedR1}A--C (i.e., increasing $T_0$ at fixed $k_s$), such as $k_s=10^0$ in panel C.
The same trend is seen moving from panel A to B to C to D, at the same location in each panel, i.e., increasing $R_1$ with $k_s$ and $T_0$ fixed.
A similar phenomenon was seen in the membrane-vortex-wake model.

%.481,.485,.4825,.485
\begin{figure}[H]
\centering
    \includegraphics[width=.481\textwidth]{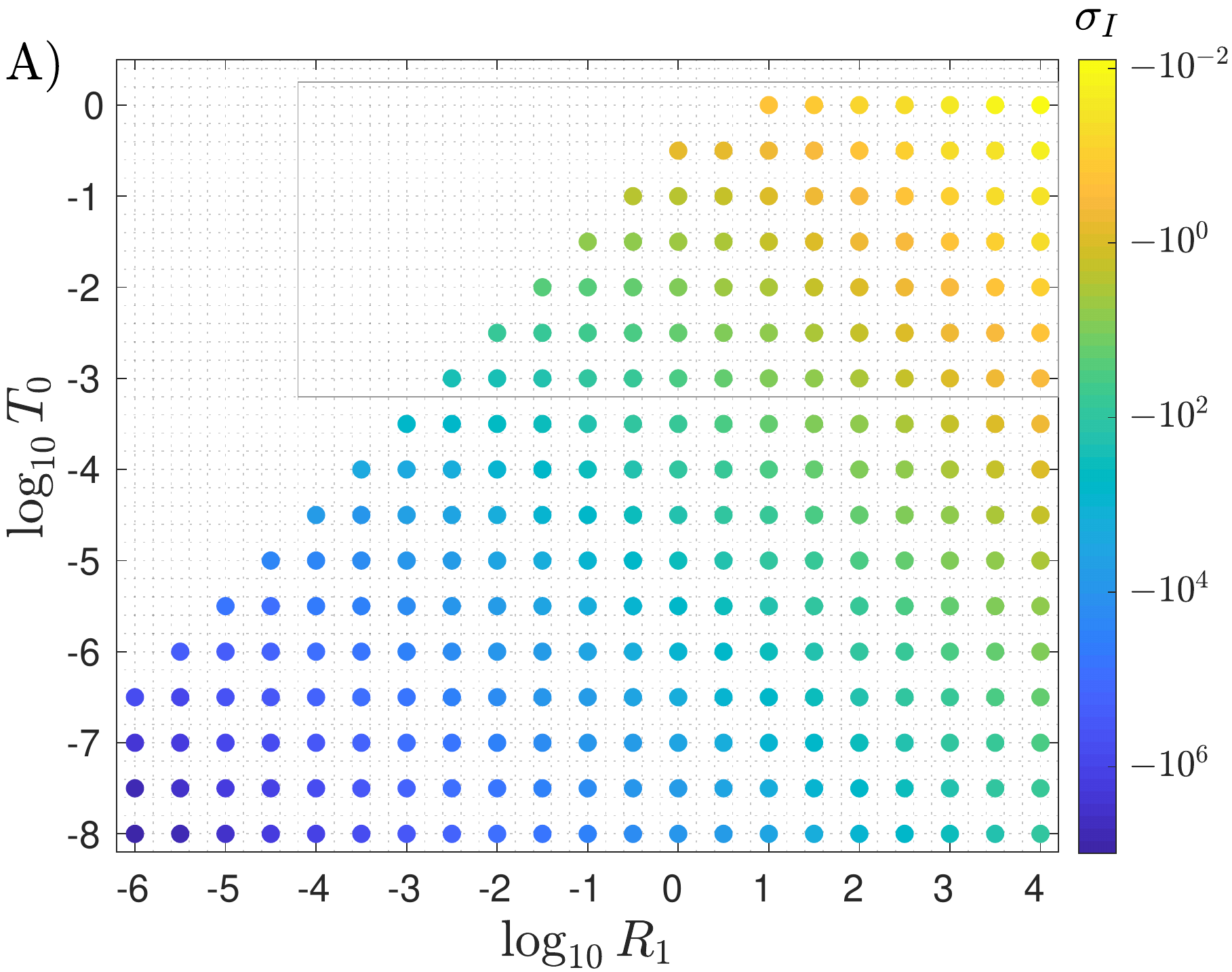}
    \includegraphics[width=.485\textwidth]{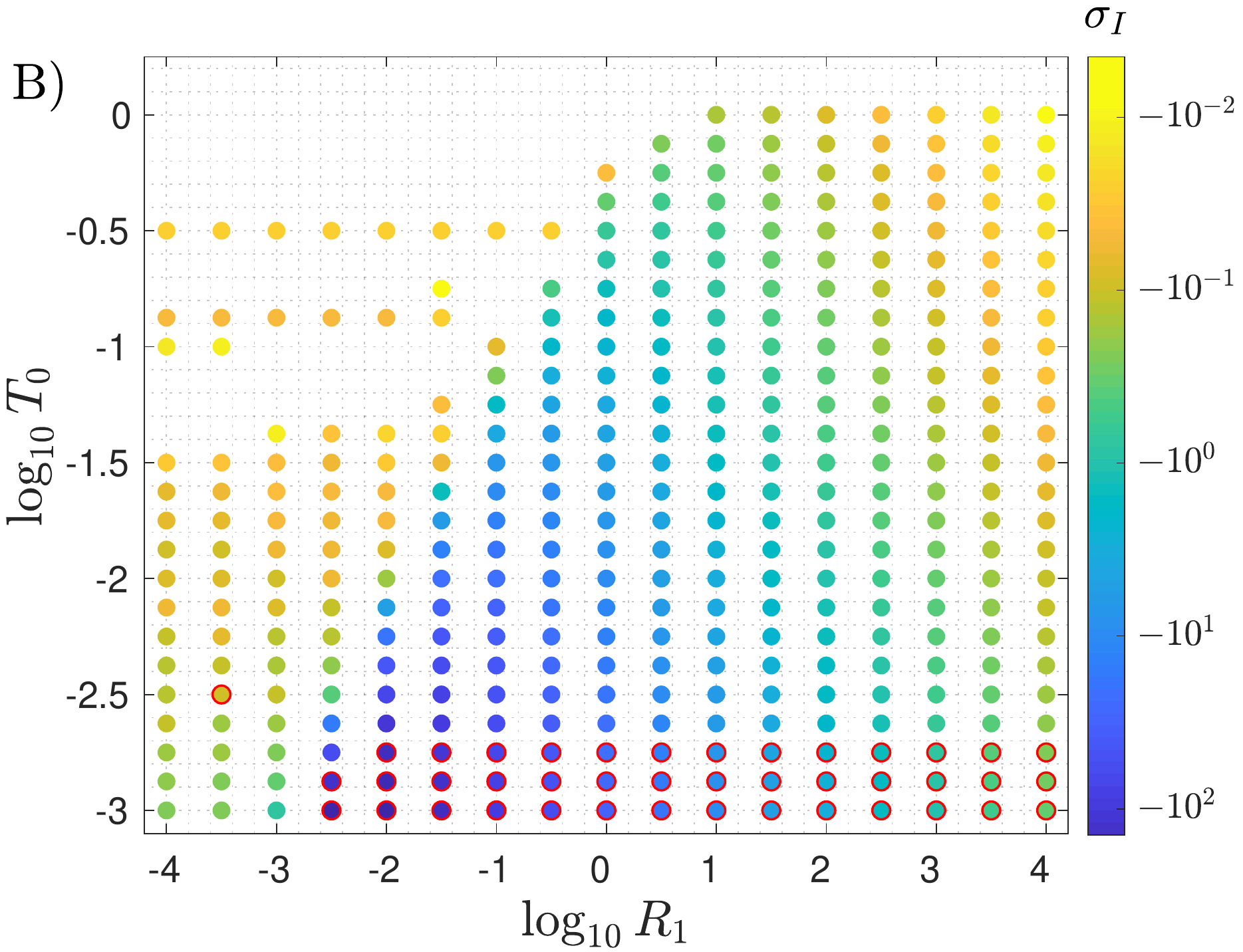}    \includegraphics[width=.4825\textwidth]{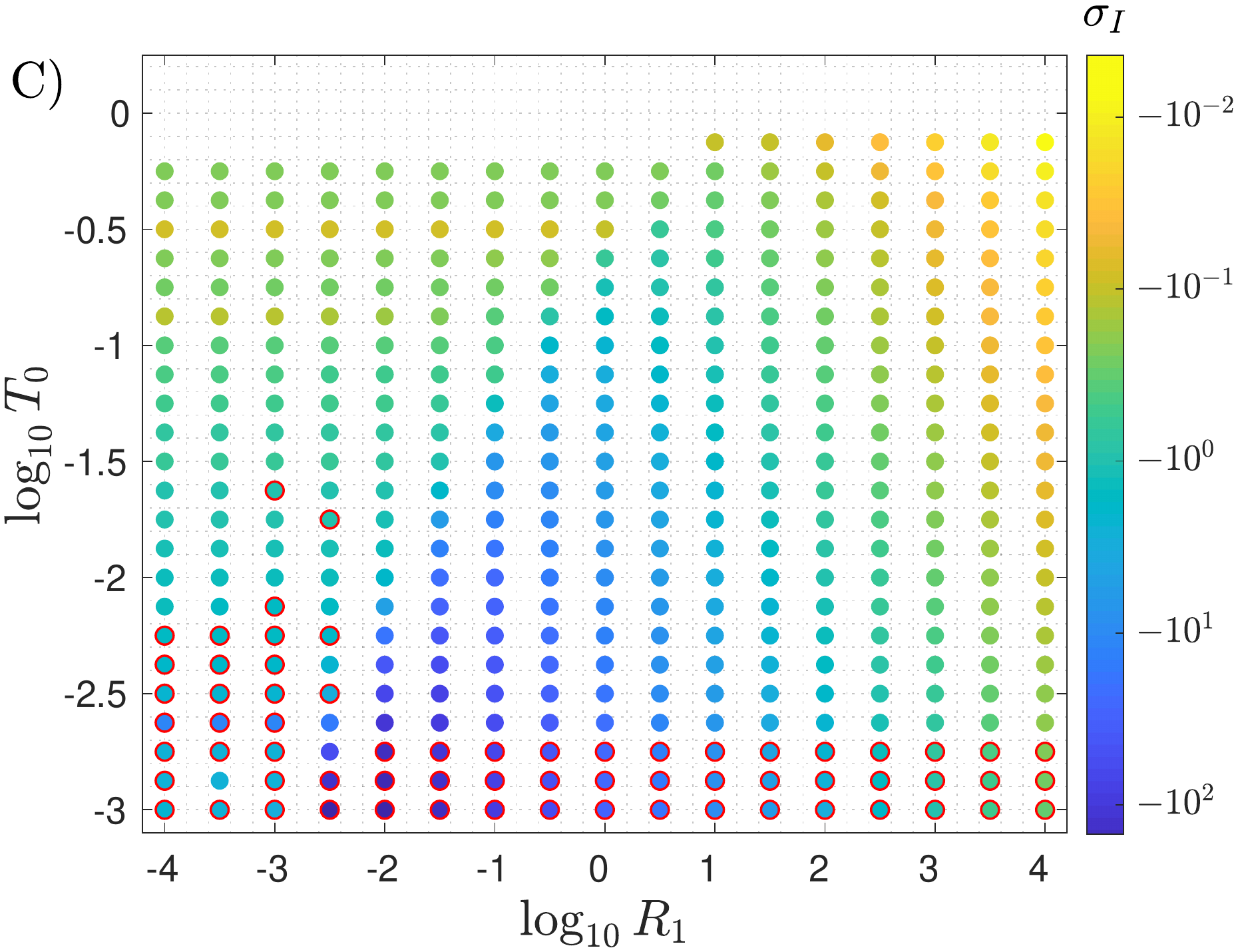}
    \includegraphics[width=.485\textwidth]{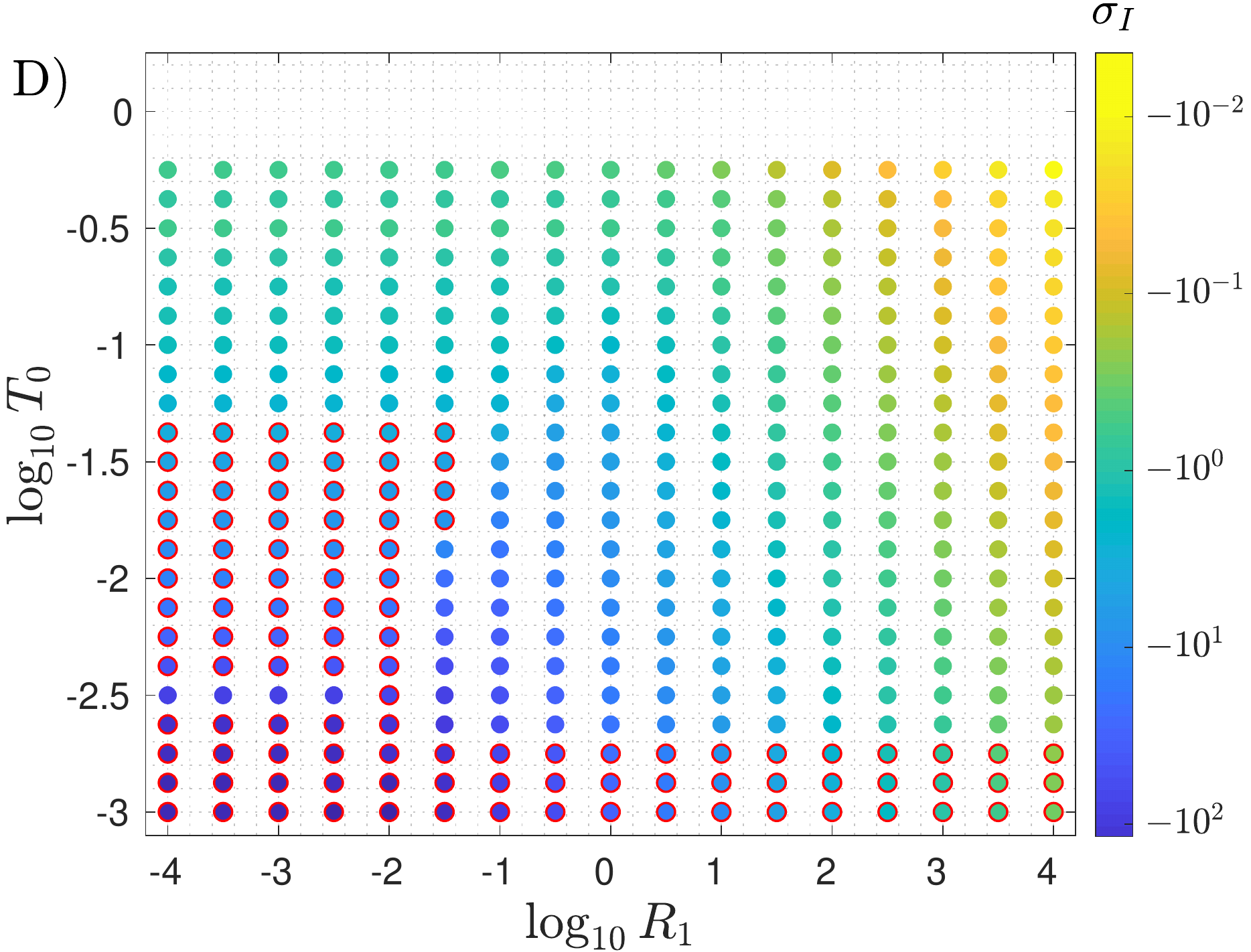}
    \caption{(Infinite, periodic membrane) The region in $R_1$--$T_0$ space in which membranes are unstable. The color of the dots in the instability region labels the imaginary part of the eigenvalues ($\sigma_\Im$) corresponding to the most unstable modes. It represents the growth rate. The springs have stiffness values of: A) $k_s=0$ (analytical result), B) $k_s=10^{-1}$, C) $k_s=10^0$, and D) $k_s=10^1$. The numerical results shown in panels B--D are with $N=2^9$. The gray rectangle in panel A indicates the region we consider in panels B--D to facilitate comparison.  The red outline on some of the colored dots indicates the cases where convergence with respect to $N$ (as defined by Eq.~\eqref{eq:relativeError}) was not obtained. }\label{fig:scatterPlotsImag}
\end{figure}

In Fig.~\ref{fig:scatterPlotsImag} we plot the imaginary parts of the most unstable eigenvalues as colored dots in the region of instability for membranes attached to a periodic array of springs with spring constants $k_s=0$ (Fig.~\ref{fig:scatterPlotsImag}A), $k_s=10^{-1}$ (Fig.~\ref{fig:scatterPlotsImag}B), $k_s=10^0$ (Fig.~\ref{fig:scatterPlotsImag}C), and $k_s=10^1$ (Fig.~\ref{fig:scatterPlotsImag}D). When $k_s = 0$, Eqs.~\eqref{eq:quadraticEigL} become decoupled scalar quadratic equations which can be solved analytically.
The resulting $\sigma_\Im$ are plotted in Fig.~\ref{fig:scatterPlotsImag}A.
In Figs.~\ref{fig:scatterPlotsImag}B--D, $k_s \neq 0$, and we use the aforementioned \textsc{Matlab} eigenvalue solver. With $2^9$ modes, the results are resolved only in a small portion of Fig.~\ref{fig:scatterPlotsImag}A---a subset of the region within the gray rectangle. The axis limits of panels B--D coincide with the gray rectangle. 
Both the analytical results in panel A and the computed results in panels B--D are much easier to obtain than in the case of the membrane-vortex-wake model, so the data in all the panels of Fig.~\ref{fig:scatterPlotsImag} are much more extensive than in Figs. \ref{fig:scatterksMinus1}, \ref{fig:scatterks0}, and 
\ref{fig:scatterks10}, a key advantage of the 
infinite-membrane model. 
%The results for $\sigma_\Im$ in Fig.~\ref{fig:scatterPlotsImag}A are obtained analytically from Eqs.~\eqref{eq:analyticSigmaks0}--\eqref{eq:kminAnalytical}. 

For this periodic problem, we see that the stability boundary at large $R_1$ plateaus, independent of the value of $k_s$, i.e., the critical pretension ($T_0$) is the same for all $R_1\gtrsim 10^1$ instead of increasing with increasing mass as in the vortex-wake model (Figs.~\ref{fig:scatterksMinus1}, \ref{fig:scatterks0}, and \ref{fig:scatterks10}). Although the stability boundaries differ at large $R_1$, here the vortex-wake model's eigenvalues are only slightly unstable [$\sigma_\Im = O(10^{-5})$] compared to neutrally stable ($\sigma_\Im = 0$) in the infinite membrane model.

We see that for smaller $R_1$ ($< 10^0$), the stability boundary in Fig.~\ref{fig:scatterPlotsImag}A is close to the diagonal line $T_0 = R_1$, and we will show this asymptotically in the next section. In panels B--D ($k_s \neq 0$),
this line is no longer the stability boundary, but is instead the location of a sharp change in $\sigma_\Im$, shown by the sharp change in colors moving across this line, particularly in panel B and less so in C and D.
%we see that at small $R_1$ and $T_0$, $\sigma_\Im\neq 0$ only if $T_0<R_1$ [Eq.~\eqref{eq:discriminantdk}]. This is in agreement with the location of the colored dots (unstable membranes) when $R_1\leq 10^0$ in Fig.~\ref{fig:scatterPlotsImag}A. 

From the colors of the dots in all the panels we see that if we fix~$R_1$ and decrease~$T_0$, the growth rate $\sigma_\Im$ becomes larger in magnitude (value in color bar at right). If we fix~$T_0$ and increase~$R_1$ above $T_0$, the growth rate $\sigma_\Im$ becomes smaller in magnitude which implies slower growth of instabilities. 

%The rectangle with the gray outline in Fig.~\ref{fig:scatterPlotsImag}A highlights the region in $R_1$--$T_0$
%space considered in the rest of the panels of Fig.~\ref{fig:scatterPlotsImag} to facilitate comparison between the results obtained for different $k_s$ values.
In Fig.~\ref{fig:scatterPlotsImag}B, $k_s=10^{-1}$
as in Fig.~\ref{fig:scatterksMinus1}A for the membrane-vortex-wake model. There
are two main points of qualitative agreement between the models in this case. One is that a lower plateau of the stability boundary occurs at small $R_1$; another is that the growth rates are much lower for $R_1<T_0$.
% In Fig.~\ref{fig:scatterPlotsImag}B, we see that 
% the growth rates $\sigma_\Im$ jump by orders of magnitude as $R_1$ increases above $T_0$.
% even though in the region of $R_1<10^{-1}$ 
% \silas{$10^{-1}$?} 
%and $T_0<R_1$ the effect of $T_0$ on $\sigma_\Im$ is not very significant, 
%$|\sigma_\Im|$ decreases significantly as soon as $T_0>R_1$. 
%\silas{ better to simply say the growth rates are orders of magnitude larger for $R_1>T_0$.}
At this $k_s$ value ($10^{-1}$) and at small $R_1$ and $T_0$ close to the stability boundary (e.g., at $T_0=10^{-0.5}$ and $10^{-0.875}$, for $R_1 \lesssim 10^{-2}$), there are also a few narrow bands of instability (lines of yellow dots) between stable regions, which was not observed in the membrane-vortex-wake model (Fig.~\ref{fig:scatterksMinus1}A). Moving to Fig.~\ref{fig:scatterPlotsImag}C, $k_s$ is increased to $10^0$, and the stable regions in panel B surrounding the isolated bands become unstable in panel C, with larger growth rates than in the bands.
Therefore, the stability boundary in panel C is almost at constant $T_0$ for all $R_1$, with a very small increase when $R_1\geq 10^1$. An upward shift in the lower plateau is also seen in the vortex-wake model with the same increase in $k_s$, moving from Fig.~\ref{fig:scatterksMinus1}A to Fig.~\ref{fig:scatterks0}A.
%The increase is not as large as that observed in the eigenvalue problem (Fig.~\ref{fig:scatterks0}) because the neutrally stable modes (with $\sigma_{\Im}$ of $O(10^{-5})$) are not captured by the infinite, periodic membrane problem considered here. \silas{this was already noted earlier}
Increasing $k_s$ further to $10^1$ (Fig.~\ref{fig:scatterPlotsImag}D) in the periodic membrane model these trends continue: the stability boundary is horizontal at~$T_{0C}\approx 10^{-0.25}$ (a factor of $\approx 3$ smaller than~$T_{0C}$ in the small-to-moderate $R_1$ region of Fig.~\ref{fig:scatterks10}), and the growth rates have increased further where $R_1 < T_0$. In Figs.~\ref{fig:scatterPlotsImag}B--D as~$T_0$ decreases,
$N=2^9$ is eventually too small to resolve the most unstable eigenmodes. These cases are shown by red outlines around the colored dots, and become more prevalent as we move from panel B to C to D. These cases correspond to
an eigenvalue relative error (as defined in Eq.~\eqref{eq:relativeError})$>3\times 10^{-2}$ (chosen somewhat arbitrarily; other values give a similar classification of nonconvergence).

%.481,.485,.485,.485
\begin{figure}[H]
\centering
    \includegraphics[width=.481\textwidth]{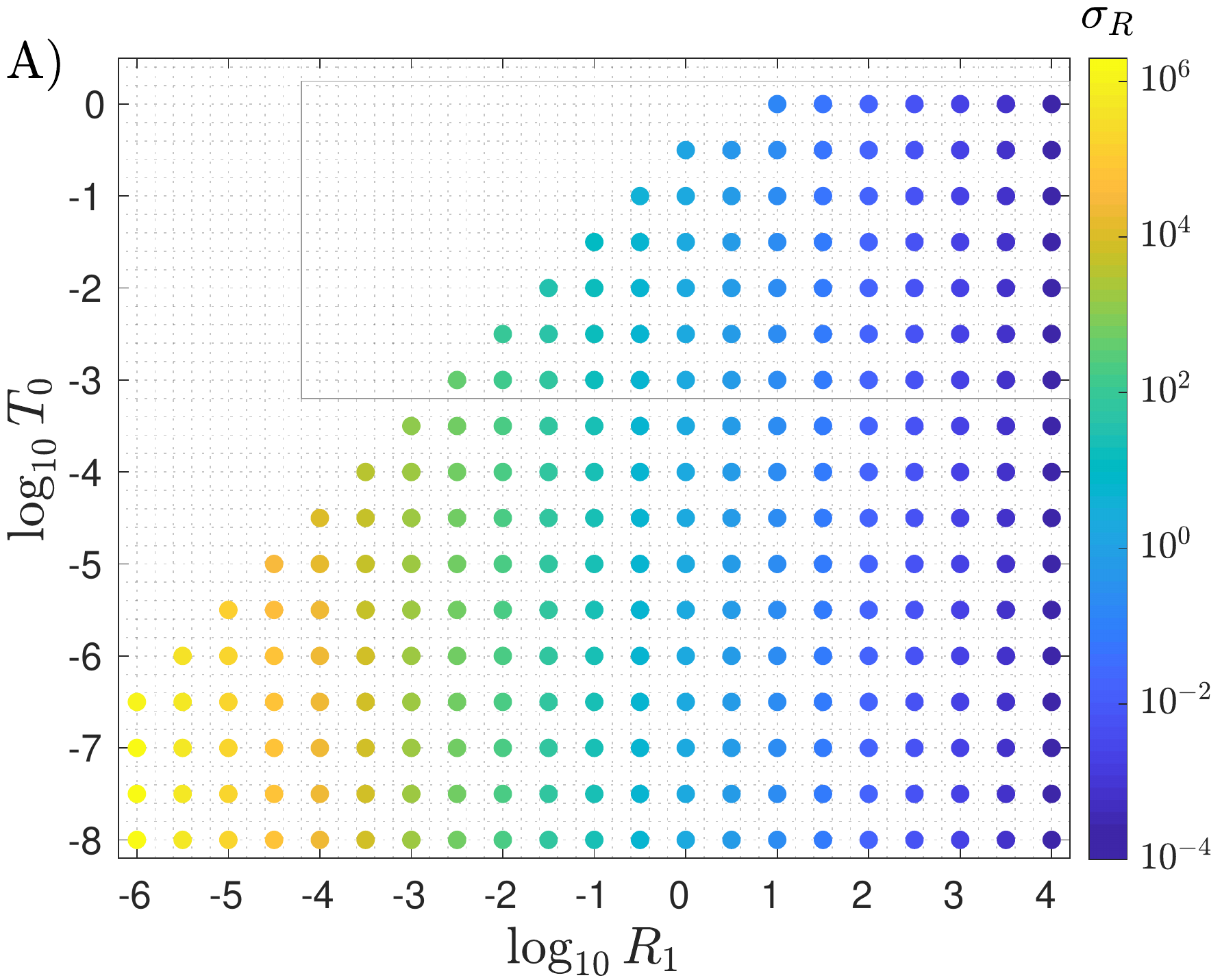}
    \includegraphics[width=.485\textwidth]{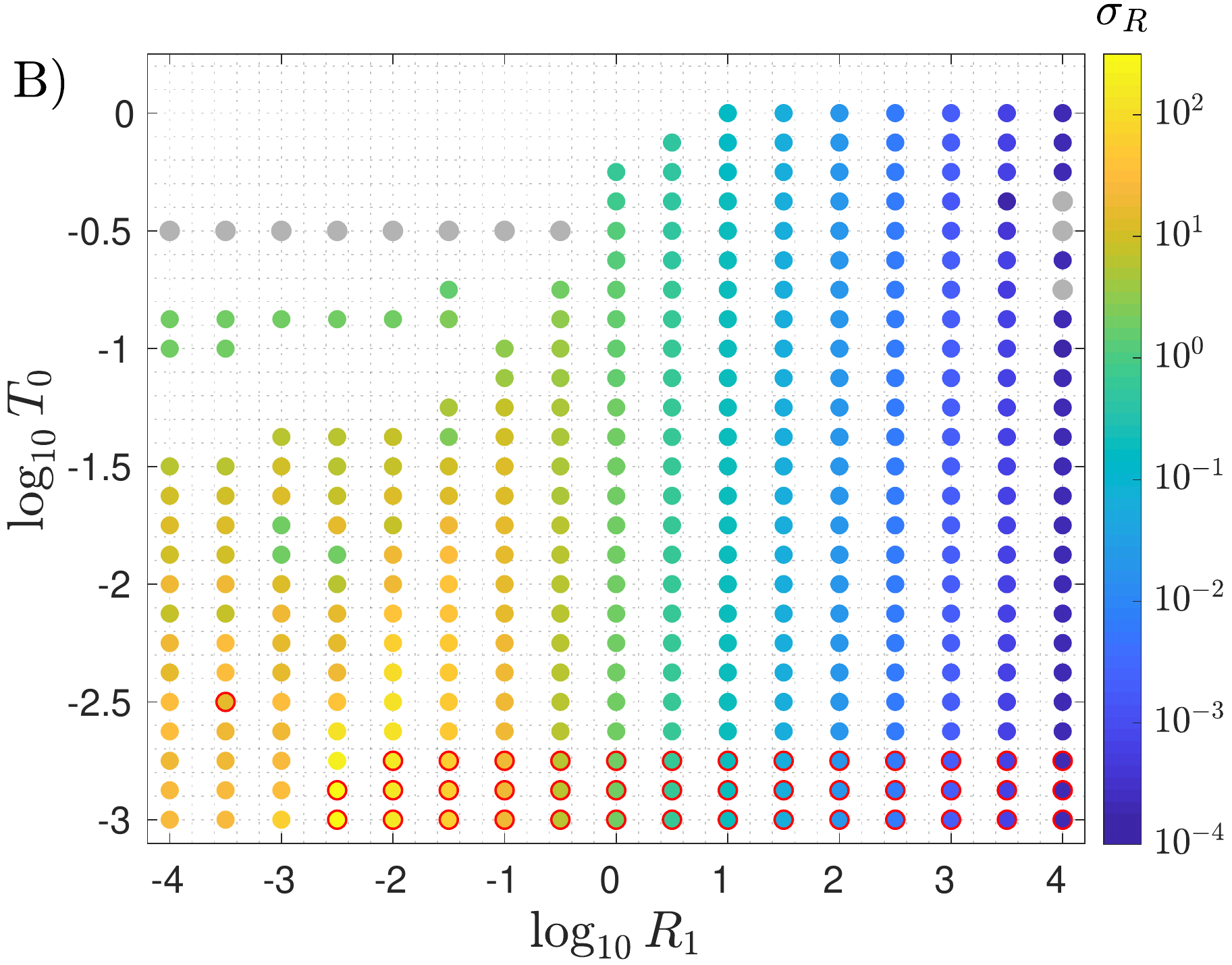}
    \includegraphics[width=.485\textwidth]{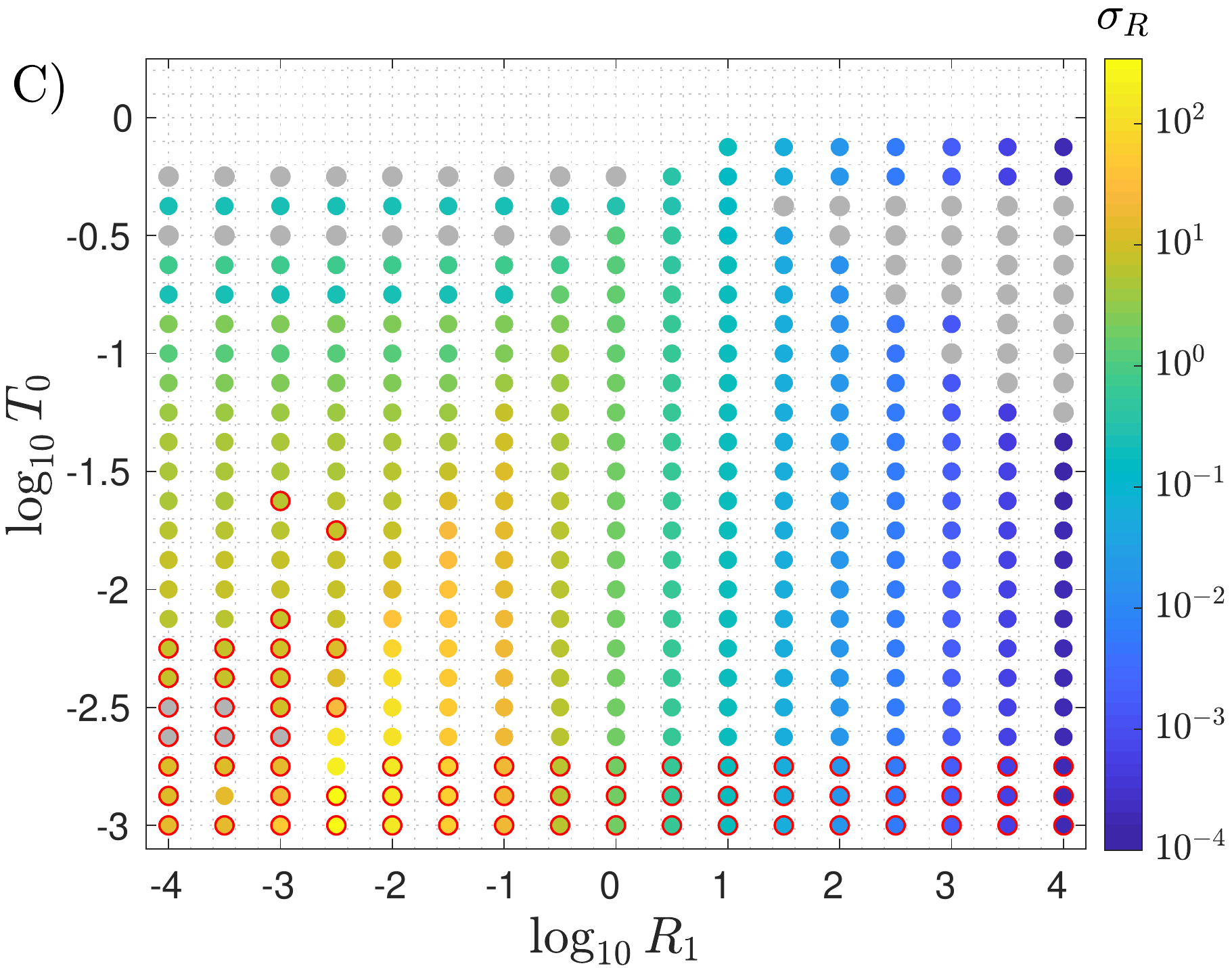}
    \includegraphics[width=.485\textwidth]{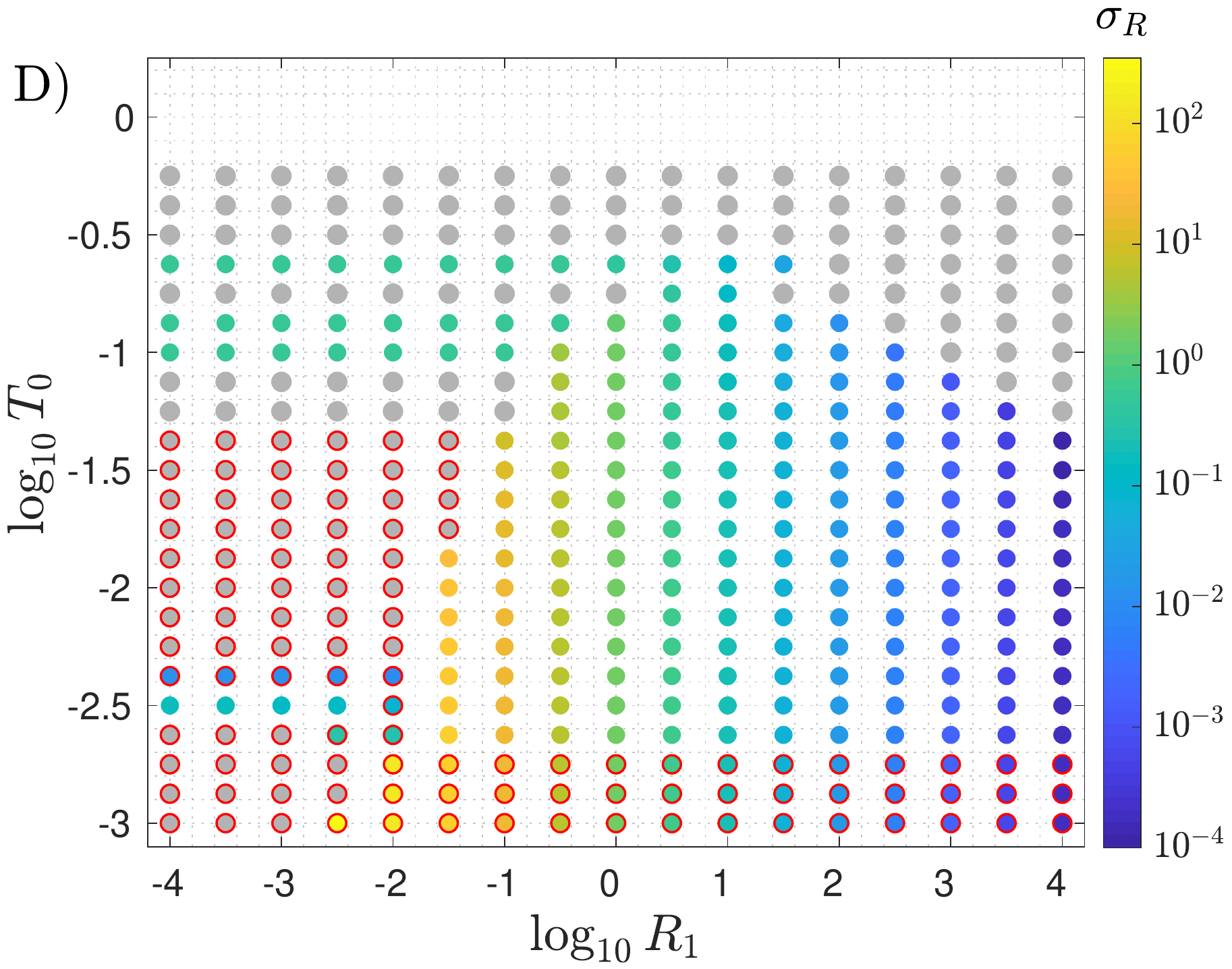}
    \caption{(Infinite, periodic membrane) The region in $R_1$--$T_0$ space in which membranes are unstable. The color of the dots in the instability region labels the real part of the eigenvalues ($\sigma_\Re$) corresponding to the most unstable modes. It represents the angular frequency. The springs have stiffness values of: A) $k_s=0$ (analytical result), B) $k_s=10^{-1}$, C) $k_s=10^0$, and D) $k_s=10^1$. The numerical results shown in panels B--D are with $N=2^9$. The gray dots correspond to modes that lose stability by divergence and have $\sigma_\Re\leq 10^{-9}$. The gray rectangle in panel A indicates the region we consider in panels B--D to facilitate comparison. The red outline on some of the colored/gray dots indicates the cases where convergence with respect to $N$ (as defined by Eq.~\eqref{eq:relativeError}) was not obtained.}\label{fig:scatterPlotsReal}
\end{figure}
In Fig.~\ref{fig:scatterPlotsReal} we plot the corresponding real parts of the eigenvalues for the most unstable modes. Increasing the spring stiffness $k_s$ introduces more divergence modes (the gray dots, $\sigma_\Re\leq 10^{-9}$). Note that this also occurs in the vortex-wake model, Figs.~\ref{fig:scatterksMinus1},~\ref{fig:scatterks0}, and~\ref{fig:scatterks10}.  $\sigma_\Re$ varies more strongly with $R_1$ than with $T_0$. There is almost no variation with $T_0$ in Fig.~\ref{fig:scatterPlotsReal}A, and little variation in panels B--D---mainly when $T_0>R_1$. Here, as $T_0$ decreases, $\sigma_\Re$ increases but non-monotonically, particularly at the isolated bands of dots in panel B that become bands of non-monotonic change in 
$\sigma_\Re$ in panels C and D, including changes between divergence (gray dots) and flutter and divergence (colored dots).
Next we discuss more quantitatively how the real and imaginary parts of the eigenvalues depend on $R_1$ and $T_0$, including asymptotic scaling laws.

%and~$T_0$, the widths of the necks relative to the maximum widths of the envelopes increases as we increase~$R_1$ from $10^0$ to $10^4$. The effect of making $R_1$ larger seems similar to making~$k_s$ smaller. 

%Also, increasing~$T_0$ (to $10^{-1.05}$ for example) makes the widths of the necks relative to the maximum widths of the envelopes increase.  Therefore, it is the size of~$k_s$ relative to~$R_1$ and $T_0$ that appears to determine how well the springs hold the membrane down. \silas{combine this with sentence about ks relative to R1; also, ``you" is too informal.} For any $k_s$ value, when both~$T_0$ and~$R_1$ are relatively large, the mode shapes \silas{large then?} tend to sinusoidal functions. Similar eigenmode shapes were obtained in the region of $R_1,T_0\gg 1$ from the vortex-wake model presented in Sec.~\ref{sec:eigenmodes}. 
%At small values of~$R_1$ ($10^{-4}$ in Fig.~\ref{fig:imagModesFixedR1}A and $10^{-1}$ in Fig.~\ref{fig:imagModesFixedR1}B), increasing the value of~$k_s$ at a fixed $T_0$ changes a mode from stable to unstable.  \silas{when you mention each of these features, could you also compare and contrast with the vortex-wake eigenmodes as appropriate? This would help justify our use of this model to some extent.} 
% \christiana{Note that in some cases the maximum deflection occurs close to the trailing edge especially at large R1. Compare with Fig. 16. Wavier at smaller T0. Blank means stable. Shifting of boundary at small R1 and increasing ks.}

%start here

\subsection{Analytical results and scaling laws in the instability region \label{sec:scalingLaws}}

In this section we find analytically the eigenvalues and the corresponding eigenmodes (sinusoidal functions)---in the special case of $k_s=0$. From these analytical solutions we derive asymptotic approximations for how the maximum growth rate, corresponding angular frequency, and dominant wave number depend on $R_1$ and $T_0$ when these parameters are small and large. We also study how the scaling laws behave when $k_s\neq 0$, where numerical solutions are required.
% In the special case of $k_s=0$ the eigenvalues can be found analytically and the corresponding eigenmodes are sinusoidal.
% From these analytical solutions we can obtain asymptotic approximations for how the maximum growth rate, corresponding angular frequency, and dominant wave number depend on $R_1$ and $T_0$ when they are small and large.
% A fundamental aspect of the problem that for each $R_1$, there is a critical $T_0$ below which the growth rate $\sigma_\Im$ increases without bound as $k$ goes to infinity. 
% We determine the curve in $R_1$--$T_0$ space for the $k_s=0$ case where the growth rate first becomes maximum for infinite $k$ (and above this curve, it is maximized for finite $k$) and superpose the curve on our plots of various $k_s$ values.
% In this special case of $k_s=0$, 

With $k_s = 0$ Eq.~\eqref{eq:quadraticEigL} reduces to 
\begin{equation}\label{eq:membraneLks0}
     \left[\left(-R_1-\frac{L}{\pi|{k}|}\right)\sigma^2-\frac{4{k}}{|{k}|}\sigma +\left(\frac{2\pi{k}}{L}\right)^2\left(T_0-\frac{L}{\pi|{k}|}\right)\right]\hat{y}_{{k}}=0,
\end{equation}
for ${k}=-\infty,\dots,-1,1,\dots,\infty$. 
%This equation can be recast in the form of Eq.~\eqref{eq:quadraticFourier}, but now $A_2$, $A_1$, and $A_0$ are all diagonal matrices. 
Solving Eq.~\eqref{eq:membraneLks0} for $\sigma$, we obtain
\begin{equation}\label{eq:analyticSigmaks0}
    \sigma = -\frac{2{k}}{|{k}|\left(R_1+L/(\pi|{k}|)\right)}\pm \sqrt{D_{{k}}}, 
\end{equation}
where
\begin{equation}\label{eq:analyticalDiscriminantks0}
      D_k := \frac{4}{\left(R_1+L/(\pi |k|)\right)^2}\left[1+\left(R_1+\frac{L}{\pi |k|}\right)\left(\frac{\pi k}{L}\right)^2\left(T_0-\frac{L}{\pi |k|}\right)\right].
\end{equation}
    % \frac{\Delta}{4a^2}= \frac{4}{\left(R_1+L/(\pi|\bar{k}|)\right)^2}\left[\left(\frac{\bar{k}}{|\bar{k}|}\right)^2+\left(R_1+\frac{L}{\pi|\bar{k}|}\right)\left(\frac{\pi \bar{k}}{L}\right)^2\left(T_0-\frac{L}{\pi|\bar{k}|}\right)\right].
% \christiana{We note that $(\bar{k}/|\bar{k}|)^2=1$ so shall we write it like that?}
%with the ${k}$ that gives the most unstable mode (i.e., the one with $\sigma_\Im$ most negative). 
The term in brackets can be written as
$(L(-R_1+T_0)+\pi R_1T_0|{k}|)$ multiplied by a positive factor. Therefore $D_k$ can be negative 
only for $T_0 < R_1$. When $R_1$ is small the
$R_1T_0$ term is negligible, so the stability boundary follows $T_0 = R_1$ as shown in Fig.~\ref{fig:scatterPlotsImag}A.

In Eq.~\eqref{eq:analyticSigmaks0} 
%we have explicitly written out the terms of the discriminant.
%We note that 
there are two possible eigenvalues for each $R_1$ and $T_0$ combination (due to the square root) that correspond to a complex-conjugate pair. We can then find ${k}$ for the most unstable mode by setting the derivative of Eq.~\eqref{eq:analyticalDiscriminantks0} with respect to ${k}$ to zero and solving for ${k}$:
% \begin{equation}
%     \bar{k}_{\max} = \frac{LR_1(R_1-5T_0)\pm LR_1\sqrt{R_1^2+14R_1T_0+T_0^2}}{4\pi R_1^2T_0}.
% \end{equation}
\begin{equation}\label{eq:kminAnalytical}
    {k}_{\max} = \pm\frac{L(R_1-5T_0)+ L\sqrt{R_1^2+14R_1T_0+T_0^2}}{4\pi R_1T_0}.
\end{equation}
Since the discriminant in Eq.~\eqref{eq:analyticalDiscriminantks0} is symmetric about ${k}=0$, we have a symmetric pair of ${k}_{\max}$ in Eq.~\eqref{eq:kminAnalytical}.
For the periodic membrane, $k_{\max}$ must be an integer, but Eq.~\eqref{eq:kminAnalytical} is not necessarily an integer. Restricting $k_{\max}$ to integer values, we find that
it is given by one of the integers nearest to the value in Eq.~\eqref{eq:kminAnalytical}.

With this model we are able to obtain asymptotic scaling laws in the instability region for a wide range of $R_1$ and $T_0$ values. Unstable membrane modes are realized when the argument of the radical in Eq.~\eqref{eq:analyticSigmaks0}  is
negative, i.e., $D_{{k}}<0$ in Eq.~\eqref{eq:analyticalDiscriminantks0}. 
We will now present the asymptotic scaling laws for $k_{\max}$,
$\sigma_\Re$, and $\sigma_\Im$ in
different limits within the instability region. As we do so,
we will refer to the summary in Table~\ref{table:scalingLaws}. We study three asymptotic regimes that correspond to moving within the unstable region of Fig.~\ref{fig:scatterPlotsImag}A (or Fig.~\ref{fig:scatterPlotsReal}A) in three different directions.
Moving rightward off to infinity, we have
$R_1\to\infty$ with fixed $T_0\leq T_{0C}$,
the first row of Table~\ref{table:scalingLaws}. Moving diagonally downward and leftward, parallel to the stability boundary, we have $R_1\to 0$ with $T_0=cR_1$, for a fixed $c$ between 0 and 1, the second row of Table~\ref{table:scalingLaws}. 
Moving vertically downward instead,
we have $T_0\to 0$ with fixed $R_1$, the
third row of Table~\ref{table:scalingLaws}.
Moving across each row, we give the asymptotic behavior of the three main quantities of interest.
%[64mm] for backslashbox
\setlength{\extrarowheight}{14pt}
\begin{table}[H]
\caption{Summary of asymptotic scalings for the dominant wavenumber $(k_{\max})$, the real part of the eigenvalue $(\sigma_{\Re})$, and the imaginary part of the eigenvalue $(\sigma_{\Im})$ in the small- and large-$R_1$ and small-$T_0$ regimes, in the instability region.}\label{table:scalingLaws}
\centering
\vspace{.1cm}
\begin{tabular}{l|c|c|c}
\hline\hline
\backslashbox[56mm]{Regimes}{Quantities} &$k_{\max}$& $\sigma_{\Re}$ &$\sigma_{\Im}$\Tstrut\Bstrut\\
\hline
\Tstrut
$R_1\to\infty$\,\, (fixed $T_0\leq T_{0C}$)
 &$\max\left(\displaystyle\frac{L}{2\pi T_0},1\right)$& $\displaystyle\frac{2}{R_1}$ & $\max\left(\displaystyle\frac{1}{\sqrt{R_1T_0}},\frac{2\pi}{L\sqrt{R_1}}\sqrt{\frac{L}{\pi}-T_0}\right)$\\
$R_1\to 0$\,\,($T_0=cR_1$, $0<c<1$)  &$\displaystyle\frac{LC}{4\pi T_0}$ & $\displaystyle\frac{2C}{R_1(C+4c)}$ & $\displaystyle\frac{\sqrt{C^3(4-4c-C)}}{2R_1(C+4c)\sqrt{c}}$  \\
$T_0\to 0$\,\, (fixed $R_1$)  &$\displaystyle\frac{L}{2\pi T_0}$ & $\displaystyle\frac{2}{R_1}$ & \,\, $\displaystyle\frac{1}{\sqrt{R_1T_0}}$
\Bstrutr\\[.35cm]
\hline\hline
\end{tabular}
\end{table}
% \begin{table}[H]
% \caption{Summary of asymptotic scalings for the dominant wavenumber $(k_{\max})$, the real part of the eigenvalue $(\sigma_{\Re})$, and the imaginary part of the eigenvalue $(\sigma_{\Im})$ in the small- and large-$R_1$ and small-$T_0$ regimes, in the instability region.}\label{table:scalingLaws}
% \centering
% \vspace{.1cm}
% \begin{tabular}{l|c|c|c}
% \hline\hline
% \backslashbox[60mm]{Regimes}{Quantities} &$k_{\max}$& $\sigma_{\Re}$ &$\sigma_{\Im}$\Tstrut\Bstrut\\
% \hline
% \Tstrut
% $R_1\to\infty$ (fixed $T_0\leq T_{0C}$) &$\max\left(L/(2\pi T_0),1\right)$& $2/R_1$ & $\max\left(1/\sqrt{R_1T_0},(2 \pi/L)\sqrt{T_0-L/\pi}1/\sqrt{R_1}\right)$\\
% $R_1\to 0$ ($T_0=cR_1$, $0<c<1$) &$LC/(4\pi T_0)$ & $2C/(R_1(C+4c))$ & $\sqrt{(C^3(C+4c-4)/c)}/(2R_1(C+4c))$  \\
% $T_0\to 0$ (fixed $R_1$)  &$L/(2\pi T_0)$ & $2/R_1$ & \,\, $1/\sqrt{R_1T_0}$
%  \Bstrutr\\
% \hline\hline
% \end{tabular}
% \end{table}
In the first column of Table~\ref{table:scalingLaws}, we
give the asymptotic forms of
$k_{\max}$ by taking the appropriate
limits in Eq.~\eqref{eq:kminAnalytical}.
In the first and third rows, we obtain
\begin{equation}\label{eq:kminR1large}
    k=k_{\max}\to \frac{L}{2\pi T_0}.
\end{equation} 
In the second row, setting $T_0=cR_1$ and
taking $R_1\to 0$, we have
\begin{equation}\label{eq:kmaxT0cR1}
    k=k_{\max}=\pm \frac{LC}{4\pi cR_1}=\pm\frac{LC}{4\pi T_0} \quad \mbox{where} \quad C=(1-5c)+\sqrt{1+14c+c^2}.
\end{equation}

\begin{figure}[H]
    \centering
    \includegraphics[width=.485\textwidth]{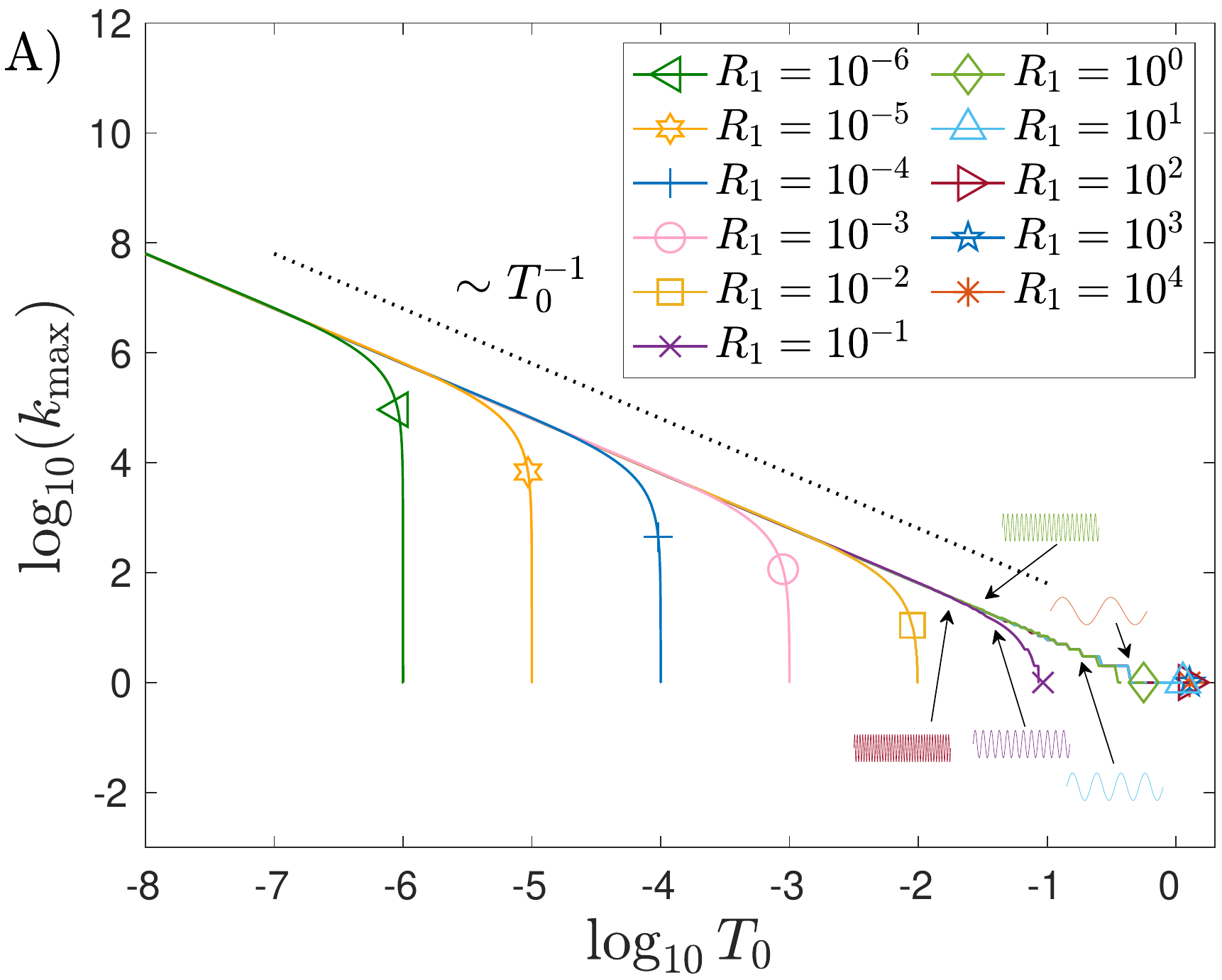}
     \includegraphics[width=.485\textwidth]{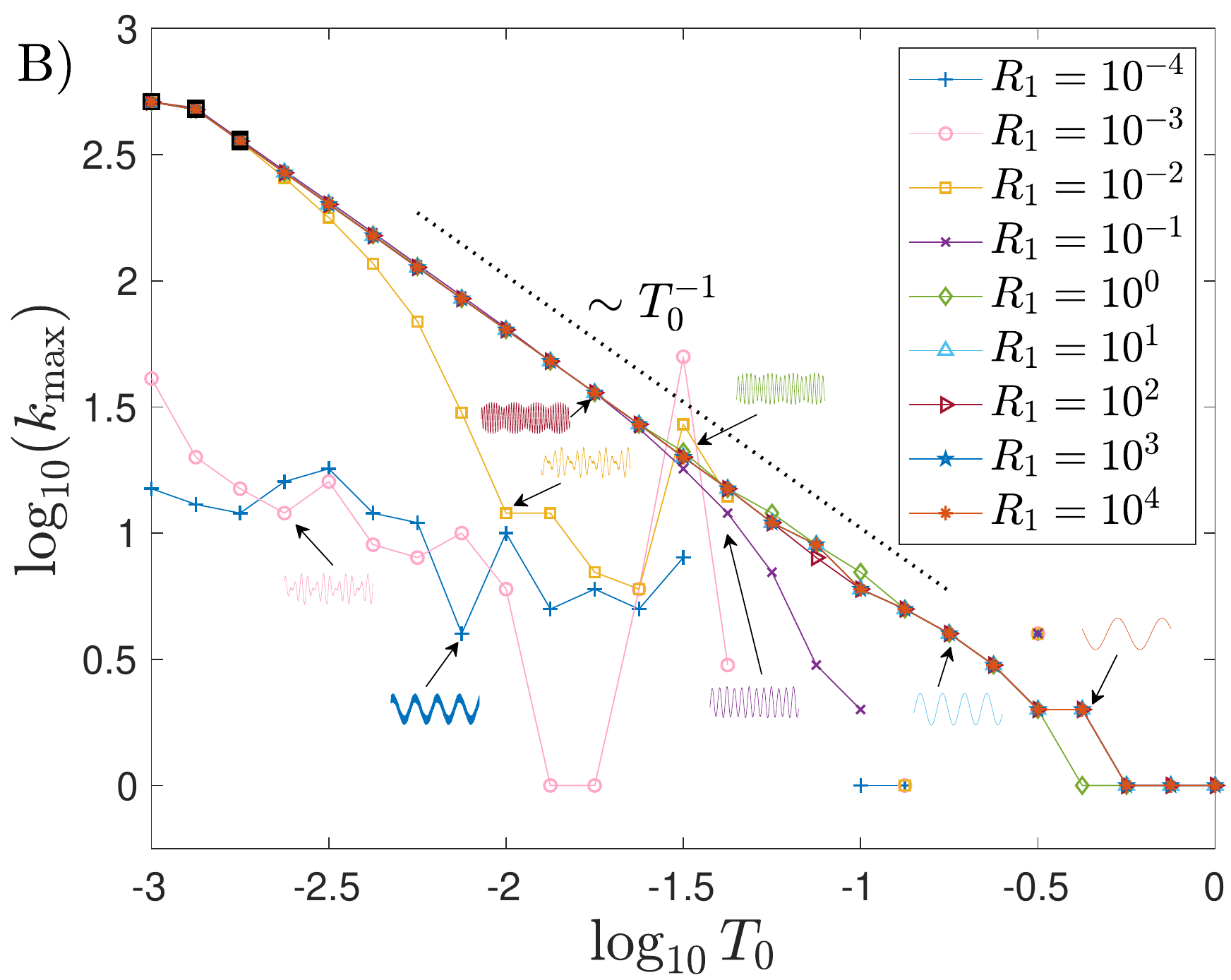}
     \includegraphics[width=.485\textwidth]{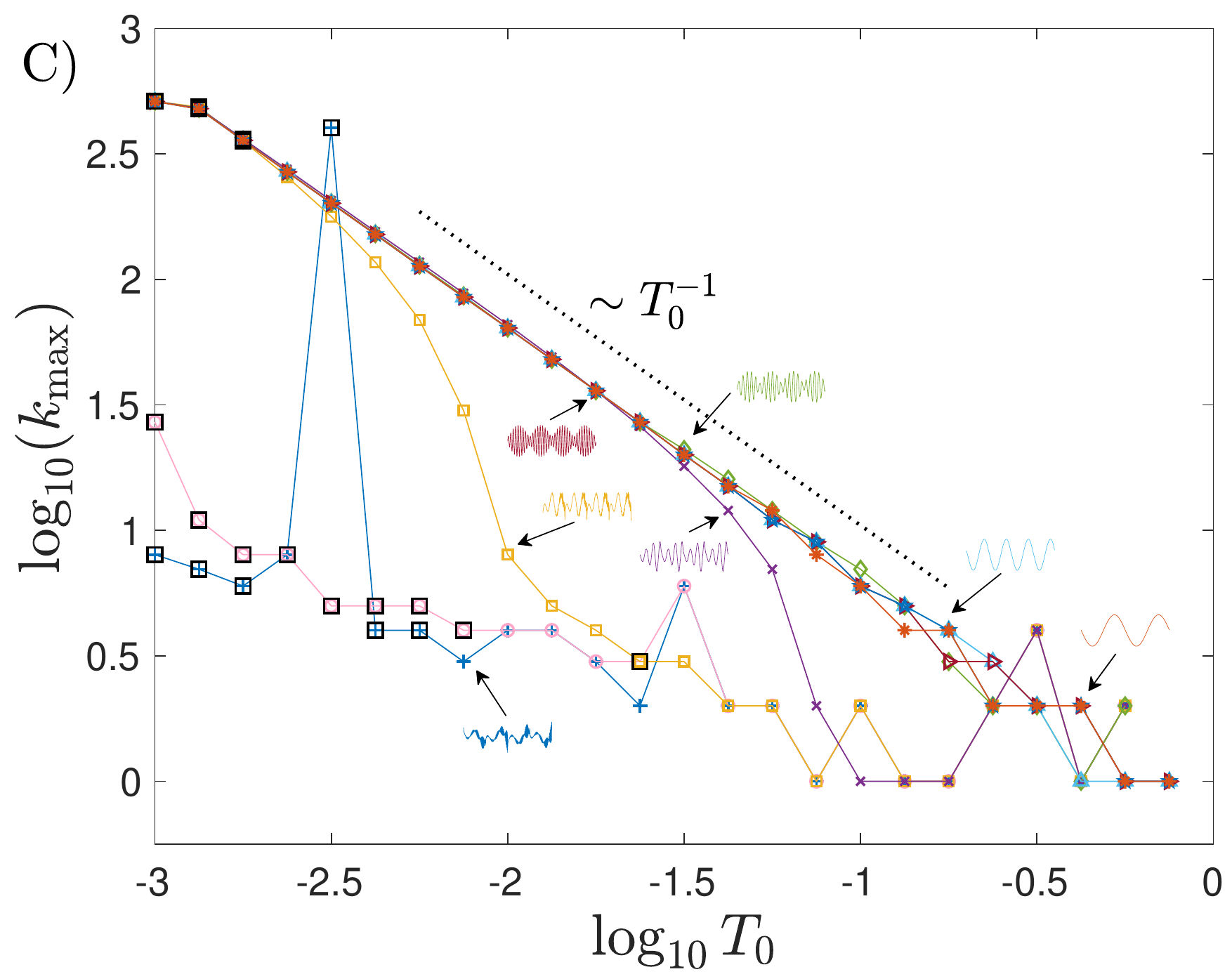}
     \includegraphics[width=.485\textwidth]{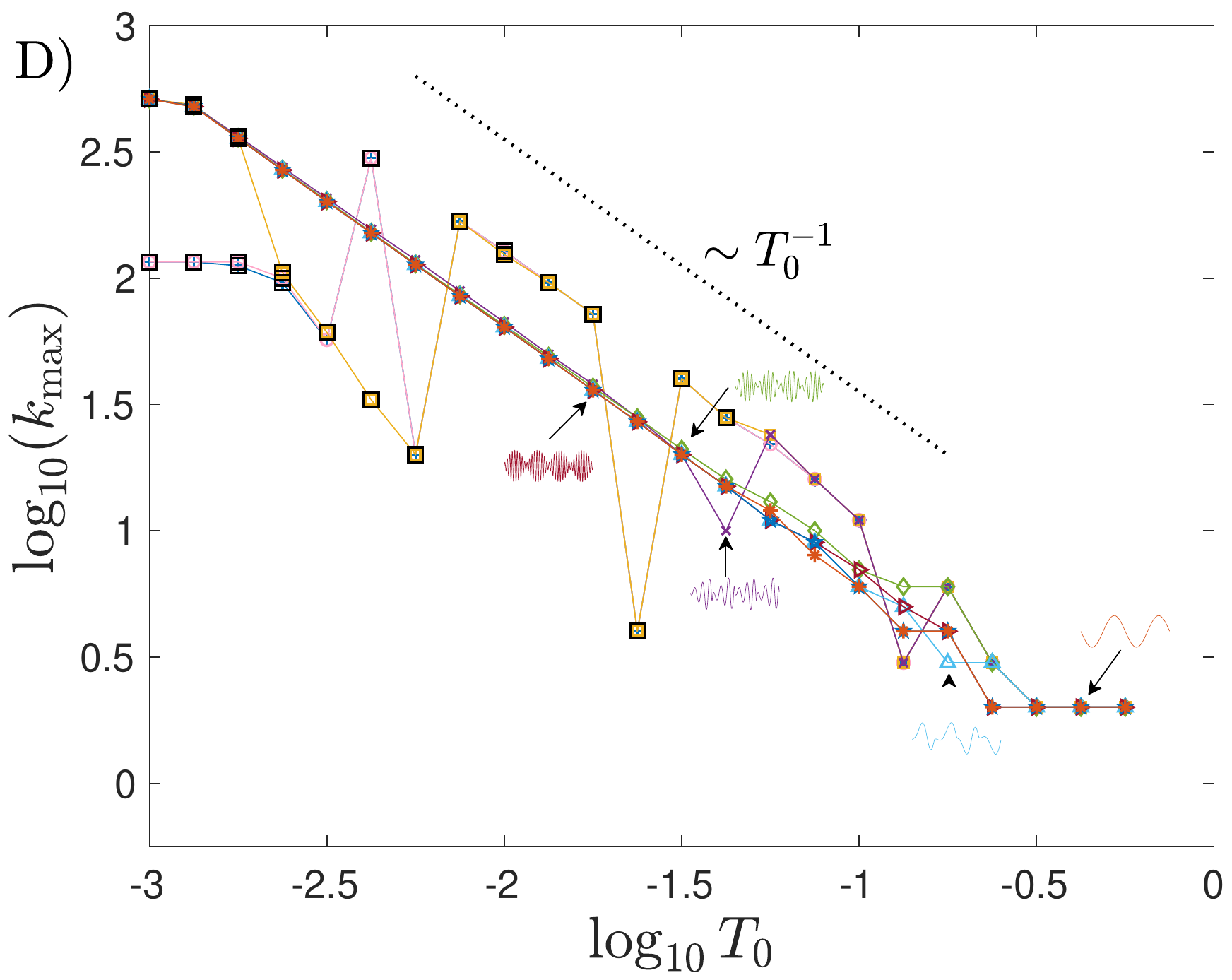}
    \caption{(Infinite, periodic membrane) Plots showing the membrane's dominant wavenumber versus $T_0$ for various fixed $R_1$ values at four values of spring constants: A) $k_s=0$ (analytical results), B) $k_s=10^{-1}$, C) $k_s=10^0$, and D) $k_s=10^1$. We show typical examples of the imaginary part of the eigenmode shapes. The dotted black line shows the scaling $T_0^{-1}$.}\label{fig:scalingLawsWavenumber}
\end{figure}

In Fig.~\ref{fig:scalingLawsWavenumber} we plot the dominant wavenumber versus $T_0$ for various fixed values of $R_1$ (one per line) and for four values of spring stiffness: $k_s=0$ (Fig.~\ref{fig:scalingLawsWavenumber}A), $k_s=10^{-1}$ (Fig.~\ref{fig:scalingLawsWavenumber}B), $k_s=10^0$ (Fig.~\ref{fig:scalingLawsWavenumber}C), and $k_s=10^1$ (Fig.~\ref{fig:scalingLawsWavenumber}D). When $k_s=0$, we have the analytical result in Eq.~\eqref{eq:kminAnalytical} (actually,
the nonzero integer closest to it, as mentioned previously). We also still assume that the membrane has period $L=4$, as in the $k_s\neq 0$ case discussed previously. In panel~A, we find that the wavenumber does not vary significantly with $R_1$ except when $R_1 \ll 1$ and we are close to the stability boundary, i.e., $T_0 \approx R_1$ for small $R_1$. The lines in panel~A with $R_1 \leq 10^{-2}$ curve downwards towards a vertical asymptote as they approach the stability boundary, but $k_{\max}$ is bounded below by 1, the endpoint of each line. 
%Here we plot each line up to $T_0$ = 0.975$R_1$, shown by the marker at the end of each line.  
%This enables better comparison between the $k_s=0$ and non-zero cases \silas{fix grammar here}.
The dotted black line in Fig.~\ref{fig:scalingLawsWavenumber}A shows that the dominant wave number for  various fixed $R_1$ values follows the scaling $T_0^{-1}$. Representative mode shapes at various $(R_1,T_0)$ pairs are shown for $x\in[0,L]$, with the colors of the modes corresponding to the value of~$R_1$. They are sinusoidal modes with wavelength that increases with $T_0$.

In panels B--D, $k_s\neq 0$, and the eigenmodes are found computationally. They are a superposition of multiple sinusoidal modes. For the most unstable mode we
define the dominant wavenumber to be that of the sinusoidal component with the largest
amplitude (the $k$ for which $|\hat{y}_k|$ is largest [see Eq.~\eqref{eq:YandGammaFourierL}]).
%\silas{don't need this sentence} By looking at its \texttt{fft} with respect to $x$ (note that this is the output of \texttt{polyeig} when solving the eigenvalue problem in Fourier space), one obtains $\hat{y}_{{k}}$ in Eq.~\eqref{eq:trunctatedyfourierL}. The ${{k}}$ with $\max |\hat{y}_{{k}}|$ is the dominant wavenumber. 
In Figs.~\ref{fig:scalingLawsWavenumber}B--D we find that at large $R_1$ ($\gtrsim 10^0$), where the
spring force is relatively less significant, the lines scale as $T_0^{-1}$ and do not vary significantly with $R_1$, similarly to the case without springs in panel A. At smaller $R_1$, the lines deviate greatly from this behavior, and do not seem to follow any specific power law.
The data points outlined with black squares are cases that are not resolved (using the same definition as for the red circles in Figs.~\ref{fig:scatterPlotsImag} and~\ref{fig:scatterPlotsReal}---when the
eigenvalue relative error [Eq.~\eqref{eq:relativeError}] $> 3\times 10^{-2}$). These occur mostly at small $T_0$, when the dominant wavenumber $k_{\max}$ is very large, so good resolution would require a larger $N$ than is feasible computationally. The deviations at
small $R_1$ coincide with changes in
the eigenmodes similar to those 
seen in Fig.~\ref{fig:imagModesFixedR1} when $R_1$ and~$T_0$ are small relative to $k_s$. In particular, the mode shapes are less sinusoidal and less symmetric than at large~$R_1$.
For example, as the spring stiffness $k_s$ increases, moving from
panel B to C to D, the envelopes of deflection for the green modes at $(R_1,T_0)=(10^0,10^{-1.5})$ and the red modes at $(R_1,T_0)=(10^2,10^{-1.75})$ are decreased near the springs at    $x=0,1,\dots,L=4$. The light blue modes at $(R_1,T_0)=(10^1,10^{-0.75})$ are sinusoidal in panels A--C but change to a non-sinusoidal shape at largest $k_s$ (panel D), and the dominant wavenumber there is also decreased compared to the sinusoidal cases in panels A--C. The orange mode (at $(R_1,T_0)=(10^4,10^{-0.375})$) has larger $R_1$ and therefore retains a sinusoidal shape even at the largest $k_s$ value.
Moving to much smaller $R_1$, such as the purple mode ($(R_1,T_0)=(10^{-1},10^{-1.375})$) we again have a transition from a sinusoidal shape at $k_s= 0$ to a shape that is less sinusoidal as $k_s$ increases (from panels B to D) and less fore-aft symmetric, with peaks of deflection just upstream of the spring locations, unlike the more fore-aft symmetric red and green shapes
at larger $R_1$.

\begin{figure}[H]
\centering
    \includegraphics[width=.485\textwidth]{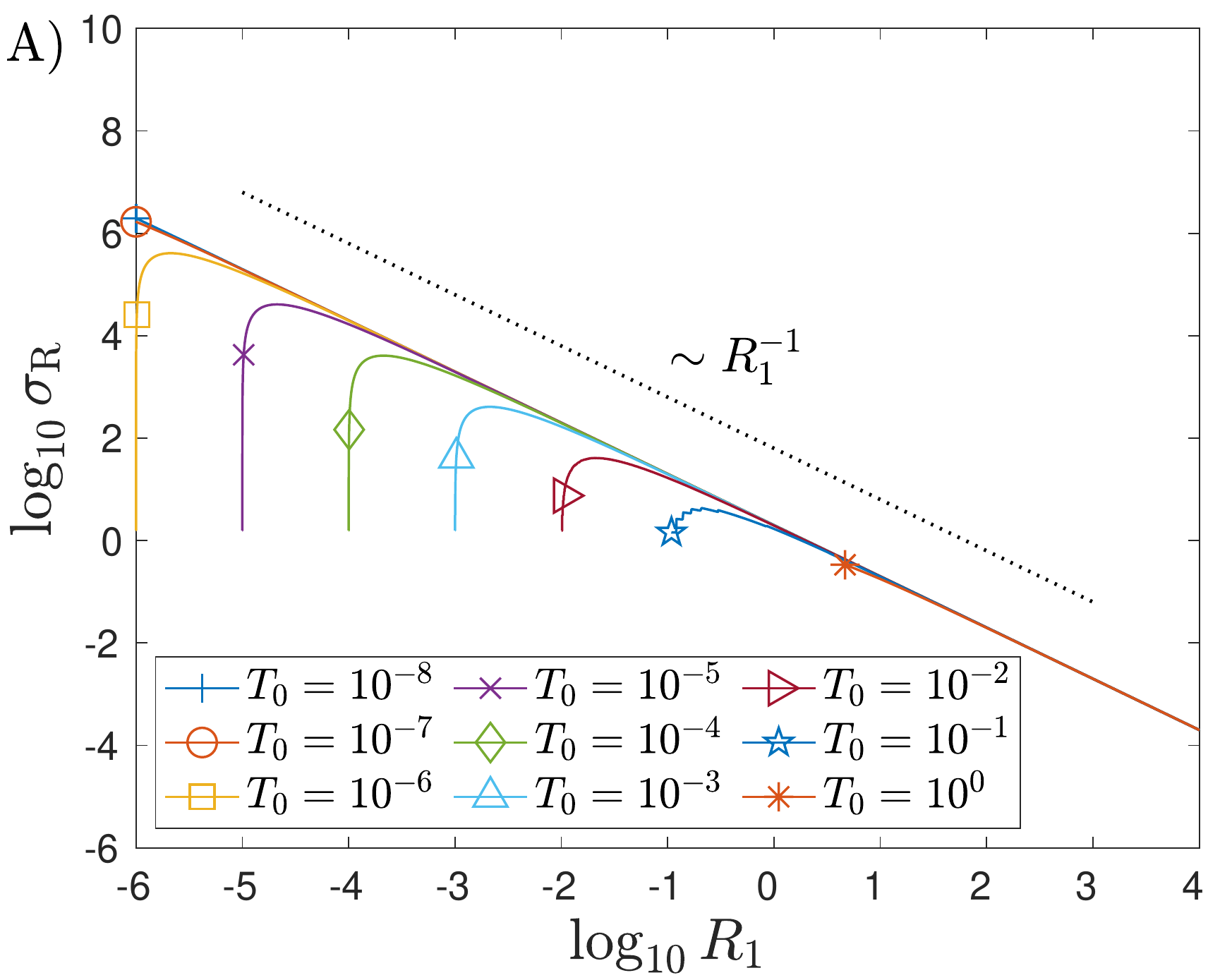}
    \includegraphics[width=.485\textwidth]{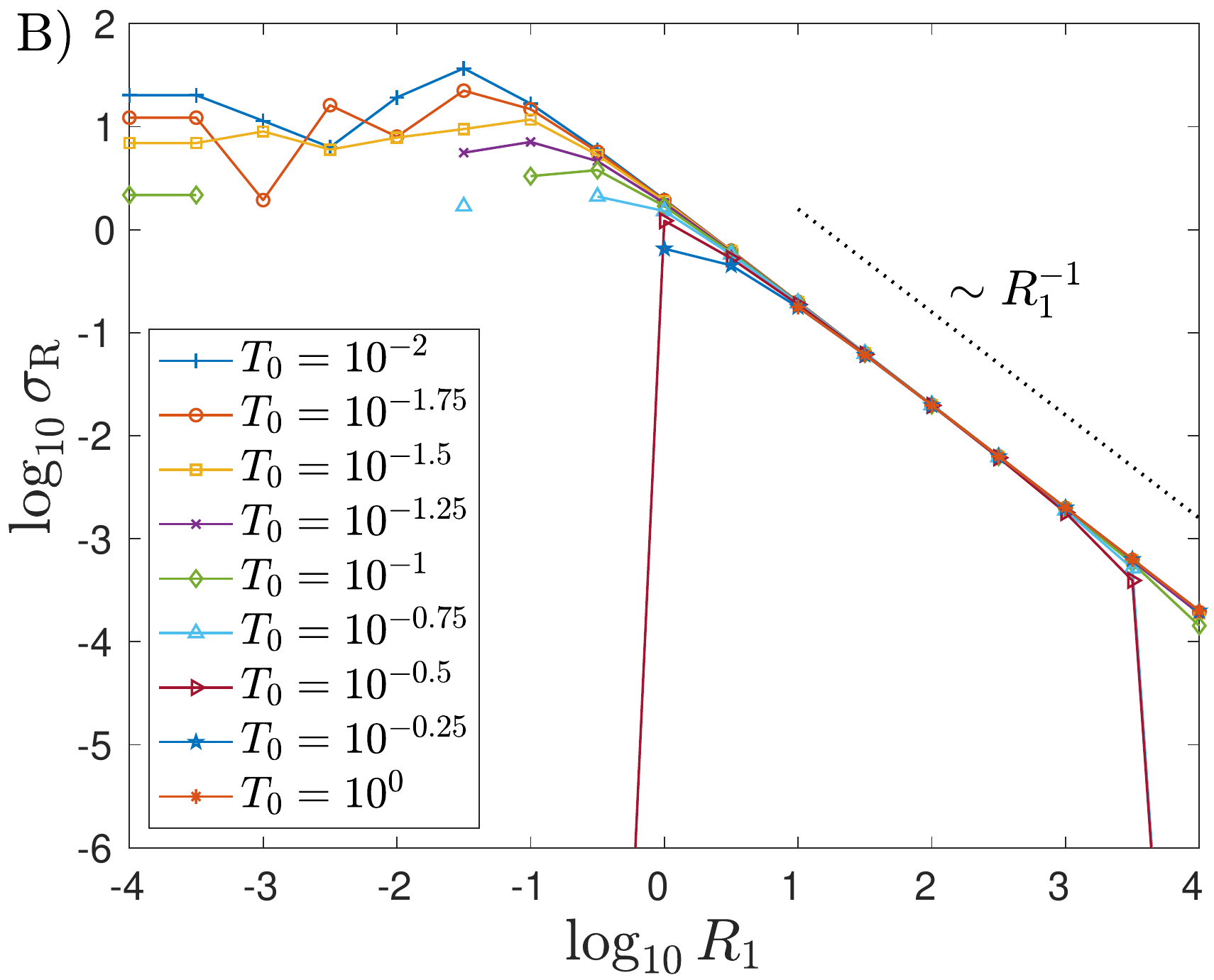}
    \includegraphics[width=.485\textwidth]{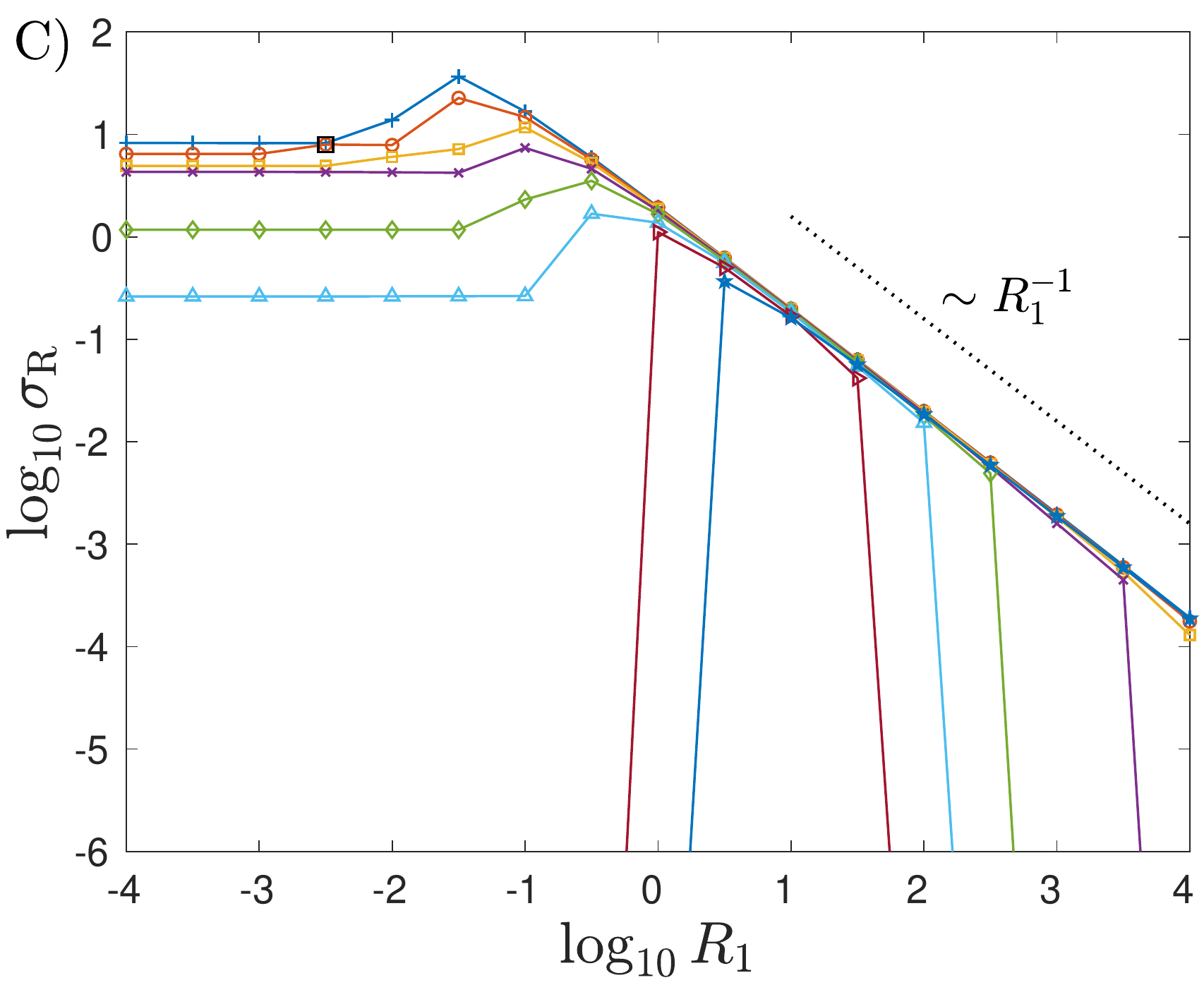}
    \includegraphics[width=.485\textwidth]{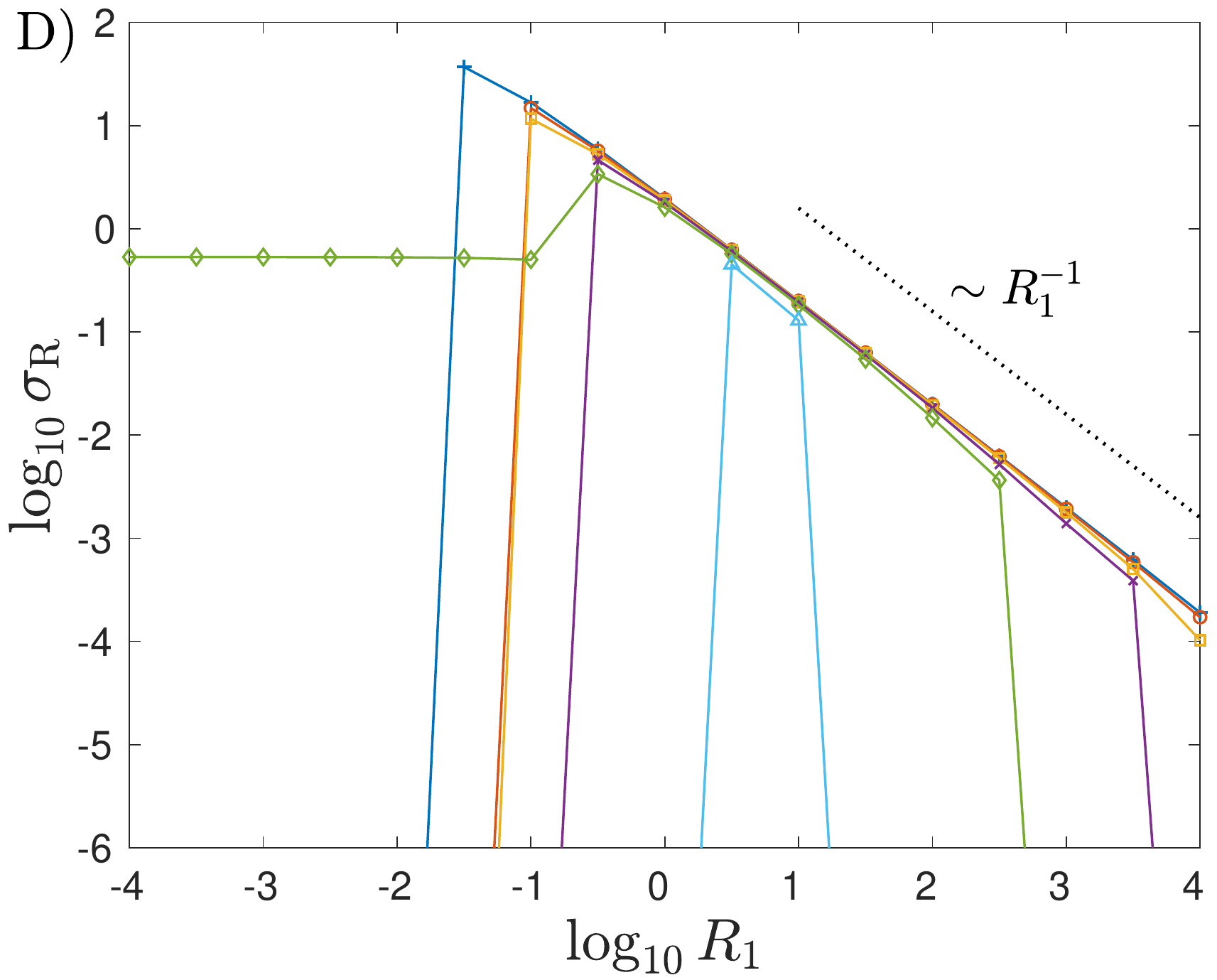}
    \caption{(Infinite, periodic membrane) Plots showing the real parts of the eigenvalues for spring constants: A) $k_s=0$ (analytical result), B) $k_s=10^{-1}$, C) $k_s=10^0$, and D) $k_s=10^1$. Panels B--D share the same legend, and result from computations with $N=2^9$. The dotted black line at moderate-to-large values of $R_1$ shows the scaling~$R_1^{-1}$.}\label{fig:plotsScalingLawsReal}
\end{figure}

We now present the real parts of the eigenvalues within the instability region, with three asymptotic behaviors given in the three rows of the second column of Table~\ref{table:scalingLaws}.
For each row, we find
the dominant behaviors of $\sigma_\Re$ by
inserting the values of $k_{\max}$ from the first column of that row into the first term on the right side of Eq.~\eqref{eq:analyticSigmaks0}, which is $\sigma_\Re$. When we take the appropriate limits for each row (shown on the left side of Table~\ref{table:scalingLaws}), we obtain the expressions for
$\sigma_\Re$ in the second column of Table~\ref{table:scalingLaws}.

% When $R_1 \to \infty$ or $T_0\to 0$, we find  from the first or third rows, respectively, of the first column of Table~\ref{table:scalingLaws}, 
%in  the first term on the right side of Eq.~\eqref{eq:analyticSigmaks0} and obtain $\sigma_\Re\sim 2/R_1$, shown in the first or third rows, respectively, of the second column of Table~\ref{table:scalingLaws}.

%In the second row, we already obtained that the dominant wavenumber behaves as Eq.~\eqref{eq:kmaxT0cR1}. In this limit of $R_1\to 0$, if we substitute the asymptotic $k_{\max}=LC/(4\pi cR_1)$ into the real part of $\sigma$ in Eq.~\eqref{eq:analyticSigmaks0} we get
%\begin{equation}
%    \sigma_\Re\to\frac{2C}{R_1(C+4c)}.
%\end{equation}

% when $R_1\to \infty$. Thus for any $T_0$,
% \begin{equation}
%     \sigma_\Re\sim R_1^{-1}\quad \mathrm{when}\,\,\,R_1\to \infty.
% \end{equation}
%From the form of~$\sigma_\Re$ in Eq.~\eqref{eq:analyticSigmaks0}, if we look at the effect of~$T_0$ it is clear that~$\sigma_\Re$ is independent of~$T_0$.
% As $c\to 0^+$, $T_0\ll R_1$ and Eq.~\eqref{eq:kmaxT0cR1} results in $k_{\max}\to\pm L/(2\pi c R_1)$. 

Fig.~\ref{fig:plotsScalingLawsReal} plots the values of the real parts of the eigenvalues~($\sigma_\Re$) with respect to the membrane mass~($R_1$) for various fixed  $T_0$ (one value per line) and for the same four spring stiffness constants as in Fig.~\ref{fig:scalingLawsWavenumber}, one per panel. 
As with Fig.~\ref{fig:scalingLawsWavenumber}A, the values in Fig.~\ref{fig:plotsScalingLawsReal}A, with $k_s = 0$, are obtained analytically through Eq.~\eqref{eq:analyticSigmaks0}, and are obtained computationally for the remaining panels. Most of the data lie nearly on the straight line given by $2/R_1$, corresponding to the first and third rows in the second column of Table~\ref{table:scalingLaws}. For each $T_0 \leq 10^{-1}$, the corresponding line curves downward and becomes nearly vertical at the stability boundary. A vertical asymptote would occur if $k_{\max}$ could decrease to 0 (as in Eq.~\eqref{eq:kminAnalytical} when $R_1 \to T_0$), but it is bounded below by 1 (as in Fig.~\ref{fig:scalingLawsWavenumber}A), and consequently $\sigma_\Re$ also has a positive lower bound at the stability boundary. 
%Near this asymptote, the value of $\sigma_\Re$ is approximately that in the second row of the second column of Table~\ref{table:scalingLaws} with $c \to 1^-$, which reaches 0 at the stability boundary ($T_0 \approx R_1$). 

When $k_s$ is increased from 0 to $10^{-1}$ we obtain different behaviors, shown in Fig.~\ref{fig:plotsScalingLawsReal}B. 
When $R_1 \gg 1$, the data follow the same $2/R_1$ behavior as in panel A for $T_0$ relatively large but below the stability boundary. At other ($R_1$, $T_0$) pairs, the springs cause different behaviors.
Disconnected lines or points are observed (e.g., at $T_0=10^{-1.25}$, $10^{-1}$, $10^{-0.75}$) where the membrane switches between being stable and unstable. These correspond to the isolated bands of unstable modes seen in Fig.~\ref{fig:scatterPlotsImag}B. 

In Fig.~\ref{fig:plotsScalingLawsReal} (panels B, C, and D), some membranes lose stability by divergence, shown by the sharp drop in some of the graphs to values below $10^{-6}$ (for example, $T_0=10^{-0.5}$ in panels B and
C and $T_0=10^{-1.25}$ and $T_0=10^{-0.75}$ in panel D). The graphs continue to the left or right $R_1$ limits of the plots with values $\approx 10^{-12}$ (not visible), indicating instability by divergence throughout these regions. Divergence occurs for ranges of small and large $R_1$ that are generally more extensive at larger $T_0$ until the stability boundary is reached. When $T_0=10^{-0.25}$ and~$10^{-0.5}$
in panel D all membranes lose stability by divergence. Therefore, the lines for these two cases do not appear in the panel. 
%This agrees with the trend of the ranges of $R_1$ with $\sigma_\Re>10^{-6}$ shrinking in length with increasing~$T_0$. 
%More specifically, in Fig.~\ref{fig:plotsScalingLawsReal}C we observe that a smaller range of~$R_1$ values causes the membrane to lose stability by flutter and divergence ($\sigma_\Re\neq 0$) when~$T_0$ is closer to the stability boundary. 
Another striking effect of $k_s \neq 0$ is the plateaus on the left sides of panels B--D, at small $R_1$. Here the values of $\sigma_\Re$ drop to a plateau instead of a vertical asymptote as in panel A.
The values of $\sigma_\Re$
for each plateau decrease with increasing $T_0$ in most cases in panels B and C.

The small square with the black outline in panel C shows a case with an eigenvalue that is 
% \silas{plural eigenvalues?}
not converged. More of these cases occur in panel D where divergence occurs (below the lower limit of the panel, and so not shown).

%In Fig.~\ref{fig:plotsScalingLawsReal}D, when $R_1\leq 10^{-1.5}$ and $T_0=10^{-1.75}$ (yellow line with square markers), $10^{-1.5}$ (orange line with circle markers) also lead to non-converged eigenvalues, so even though the real part of the eigenvalues being less than $10^{-6}$ indicates loss of stability through divergence we are not sure if that is indeed the real behavior.

%When $R_1\lesssim 10^{-1}$, the effect of~$T_0$ becomes more significant; increasing~$T_0$ in general decreases $\sigma_\Re$ (as can be seen mainly from Fig.~\ref{fig:plotsScalingLawsReal}C with $k_s=10^0$). 

%In all cases, the effect of~$T_0$ on $\sigma_\Re$ is insignificant when $R_1\gtrsim 10^{-1}$, unless the membrane loses stability by divergence in which case $\sigma_\Re<10^{-6}$ (almost vertical lines that drop below $10^{-6}$). This is in agreement with the special case of $k_s=0$ 
% (i.e., spring with stiffness equal to zero)
%in which the real part of the eigenvalue shown in Eq.~\eqref{eq:analyticSigmaks0} is independent of~$T_0$. The dotted black line in Fig.~\ref{fig:plotsScalingLawsReal}A shows that $\sigma_\Re$ follows the scaling $R_1^{-1}$ for large $R_1$ and any $T_0$ and when $R_1$ is small but $T_0< R_1$.
% \silas{in panel A}%.455

\begin{figure}[H]
\centering
    \includegraphics[width=.485\textwidth]{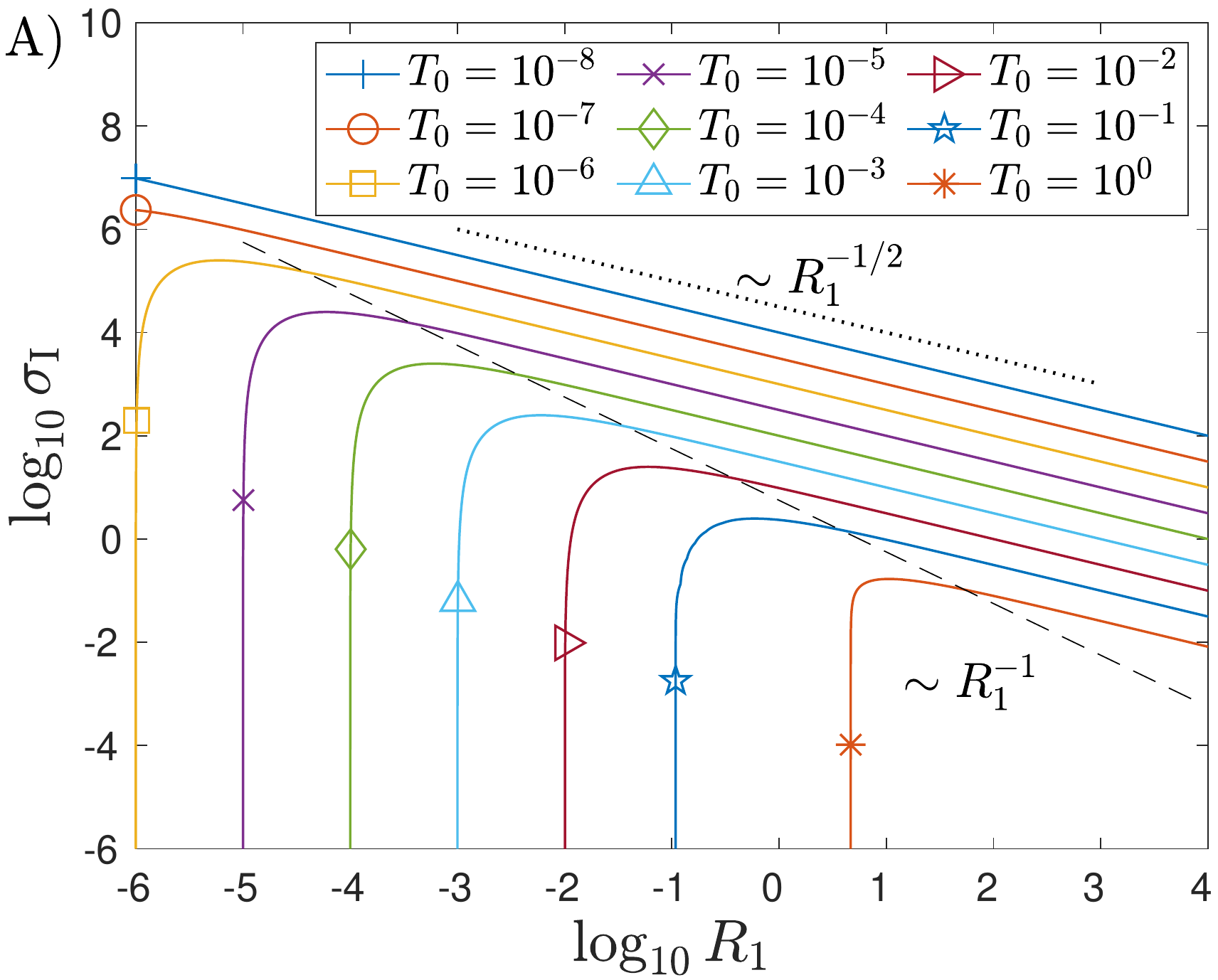}
    \includegraphics[width=.485\textwidth]{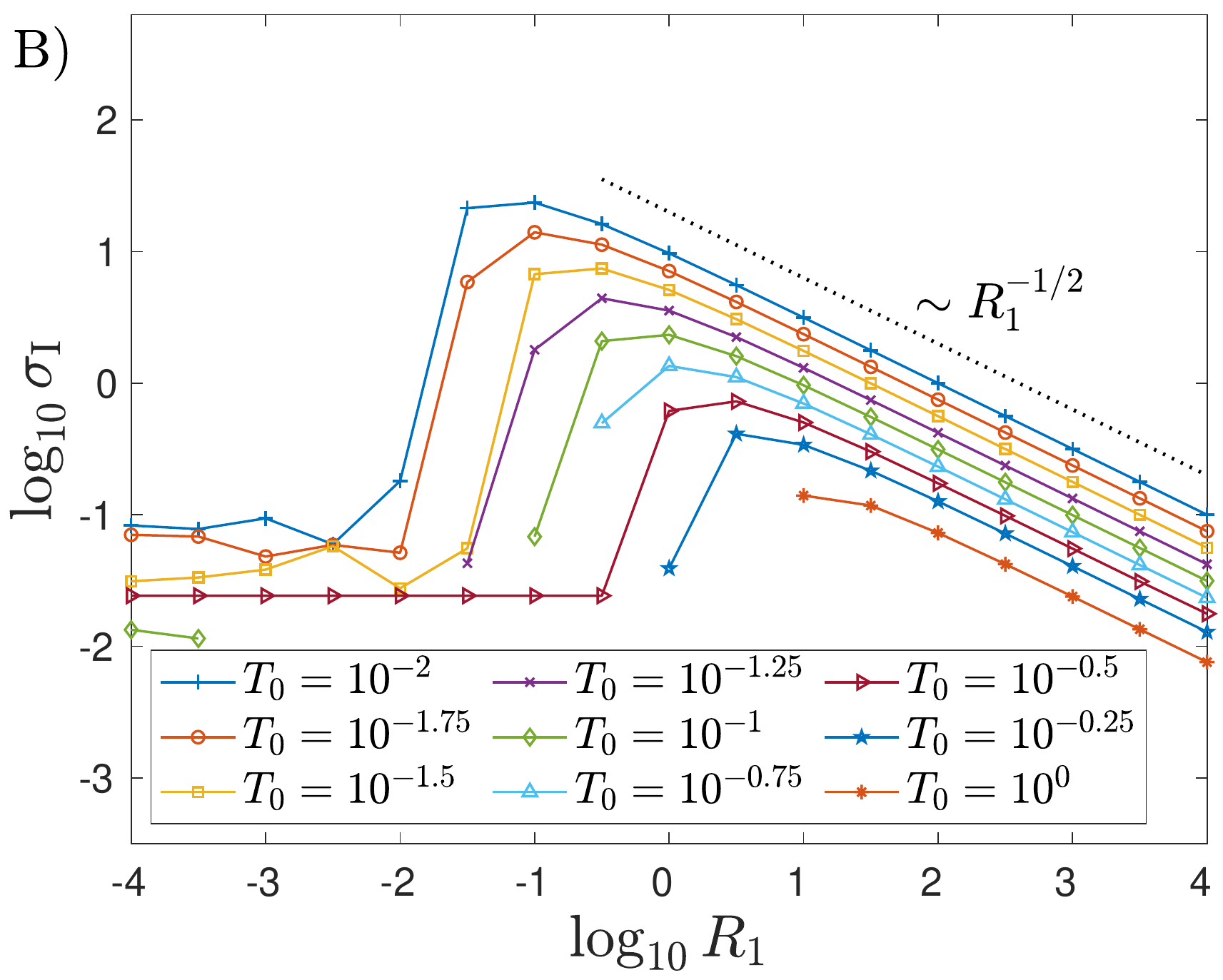}
    \includegraphics[width=.485\textwidth]{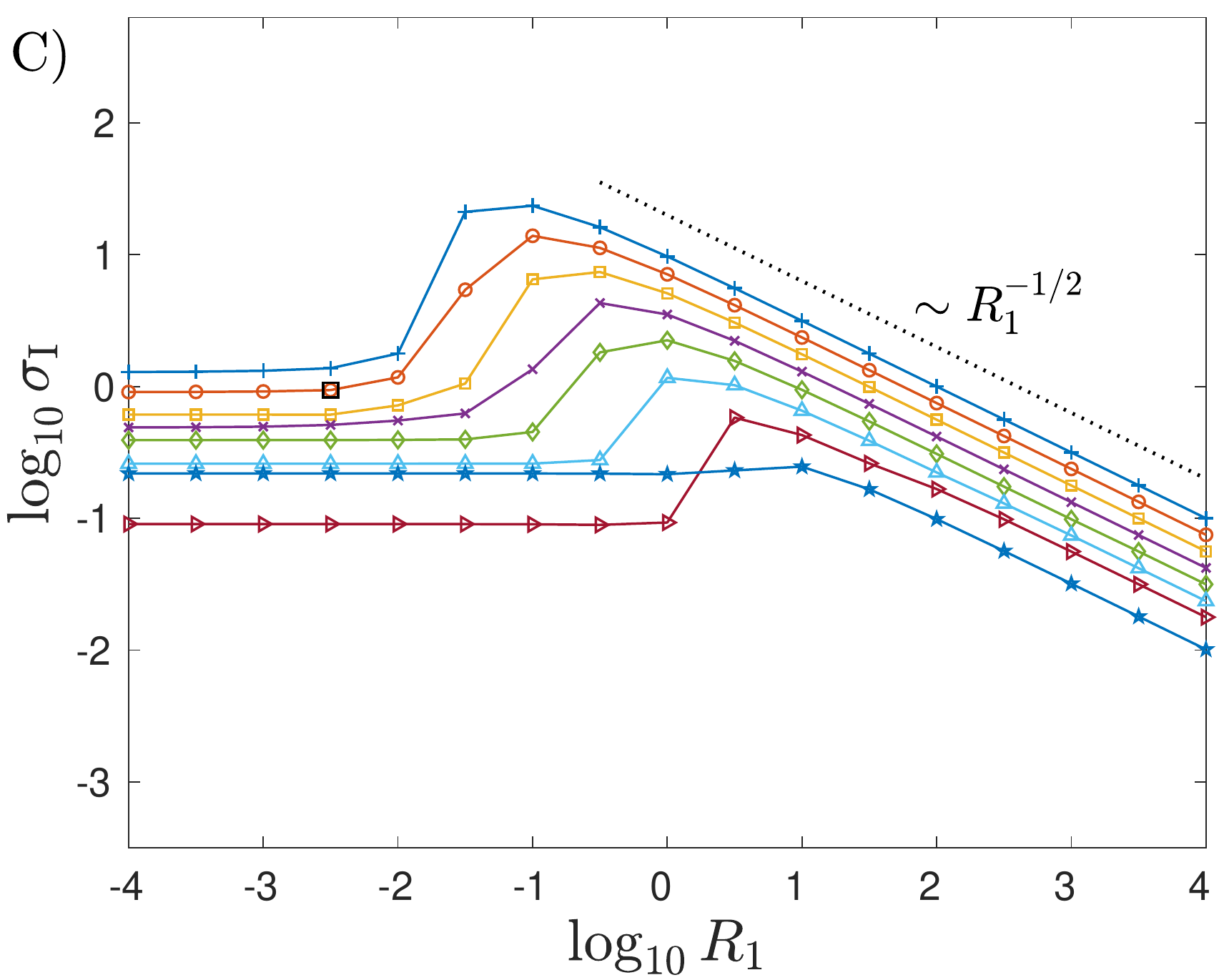}
    \includegraphics[width=.485\textwidth]{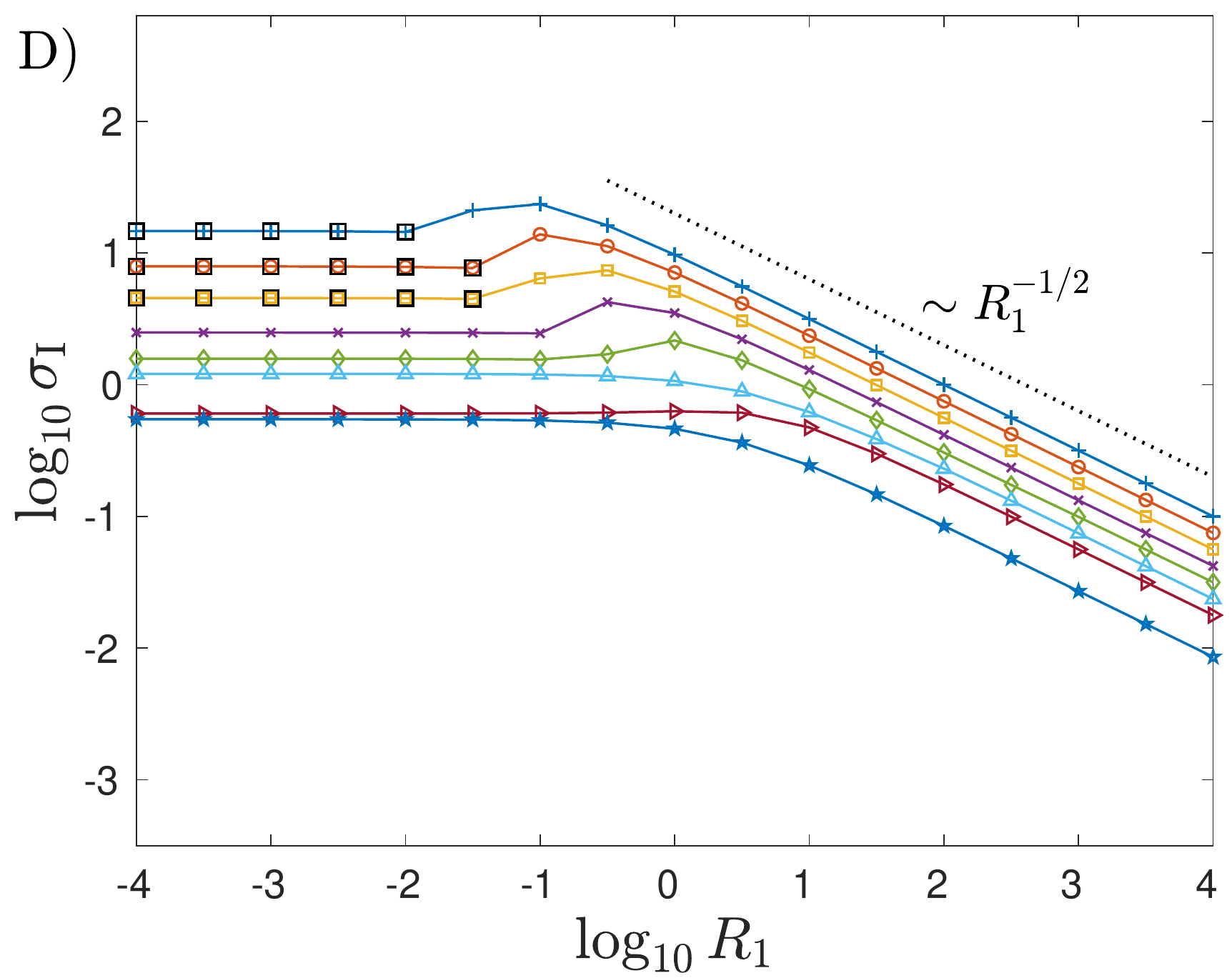}
\caption{(Infinite, periodic membrane) Plots showing the  imaginary parts of the eigenvalues for spring constants: A) $k_s=0$ (analytical result), B) $k_s=10^{-1}$, C) $k_s=10^0$, and D) $k_s=10^1$. Panels B--D share the same legend, and for the numerical results shown we use $N=2^9$. The dotted black line at moderate-to-large values of $R_1$ shows the scaling~$R_1^{-1/2}$.}\label{fig:plotsScalingLawsImag}
\end{figure}

Finally, we present the imaginary parts of the eigenvalues in the unstable region and investigate the same three asymptotic regimes as for the other two quantities in Table~\ref{table:scalingLaws}. For each regime, we derive the dominant behaviors of $\sigma_\Im$ by substituting the $k_{\max}$ values shown in the first column of Table~\ref{table:scalingLaws} in the second term on the right side of Eq.~\eqref{eq:analyticSigmaks0}, which is $\pm i\sigma_\Im$ if $D_k <0$, i.e., the mode is unstable.

In Fig.~\ref{fig:plotsScalingLawsImag} we plot the imaginary parts of the eigenvalues $(\sigma_\Im)$
versus the membrane mass ($R_1$) for various fixed values of $T_0$ and for the same spring stiffness constants, one per panel, as in Figs.~\ref{fig:scalingLawsWavenumber} and~\ref{fig:plotsScalingLawsReal}. In Fig.~\ref{fig:plotsScalingLawsImag}A at large $R_1$ for fixed $T_0$, $\sigma_\Im$ follows the $R_1^{-1/2}$ scaling shown by the dotted line. The lines are equispaced at large $R_1$,
consistent with the scaling $T_0^{-1/2}$
for fixed $R_1$. Both behaviors are consistent with the asymptotic scaling law $\sigma_\Im \sim 1/\sqrt{R_1T_0}$ at large $R_1$ or at small $T_0$, the first and third rows, respectively, of the third column of Table~\ref{table:scalingLaws}. 
%Even though the dotted line indicates the scaling $R_1^{-1/2}$, the equispaced lines in Fig.~\ref{fig:plotsScalingLawsImag}A infer that there also exists an asymptotic scaling law with respect to the dimensionless pretension~$T_0$. In fact, at small $T_0$ with fixed $R_1$,  $\sigma_\Im$ behaves again as $1/\sqrt{R_1T_0}$ (see third row in the third column of Table~\ref{table:scalingLaws}). 
As in Fig.~\ref{fig:plotsScalingLawsReal}A, each line curves downward to a vertical asymptote as it approaches the stability boundary at a certain $R_1$ value. The dashed line
shows the $R_1^{-1}$ scaling of $\sigma_\Im$ when $T_0 = cR_1$, $0 < c < 1$ and $R_1 \to 0$,
given analytically in the second row of the third column of Table~\ref{table:scalingLaws}.

Panels B--D show the results with three nonzero $k_s$ values, and have
many similarities with the corresponding results for $\sigma_\Re$ (Figs.~\ref{fig:plotsScalingLawsReal}B--D). For example, the lines end in panel B where the membrane switches between being stable and unstable.
%One of them is the ($R_1,T_0$)-location of disconnected lines or points seen in panel~B of each of the two figures where
Another similarity, when $k_s \neq 0$, is that $\sigma_\Im$ plateaus on the left sides of panels B--D, at small $R_1$. Here, when $T_0\to R_1^-$ the lines of $\sigma_\Im$ initially curve downward (but not towards a vertical asymptote as in Fig.~\ref{fig:plotsScalingLawsImag}A) and then tend to a constant value at small $R_1$ in most cases. 
%when $R_1<T_0$ with $T_0\leq 10^{-0.75}$. 
These lines curve downward less sharply as $k_s$ increases, and the region of downward curving disappears completely in some cases in panel D.
Another qualitative similarity with Fig.~\ref{fig:plotsScalingLawsReal} is that the growth rate $|\sigma_\Im|$ decreases with increasing~$T_0$ in most cases.
As previously, the small squares with the black outline in Figs.~\ref{fig:plotsScalingLawsImag}C and~\ref{fig:plotsScalingLawsImag}D correspond to $(R_1,T_0)$ pairs where the eigenvalue is not converged with respect to~$N$ but we still include them here to distinguish them from stable membranes where a marker is omitted altogether.

%This tendency of the lines to initially curve downward and then plateau, goes away with increasing $T_0$ as $k_s$ is increased to $10^0$ (panel C) and to $10^1$ (panel D). For example, if we consider $T_0=10^{-0.75}$ (light blue line), we see that   $\sigma_\Im$ curves down before it plateaus in panels B and C, but not in panel D. 

% In Figs.~\ref{fig:plotsScalingLawsImag}B--D the scaling of $\sigma_\Im\sim R_1^{-1/2}$ is not observed for small values of $R_1$. Instead, $\sigma_\Im$ plateaus. This was also observed in Figs.~\ref{fig:plotsScalingLawsReal}B--D for the real part of the eigenvalues.
% \silas{changes to plateaus at small $R_1$}
% In Fig.~\ref{fig:plotsScalingLawsImag}B there is more variation 
% % \silas{compared to what?} 
% in $\sigma_\Im$ with $R_1$ for a fixed $T_0$ compared to Figs.~\ref{fig:plotsScalingLawsImag}C and~\ref{fig:plotsScalingLawsImag}D where the lines for a fixed $T_0$ are more horizontal. 

There is no indication in Figs.~\ref{fig:plotsScalingLawsImag}B--D of a switch in behavior corresponding to the
changes from divergence with flutter to divergence without flutter shown by the sudden drops in $\sigma_\Re$ in Figs.~\ref{fig:plotsScalingLawsReal}B--D. In other words, the imaginary parts of the eigenvalues change smoothly despite the sharp changes in the real parts.
%of the eigenvalue from a nonzero value to zero (computationally, $\sigma_\Re\approx 10^{-12}$).
The orange line with asterisks, $T_0=10^0$, is not present in Figs.~\ref{fig:plotsScalingLawsImag}C and~\ref{fig:plotsScalingLawsImag}D
because the critical $T_0$ for instability drops below $10^0$ as $k_s$ increases to $10^0$ and $10^1$.

% Instead, for a fixed $T_0$ it appears that~$\sigma_\Im$ is independent of $R_1$ (horizontal lines in Figs.~\ref{fig:plotsScalingLawsImag}C and~\ref{fig:plotsScalingLawsImag}D). 

%%%%%%%%%%%%%%%%%%%%%%%%%%%
\section{Summary and conclusions}\label{sec:conclusions}

In this work we have studied the flutter instability of thin membranes whose leading and trailing edges are attached to inextensible rods of length $R$ and Hookean springs of stiffness constant $k_s$. We looked at different parts of the four-dimensional parameter spaces $(R_1,R_3,T_0,R)$ and $(R_1,R_3,T_0,k_s)$. We found that when membranes are attached to rods with small length $R$ or to springs of  moderate-to-large stiffness $k_s$, they exhibit large (but physically reasonable) deflections that converge
to states of steady deflection with single humps that are almost fore-aft symmetric. When $R$ is moderate-to-large and $k_s$ small, we find a wide range of unsteady dynamics, somewhat similar to those seen
in studies of flapping plates or flags (or fixed-free and free-free membranes in~\citet{mavroyiakoumou2020large}). In either of the two regimes, deflections scale as $R_3^{-1/2}$, when the stretching modulus $R_3$ is large. The large-amplitude dynamics depend most strongly on the membrane mass density $R_1$ and less strongly on the pretension $T_0$. At the largest $R_1$ studied we find the smallest oscillation frequencies and largest membrane deflections corresponding to somewhat chaotic and asymmetrical membrane motions. Here the dominant time period scales as $R_1^{1/2}$. As $R_1$
decreases, the membrane motions become more periodic and symmetrical, and with larger spatial frequency components (sharper curvatures and more zero crossings). At
$R_1\lesssim 10^{-1.25}$ the motions become more chaotic again, with much finer spatial features
that are difficult to resolve numerically and so a finer mesh on the membrane is required there.
Our study shows that the boundary conditions (inextensible rods and vertical Hookean springs) allow for a smooth transition between types of membrane dynamics that were observed when both membrane ends are fixed at zero deflection or when one or both ends are free to move in the vertical direction.

To study the onset of membrane instability and small-amplitude membrane motions, we used a linearized model and a nonlinear eigenvalue solver---similar to the one in~\citet{mavroyiakoumou2021eigenmode}.
In this regime, the nonlinear $R_3$ term in Eq.~\eqref{eq:membrane} is negligible so we characterized the different types of motions with respect to the other two key dimensionless parameters---membrane mass and pretension. In the small-amplitude model we focused on the vertical Hookean springs, equivalent to inextensible rods %although one could have looked at the inextensible rods instead and the dynamics would be related through 
via $1/R=k_s/T_0$. 
When membrane inertia and pretension dominate fluid pressure forces, the eigenmodes tend toward neutrally stable sinusoidal functions. As we increase~$k_s$, we transition from membranes that resemble the free-free case to membranes that resemble the fixed-fixed case. There are roughly two regimes: small membrane density, where divergence occurs and the most unstable mode becomes more fore-aft asymmetric as one moves further into the instability region, and large membrane density, where flutter and divergence occur with approximately sinusoidal modes. In both regimes, the most unstable modes become wavier at smaller $T_0$, akin to the most unstable beam modes at smaller bending rigidity in~\cite{alben2008ffi}. The stability boundaries with $k_s=10^{-1}$ and $10^0$ are very similar at
large membrane densities, showing an upward slope for $R_1$. This upward slope for $R_1$ is also seen with $k_s=10^1$ but it starts at a larger $R_1$.
 
To derive asymptotic scaling laws theoretically, we introduced a simplified model with spatially periodic solutions by assuming that the membrane extends to infinity upstream and downstream and is tethered by an infinite, periodic array of Hookean springs, all with stiffness $k_s$. This model corresponds to a standard eigenvalue problem, and is much faster to compute than the nonlinear eigenvalue problem of the membrane-vortex-wake model. We can thus study much wider ranges of the key parameters $R_1$, $T_0$, and $k_s$. When $k_s = 0$ we can compute asymptotic scaling laws for the real and imaginary parts of the eigenvalues, and the dominant wave number of the most unstable eigenmodes. We find that 
as $R_1$ increases from small to large, the dominant wave number scaling varies from $R_1^{-1}$ to $R_1^0$ for the periodic membrane within the instability region. In the large amplitude simulations, the time-averaged number of extrema of deflection also changes from $R_1^{-1}$ to $R_1^0$ scalings as $R_1$ increases from small to large. For the periodic membrane, the frequency $\sigma_\Re$ scales as $R_1^{-1}$ at both small and large $R_1$, while the large-amplitude dominant frequency transitions scales as $R_1^{-5/6}$ and $R_1^{-1/2}$, respectively. At small $R_1$, the large-amplitude results are mostly independent of $T_0$ within the instability region, while the periodic membrane results do depend on $T_0$. For the periodic membrane, we also considered the small-amplitude growth rate $\sigma_\Im$.  At large $R_1$, it decays as $R_1^{-1/2}$ for a fixed $T_0$; at small $R_1$ and $T_0 = c R_1$ for $0 < c < 1$, it decays as $R_1^{-1}$. When $k_s$ is increased to a nonzero value, both $\sigma_\Re$ and $\sigma_\Im$
plateau at small $R_1$.

There are qualitative similarities in the  shapes of the stability boundaries for the periodic membrane and membrane-vortex-wake models. At small $R_1$, the stability boundaries have a plateau at a certain $T_0$ value, that decreases as $k_s$
decreases.  At large $R_1$, the periodic membrane has a flat stability boundary, while that with the vortex wake is upward sloping, corresponding to unstable modes at larger $T_0$, albeit with very slow growth rates.
At all $R_1$, as $k_s$ increases divergence modes become more common near the stability boundary in both models.

The membrane modes from the two models also share many features. For example, the mode shapes become wavier at smaller $T_0$ in both models. Additionally, by tracking the eigenmodes across the three parameter space of $R_1$, $T_0$, and $k_s$, we found that at larger $R_1$, the modes are more sinusoidal and fore-aft symmetric in both models. At small-to-moderate $R_1$, the modes are more asymmetric, with peak deflections shifted downstream.

%%%%%%%%%%%%%%%%%%%%%%%%%%%%%%%%%
\section*{Acknowledgements}

We acknowledge support from the NSF Mathematical Biology program under Award No.\ DMS-1811889.
%%%%%%%%%%%%%%%%%%%%%%%%%%%%%%%%%

\appendix

\section{Numerical method for determining the set of eigenvalues and eigenmodes for each membrane}\label{app:numericalEigenmodes}

We solve the nonlinear eigenvalue problem iteratively by the same method  as in~\citet{mavroyiakoumou2021eigenmode} but we include a brief description here for completeness.

At each iteration, we have an approximation~$\sigma_0$ to a given eigenvalue $\sigma$. We approximate the equations as a quadratic eigenvalue problem:
\begin{equation}\label{eq:quadevalue}
[\sigma^2 A_2+\sigma A_1+A_0(\sigma_0)]w=0,
\end{equation}
where the matrices $A_2$, $A_1$, $A_0$ are known from Eqs.~\eqref{eq:Ymembrane}, \eqref{eq:2intB}, and $g(x)=V(x)/\sqrt{1-x^2}$. The eigenvector consists $w$ consists of: (a) values of the eigenmodes, defined as $Y$ on the Chebyshev grid $\{x_j=\cos\theta_j,\,\theta_j=(j-1)\pi /m,\, j=1,\dots,m+1\}$ and (b) the scalar $\Gamma_0$. The term $A_0(\sigma)w$ includes the exponential integral involving~$\sigma$ in Eq.~\eqref{eq:2intB} as well as terms that are constant in $\sigma$. In the exponential integral, $\sigma$ is fixed at $\sigma_0$, the value of $\sigma$ from the previous iteration, resulting in the quadratic eigenvalue problem [Eq.~\eqref{eq:quadevalue}], which is solved using \texttt{polyeig} function in \textsc{Matlab}. Eq.~\eqref{eq:quadevalue} has $2m+4$ eigenvalue solutions. As in~\citet{alben2008ffi,mavroyiakoumou2021eigenmode}, we define an error function as the difference between $\sigma_0$ and the eigenvalue (out of the $2m+4$ possibilities) closest to it. We also compute the derivatives of the error function (i.e.,\ the Jacobian matrix) with respect to $\sigma_{\Re}$ and $\sigma_{\Im}$ using finite differences at the initial iterate, and update it at subsequent iterates using Broyden's approximate formula. The error function and Jacobian define the search direction (via Newton's formula) for the next iterate. With this approach we obtain superlinear convergence to a given eigenvalue. By using a wide range of initial guesses, we obtain convergence to various eigenvalues and corresponding eigenmodes. More details about the numerical method can be found in~\citet{mavroyiakoumou2021eigenmode}.

\bibliographystyle{elsarticle-harv}
\biboptions{authoryear}
\bibliography{biblio.bib} 

\begin{thebibliography}{64}
\expandafter\ifx\csname natexlab\endcsname\relax\def\natexlab#1{#1}\fi
\providecommand{\url}[1]{\texttt{#1}}
\providecommand{\href}[2]{#2}
\providecommand{\path}[1]{#1}
\providecommand{\DOIprefix}{doi:}
\providecommand{\ArXivprefix}{arXiv:}
\providecommand{\URLprefix}{URL: }
\providecommand{\Pubmedprefix}{pmid:}
\providecommand{\doi}[1]{\href{http://dx.doi.org/#1}{\path{#1}}}
\providecommand{\Pubmed}[1]{\href{pmid:#1}{\path{#1}}}
\providecommand{\bibinfo}[2]{#2}
\ifx\xfnm\relax \def\xfnm[#1]{\unskip,\space#1}\fi
%Type = Article
\bibitem[{Alben(2008)}]{alben2008ffi}
\bibinfo{author}{Alben, S.}, \bibinfo{year}{2008}.
\newblock \bibinfo{title}{{The flapping-flag instability as a nonlinear
  eigenvalue problem}}.
\newblock \bibinfo{journal}{Physics of Fluids} \bibinfo{volume}{20},
  \bibinfo{pages}{104106}.
%Type = Article
\bibitem[{Alben(2009)}]{alben2009simulating}
\bibinfo{author}{Alben, S.}, \bibinfo{year}{2009}.
\newblock \bibinfo{title}{Simulating the dynamics of flexible bodies and vortex
  sheets}.
\newblock \bibinfo{journal}{Journal of Computational Physics}
  \bibinfo{volume}{228}, \bibinfo{pages}{2587--2603}.
%Type = Article
\bibitem[{Alben(2015)}]{alben2015flag}
\bibinfo{author}{Alben, S.}, \bibinfo{year}{2015}.
\newblock \bibinfo{title}{Flag flutter in inviscid channel flow}.
\newblock \bibinfo{journal}{Physics of Fluids} \bibinfo{volume}{27},
  \bibinfo{pages}{033603}.
%Type = Article
\bibitem[{Alben and Shelley(2008)}]{alben2008flapping}
\bibinfo{author}{Alben, S.}, \bibinfo{author}{Shelley, M.J.},
  \bibinfo{year}{2008}.
\newblock \bibinfo{title}{Flapping states of a flag in an inviscid fluid:
  bistability and the transition to chaos}.
\newblock \bibinfo{journal}{Phys. Rev. Lett.} \bibinfo{volume}{100},
  \bibinfo{pages}{074301}.
%Type = Article
\bibitem[{Arb{\'o}s-Torrent et~al.(2013)Arb{\'o}s-Torrent, Ganapathisubramani
  and Palacios}]{arbos2013leading}
\bibinfo{author}{Arb{\'o}s-Torrent, S.}, \bibinfo{author}{Ganapathisubramani,
  B.}, \bibinfo{author}{Palacios, R.}, \bibinfo{year}{2013}.
\newblock \bibinfo{title}{Leading-and trailing-edge effects on the
  aeromechanics of membrane aerofoils}.
\newblock \bibinfo{journal}{J. Fluids and Struct.} \bibinfo{volume}{38},
  \bibinfo{pages}{107--126}.
%Type = Article
\bibitem[{Argentina and Mahadevan(2005)}]{argentina2005fluid}
\bibinfo{author}{Argentina, M.}, \bibinfo{author}{Mahadevan, L.},
  \bibinfo{year}{2005}.
\newblock \bibinfo{title}{Fluid-flow-induced flutter of a flag}.
\newblock \bibinfo{journal}{Proceedings of the National Academy of Sciences}
  \bibinfo{volume}{102}, \bibinfo{pages}{1829--1834}.
%Type = Article
\bibitem[{Carrier(1945)}]{carrier1945non}
\bibinfo{author}{Carrier, G.F.}, \bibinfo{year}{1945}.
\newblock \bibinfo{title}{On the non-linear vibration problem of the elastic
  string}.
\newblock \bibinfo{journal}{Quarterly of Applied Mathematics}
  \bibinfo{volume}{3}, \bibinfo{pages}{157--165}.
%Type = Article
\bibitem[{Cheney et~al.(2015)Cheney, Konow, Bearnot and
  Swartz}]{cheney2015wrinkle}
\bibinfo{author}{Cheney, J.A.}, \bibinfo{author}{Konow, N.},
  \bibinfo{author}{Bearnot, A.}, \bibinfo{author}{Swartz, S.M.},
  \bibinfo{year}{2015}.
\newblock \bibinfo{title}{A wrinkle in flight: the role of elastin fibres in
  the mechanical behaviour of bat wing membranes}.
\newblock \bibinfo{journal}{Journal of the Royal Society Interface}
  \bibinfo{volume}{12}, \bibinfo{pages}{20141286}.
%Type = Book
\bibitem[{Colgate(1996)}]{colgate1996fundamentals}
\bibinfo{author}{Colgate, S.}, \bibinfo{year}{1996}.
\newblock \bibinfo{title}{Fundamentals of Sailing, Cruising, and Racing}.
\newblock \bibinfo{publisher}{WW Norton \& Company}, \bibinfo{address}{New
  York}.
%Type = Article
\bibitem[{Connell and Yue(2007)}]{connell2007fdf}
\bibinfo{author}{Connell, B.S.H.}, \bibinfo{author}{Yue, D.K.P.},
  \bibinfo{year}{2007}.
\newblock \bibinfo{title}{Flapping dynamics of a flag in a uniform stream}.
\newblock \bibinfo{journal}{J. Fluid Mech.} \bibinfo{volume}{581},
  \bibinfo{pages}{33--67}.
%Type = Article
\bibitem[{Das et~al.(2020a)Das, Breuer and Mathai}]{das2020nonlinear}
\bibinfo{author}{Das, A.}, \bibinfo{author}{Breuer, K.S.},
  \bibinfo{author}{Mathai, V.}, \bibinfo{year}{2020}a.
\newblock \bibinfo{title}{Nonlinear modeling and characterization of ultrasoft
  silicone elastomers}.
\newblock \bibinfo{journal}{Applied Physics Letters} \bibinfo{volume}{116},
  \bibinfo{pages}{203702}.
%Type = Inproceedings
\bibitem[{Das et~al.(2020b)Das, Mathai and Breuer}]{das2020deformation}
\bibinfo{author}{Das, A.}, \bibinfo{author}{Mathai, V.},
  \bibinfo{author}{Breuer, K.}, \bibinfo{year}{2020}b.
\newblock \bibinfo{title}{Deformation, forces, and flows associated with
  extremely compliant membrane disks}, in: \bibinfo{booktitle}{AIAA Scitech
  2020 Forum}, p. \bibinfo{pages}{1049}.
%Type = Article
\bibitem[{Eloy et~al.(2008)Eloy, Lagrange, Souilliez and
  Schouveiler}]{eloy2008aeroelastic}
\bibinfo{author}{Eloy, C.}, \bibinfo{author}{Lagrange, R.},
  \bibinfo{author}{Souilliez, C.}, \bibinfo{author}{Schouveiler, L.},
  \bibinfo{year}{2008}.
\newblock \bibinfo{title}{Aeroelastic instability of cantilevered flexible
  plates in uniform flow}.
\newblock \bibinfo{journal}{J. Fluid Mech.} \bibinfo{volume}{611},
  \bibinfo{pages}{97–106}.
\newblock \DOIprefix\doi{10.1017/S002211200800284X}.
%Type = Article
\bibitem[{Eloy et~al.(2007)Eloy, Souilliez and Schouveiler}]{eloy2007flutter}
\bibinfo{author}{Eloy, C.}, \bibinfo{author}{Souilliez, C.},
  \bibinfo{author}{Schouveiler, L.}, \bibinfo{year}{2007}.
\newblock \bibinfo{title}{Flutter of a rectangular plate}.
\newblock \bibinfo{journal}{J. Fluids and Struct.} \bibinfo{volume}{23},
  \bibinfo{pages}{904--919}.
%Type = Book
\bibitem[{Farlow(1993)}]{farlow1993partial}
\bibinfo{author}{Farlow, S.J.}, \bibinfo{year}{1993}.
\newblock \bibinfo{title}{Partial differential equations for scientists and
  engineers}.
\newblock \bibinfo{publisher}{Courier Corporation}, \bibinfo{address}{New
  York}.
%Type = Article
\bibitem[{Gordnier and Attar(2014)}]{gordnier2014impact}
\bibinfo{author}{Gordnier, R.E.}, \bibinfo{author}{Attar, P.J.},
  \bibinfo{year}{2014}.
\newblock \bibinfo{title}{Impact of flexibility on the aerodynamics of an
  aspect ratio two membrane wing}.
\newblock \bibinfo{journal}{J. Fluids and Struct.} \bibinfo{volume}{45},
  \bibinfo{pages}{138--152}.
%Type = Book
\bibitem[{Graff(1975)}]{graff1975wave}
\bibinfo{author}{Graff, K.F.}, \bibinfo{year}{1975}.
\newblock \bibinfo{title}{Wave motion in elastic solids}.
\newblock \bibinfo{publisher}{Oxford University Press},
  \bibinfo{address}{Oxford, England}.
%Type = Article
\bibitem[{Haruo(1975)}]{haruo1975flutter}
\bibinfo{author}{Haruo, K.}, \bibinfo{year}{1975}.
\newblock \bibinfo{title}{Flutter of hanging roofs and curved membrane roofs}.
\newblock \bibinfo{journal}{International Journal of Solids and Structures}
  \bibinfo{volume}{11}, \bibinfo{pages}{477--492}.
%Type = Article
\bibitem[{Hooke(1678)}]{hooke1678potentia}
\bibinfo{author}{Hooke, R.}, \bibinfo{year}{1678}.
\newblock \bibinfo{title}{{De Potentia Restitutiva, or of Spring. Explaining
  the Power of Springing Bodies}}.
\newblock \bibinfo{journal}{Royal Society London} , \bibinfo{pages}{1--56}.
%Type = Article
\bibitem[{Hu et~al.(2008)Hu, Tamai and Murphy}]{hu2008flexible}
\bibinfo{author}{Hu, H.}, \bibinfo{author}{Tamai, M.}, \bibinfo{author}{Murphy,
  J.T.}, \bibinfo{year}{2008}.
\newblock \bibinfo{title}{Flexible-membrane airfoils at low {R}eynolds
  numbers}.
\newblock \bibinfo{journal}{Journal of Aircraft} \bibinfo{volume}{45},
  \bibinfo{pages}{1767--1778}.
%Type = Article
\bibitem[{Jaworski(2012)}]{jaworski2012thrust}
\bibinfo{author}{Jaworski, J.W.}, \bibinfo{year}{2012}.
\newblock \bibinfo{title}{Thrust and aerodynamic forces from an oscillating
  leading edge flap}.
\newblock \bibinfo{journal}{AIAA journal} \bibinfo{volume}{50},
  \bibinfo{pages}{2928--2931}.
%Type = Article
\bibitem[{Jaworski and Gordnier(2012)}]{jaworski2012high}
\bibinfo{author}{Jaworski, J.W.}, \bibinfo{author}{Gordnier, R.E.},
  \bibinfo{year}{2012}.
\newblock \bibinfo{title}{High-order simulations of low {R}eynolds number
  membrane airfoils under prescribed motion}.
\newblock \bibinfo{journal}{J. Fluids and Struct.} \bibinfo{volume}{31},
  \bibinfo{pages}{49--66}.
%Type = Article
\bibitem[{Kashy et~al.(1997)Kashy, Johnson, McIntyre and
  Wolfe}]{kashy1997transverse}
\bibinfo{author}{Kashy, E.}, \bibinfo{author}{Johnson, D.A.},
  \bibinfo{author}{McIntyre, J.}, \bibinfo{author}{Wolfe, S.L.},
  \bibinfo{year}{1997}.
\newblock \bibinfo{title}{Transverse standing waves in a string with free
  ends}.
\newblock \bibinfo{journal}{American Journal of Physics} \bibinfo{volume}{65},
  \bibinfo{pages}{310--313}.
%Type = Book
\bibitem[{Kimball(2009)}]{kimball2009physics}
\bibinfo{author}{Kimball, J.}, \bibinfo{year}{2009}.
\newblock \bibinfo{title}{Physics of sailing}.
\newblock \bibinfo{publisher}{CRC Press}, \bibinfo{address}{Boca Raton, FL}.
%Type = Article
\bibitem[{Knudson(1991)}]{knudson1991recent}
\bibinfo{author}{Knudson, W.C.}, \bibinfo{year}{1991}.
\newblock \bibinfo{title}{Recent advances in the field of long span tension
  structures}.
\newblock \bibinfo{journal}{Engineering Structures} \bibinfo{volume}{13},
  \bibinfo{pages}{164--177}.
%Type = Article
\bibitem[{Kornecki et~al.(1976)Kornecki, Dowell and O'Brien}]{kornecki1976ait}
\bibinfo{author}{Kornecki, A.}, \bibinfo{author}{Dowell, E.H.},
  \bibinfo{author}{O'Brien, J.}, \bibinfo{year}{1976}.
\newblock \bibinfo{title}{{On the aeroelastic instability of two-dimensional
  panels in uniform incompressible flow}}.
\newblock \bibinfo{journal}{J. Sound Vibration} \bibinfo{volume}{47},
  \bibinfo{pages}{163--178}.
%Type = Article
\bibitem[{Le~Ma{\^\i}tre et~al.(1999)Le~Ma{\^\i}tre, Huberson and
  De~Cursi}]{le1999unsteady}
\bibinfo{author}{Le~Ma{\^\i}tre, O.}, \bibinfo{author}{Huberson, S.},
  \bibinfo{author}{De~Cursi, E.S.}, \bibinfo{year}{1999}.
\newblock \bibinfo{title}{Unsteady model of sail and flow interaction}.
\newblock \bibinfo{journal}{J. Fluids and Struct.} \bibinfo{volume}{13},
  \bibinfo{pages}{37--59}.
%Type = Article
\bibitem[{Lian and Shyy(2005)}]{lian2005numerical}
\bibinfo{author}{Lian, Y.}, \bibinfo{author}{Shyy, W.}, \bibinfo{year}{2005}.
\newblock \bibinfo{title}{Numerical simulations of membrane wing aerodynamics
  for micro air vehicle applications}.
\newblock \bibinfo{journal}{Journal of Aircraft} \bibinfo{volume}{42},
  \bibinfo{pages}{865--873}.
%Type = Article
\bibitem[{Manela and Weidenfeld(2017)}]{manela2017hanging}
\bibinfo{author}{Manela, A.}, \bibinfo{author}{Weidenfeld, M.},
  \bibinfo{year}{2017}.
\newblock \bibinfo{title}{The `hanging flag' problem: on the heaving motion of
  a thin filament in the limit of small flexural stiffness}.
\newblock \bibinfo{journal}{J. Fluid Mech.} \bibinfo{volume}{829},
  \bibinfo{pages}{190--213}.
%Type = Article
\bibitem[{Mavroyiakoumou and Alben(2020)}]{mavroyiakoumou2020large}
\bibinfo{author}{Mavroyiakoumou, C.}, \bibinfo{author}{Alben, S.},
  \bibinfo{year}{2020}.
\newblock \bibinfo{title}{Large-amplitude membrane flutter in inviscid flow}.
\newblock \bibinfo{journal}{J. Fluid Mech.} \bibinfo{volume}{891},
  \bibinfo{pages}{A23}.
\newblock \DOIprefix\doi{10.1017/jfm.2020.153}.
%Type = Article
\bibitem[{Mavroyiakoumou and Alben(2021)}]{mavroyiakoumou2021eigenmode}
\bibinfo{author}{Mavroyiakoumou, C.}, \bibinfo{author}{Alben, S.},
  \bibinfo{year}{2021}.
\newblock \bibinfo{title}{Eigenmode analysis of membrane stability in inviscid
  flow}.
\newblock \bibinfo{journal}{Phys. Rev. Fluids} \bibinfo{volume}{6},
  \bibinfo{pages}{043901}.
\newblock \URLprefix
  \url{https://link.aps.org/doi/10.1103/PhysRevFluids.6.043901},
  \DOIprefix\doi{10.1103/PhysRevFluids.6.043901}.
%Type = Article
\bibitem[{Michelin et~al.(2008)Michelin, Smith and Glover}]{michelin2008vortex}
\bibinfo{author}{Michelin, S.}, \bibinfo{author}{Smith, S.G.L.},
  \bibinfo{author}{Glover, B.J.}, \bibinfo{year}{2008}.
\newblock \bibinfo{title}{Vortex shedding model of a flapping flag}.
\newblock \bibinfo{journal}{J. Fluid Mech.} \bibinfo{volume}{617},
  \bibinfo{pages}{1--10}.
%Type = Article
\bibitem[{Narasimha(1968)}]{narasimha1968non}
\bibinfo{author}{Narasimha, R.}, \bibinfo{year}{1968}.
\newblock \bibinfo{title}{Non-linear vibration of an elastic string}.
\newblock \bibinfo{journal}{Journal of Sound and Vibration}
  \bibinfo{volume}{8}, \bibinfo{pages}{134--146}.
%Type = Inproceedings
\bibitem[{Nardini et~al.(2018)Nardini, Illingworth and
  Sandberg}]{nardini2018reduced}
\bibinfo{author}{Nardini, M.}, \bibinfo{author}{Illingworth, S.J.},
  \bibinfo{author}{Sandberg, R.D.}, \bibinfo{year}{2018}.
\newblock \bibinfo{title}{Reduced-order modeling for fluid-structure
  interaction of membrane wings at low and moderate {R}eynolds numbers}, in:
  \bibinfo{booktitle}{2018 AIAA Aerospace Sciences Meeting}, p.
  \bibinfo{pages}{1544}.
%Type = Book
\bibitem[{Nayfeh and Pai(2008)}]{nayfeh2008linear}
\bibinfo{author}{Nayfeh, A.H.}, \bibinfo{author}{Pai, P.F.},
  \bibinfo{year}{2008}.
\newblock \bibinfo{title}{Linear and nonlinear structural mechanics}.
\newblock \bibinfo{publisher}{John Wiley \& Sons}, \bibinfo{address}{New York}.
%Type = Article
\bibitem[{Newman and Low(1984)}]{newman1984two}
\bibinfo{author}{Newman, B.G.}, \bibinfo{author}{Low, H.T.},
  \bibinfo{year}{1984}.
\newblock \bibinfo{title}{Two-dimensional impervious sails: experimental
  results compared with theory}.
\newblock \bibinfo{journal}{J. Fluid Mech.} \bibinfo{volume}{144},
  \bibinfo{pages}{445--462}.
%Type = Article
\bibitem[{Newman and Paidoussis(1991)}]{newman1991stability}
\bibinfo{author}{Newman, B.G.}, \bibinfo{author}{Paidoussis, M.P.},
  \bibinfo{year}{1991}.
\newblock \bibinfo{title}{The stability of two-dimensional membranes in
  streaming flow}.
\newblock \bibinfo{journal}{J. Fluids and Struct.} \bibinfo{volume}{5},
  \bibinfo{pages}{443--454}.
%Type = Article
\bibitem[{Pepper and Maydew(1971)}]{pepper1971aerodynamic}
\bibinfo{author}{Pepper, W.B.}, \bibinfo{author}{Maydew, R.C.},
  \bibinfo{year}{1971}.
\newblock \bibinfo{title}{Aerodynamic decelerators-an engineering review}.
\newblock \bibinfo{journal}{Journal of Aircraft} \bibinfo{volume}{8},
  \bibinfo{pages}{3--19}.
%Type = Inproceedings
\bibitem[{Piquee et~al.(2018)Piquee, L{\'o}pez, Breitsamter, W{\"u}chner and
  Bletzinger}]{piquee2018aerodynamic}
\bibinfo{author}{Piquee, J.}, \bibinfo{author}{L{\'o}pez, I.},
  \bibinfo{author}{Breitsamter, C.}, \bibinfo{author}{W{\"u}chner, R.},
  \bibinfo{author}{Bletzinger, K.U.}, \bibinfo{year}{2018}.
\newblock \bibinfo{title}{Aerodynamic characteristics of an elasto-flexible
  membrane wing based on experimental and numerical investigations}, in:
  \bibinfo{booktitle}{2018 Applied Aerodynamics Conference}, p.
  \bibinfo{pages}{3338}.
%Type = Article
\bibitem[{Rojratsirikul et~al.(2011)Rojratsirikul, Genc, Wang and
  Gursul}]{rojratsirikul2011flow}
\bibinfo{author}{Rojratsirikul, P.}, \bibinfo{author}{Genc, M.S.},
  \bibinfo{author}{Wang, Z.}, \bibinfo{author}{Gursul, I.},
  \bibinfo{year}{2011}.
\newblock \bibinfo{title}{Flow-induced vibrations of low aspect ratio
  rectangular membrane wings}.
\newblock \bibinfo{journal}{J. Fluids and Struct.} \bibinfo{volume}{27},
  \bibinfo{pages}{1296--1309}.
%Type = Article
\bibitem[{Rojratsirikul et~al.(2010)Rojratsirikul, Wang and
  Gursul}]{rojratsirikul2010effect}
\bibinfo{author}{Rojratsirikul, P.}, \bibinfo{author}{Wang, Z.},
  \bibinfo{author}{Gursul, I.}, \bibinfo{year}{2010}.
\newblock \bibinfo{title}{Effect of pre-strain and excess length on unsteady
  fluid--structure interactions of membrane airfoils}.
\newblock \bibinfo{journal}{J. Fluids and Struct.} \bibinfo{volume}{26},
  \bibinfo{pages}{359--376}.
%Type = Inproceedings
\bibitem[{Schomberg et~al.(2018)Schomberg, Gerland, Liese, W{\"u}nsch and
  Ruetten}]{schomberg2018transition}
\bibinfo{author}{Schomberg, T.}, \bibinfo{author}{Gerland, F.},
  \bibinfo{author}{Liese, F.}, \bibinfo{author}{W{\"u}nsch, O.},
  \bibinfo{author}{Ruetten, M.}, \bibinfo{year}{2018}.
\newblock \bibinfo{title}{Transition manipulation by the use of an
  electrorheologically driven membrane}, in: \bibinfo{booktitle}{2018 Flow
  Control Conference}, p. \bibinfo{pages}{3213}.
%Type = Article
\bibitem[{Shelley et~al.(2005)Shelley, Vandenberghe and
  Zhang}]{shelley2005heavy}
\bibinfo{author}{Shelley, M.J.}, \bibinfo{author}{Vandenberghe, N.},
  \bibinfo{author}{Zhang, J.}, \bibinfo{year}{2005}.
\newblock \bibinfo{title}{Heavy flags undergo spontaneous oscillations in
  flowing water}.
\newblock \bibinfo{journal}{Physical Review Letters} \bibinfo{volume}{94},
  \bibinfo{pages}{094302}.
%Type = Article
\bibitem[{Shelley and Zhang(2011)}]{shelley2011flapping}
\bibinfo{author}{Shelley, M.J.}, \bibinfo{author}{Zhang, J.},
  \bibinfo{year}{2011}.
\newblock \bibinfo{title}{Flapping and bending bodies interacting with fluid
  flows}.
\newblock \bibinfo{journal}{Annual Review of Fluid Mechanics}
  \bibinfo{volume}{43}, \bibinfo{pages}{449--465}.
%Type = Article
\bibitem[{Song et~al.(2008)Song, Tian, Israeli, Galvao, Bishop, Swartz and
  Breuer}]{song2008aeromechanics}
\bibinfo{author}{Song, A.}, \bibinfo{author}{Tian, X.},
  \bibinfo{author}{Israeli, E.}, \bibinfo{author}{Galvao, R.},
  \bibinfo{author}{Bishop, K.}, \bibinfo{author}{Swartz, S.},
  \bibinfo{author}{Breuer, K.}, \bibinfo{year}{2008}.
\newblock \bibinfo{title}{Aeromechanics of membrane wings with implications for
  animal flight}.
\newblock \bibinfo{journal}{AIAA journal} \bibinfo{volume}{46},
  \bibinfo{pages}{2096}.
%Type = Article
\bibitem[{Stanford et~al.(2008)Stanford, Ifju, Albertani and
  Shyy}]{stanford2008fixed}
\bibinfo{author}{Stanford, B.}, \bibinfo{author}{Ifju, P.},
  \bibinfo{author}{Albertani, R.}, \bibinfo{author}{Shyy, W.},
  \bibinfo{year}{2008}.
\newblock \bibinfo{title}{Fixed membrane wings for micro air vehicles:
  Experimental characterization, numerical modeling, and tailoring}.
\newblock \bibinfo{journal}{Progress in Aerospace Sciences}
  \bibinfo{volume}{44}, \bibinfo{pages}{258--294}.
%Type = Article
\bibitem[{Stein et~al.(2000)Stein, Benney, Kalro, Tezduyar, Leonard and
  Accorsi}]{stein2000parachute}
\bibinfo{author}{Stein, K.}, \bibinfo{author}{Benney, R.},
  \bibinfo{author}{Kalro, V.}, \bibinfo{author}{Tezduyar, T.E.},
  \bibinfo{author}{Leonard, J.}, \bibinfo{author}{Accorsi, M.},
  \bibinfo{year}{2000}.
\newblock \bibinfo{title}{Parachute fluid--structure interactions: 3-d
  computation}.
\newblock \bibinfo{journal}{Computer Methods in Applied Mechanics and
  Engineering} \bibinfo{volume}{190}, \bibinfo{pages}{373--386}.
%Type = Article
\bibitem[{Sun et~al.(2018)Sun, Wang, Zhang and Ye}]{sun2018bifurcations}
\bibinfo{author}{Sun, X.}, \bibinfo{author}{Wang, S.Z.},
  \bibinfo{author}{Zhang, J.Z.}, \bibinfo{author}{Ye, Z.H.},
  \bibinfo{year}{2018}.
\newblock \bibinfo{title}{Bifurcations of vortex-induced vibrations of a fixed
  membrane wing at {Re} $\leq 1000$}.
\newblock \bibinfo{journal}{Nonlinear Dynamics} \bibinfo{volume}{91},
  \bibinfo{pages}{2097--2112}.
%Type = Article
\bibitem[{Sunny et~al.(2014)Sunny, Sultan and Kapania}]{sunny2014optimal}
\bibinfo{author}{Sunny, M.R.}, \bibinfo{author}{Sultan, C.},
  \bibinfo{author}{Kapania, R.K.}, \bibinfo{year}{2014}.
\newblock \bibinfo{title}{Optimal energy harvesting from a membrane attached to
  a tensegrity structure}.
\newblock \bibinfo{journal}{AIAA journal} \bibinfo{volume}{52},
  \bibinfo{pages}{307--319}.
%Type = Article
\bibitem[{Swartz et~al.(1996)Swartz, Groves, Kim and
  Walsh}]{swartz1996mechanical}
\bibinfo{author}{Swartz, S.M.}, \bibinfo{author}{Groves, M.S.},
  \bibinfo{author}{Kim, H.D.}, \bibinfo{author}{Walsh, W.R.},
  \bibinfo{year}{1996}.
\newblock \bibinfo{title}{Mechanical properties of bat wing membrane skin}.
\newblock \bibinfo{journal}{Journal of Zoology} \bibinfo{volume}{239},
  \bibinfo{pages}{357--378}.
%Type = Article
\bibitem[{Sygulski(1996)}]{sygulski1996dynamic}
\bibinfo{author}{Sygulski, R.}, \bibinfo{year}{1996}.
\newblock \bibinfo{title}{Dynamic stability of pneumatic structures in wind:
  theory and experiment}.
\newblock \bibinfo{journal}{J. Fluids and Struct.} \bibinfo{volume}{10},
  \bibinfo{pages}{945--963}.
%Type = Article
\bibitem[{Sygulski(1997)}]{sygulski1997numerical}
\bibinfo{author}{Sygulski, R.}, \bibinfo{year}{1997}.
\newblock \bibinfo{title}{Numerical analysis of membrane stability in air
  flow}.
\newblock \bibinfo{journal}{Journal of Sound and Vibration}
  \bibinfo{volume}{201}, \bibinfo{pages}{281--292}.
%Type = Article
\bibitem[{Sygulski(2007)}]{sygulski2007stability}
\bibinfo{author}{Sygulski, R.}, \bibinfo{year}{2007}.
\newblock \bibinfo{title}{Stability of membrane in low subsonic flow}.
\newblock \bibinfo{journal}{Inter. J. of Non-Lin. Mech.} \bibinfo{volume}{42},
  \bibinfo{pages}{196--202}.
%Type = Article
\bibitem[{Taneda(1968)}]{Taneda_JPhysSocJpn_1968}
\bibinfo{author}{Taneda, S.}, \bibinfo{year}{1968}.
\newblock \bibinfo{title}{Waving motions of flags}.
\newblock \bibinfo{journal}{J. Phys. Soc. Jpn} \bibinfo{volume}{24},
  \bibinfo{pages}{392--401}.
%Type = Article
\bibitem[{Tiomkin and Raveh(2017)}]{tiomkin2017stability}
\bibinfo{author}{Tiomkin, S.}, \bibinfo{author}{Raveh, D.E.},
  \bibinfo{year}{2017}.
\newblock \bibinfo{title}{On the stability of two-dimensional membrane wings}.
\newblock \bibinfo{journal}{J. Fluids and Struct.} \bibinfo{volume}{71},
  \bibinfo{pages}{143--163}.
%Type = Article
\bibitem[{Tregidgo et~al.(2013)Tregidgo, Wang and
  Gursul}]{tregidgo2013unsteady}
\bibinfo{author}{Tregidgo, L.}, \bibinfo{author}{Wang, Z.},
  \bibinfo{author}{Gursul, I.}, \bibinfo{year}{2013}.
\newblock \bibinfo{title}{Unsteady fluid--structure interactions of a pitching
  membrane wing}.
\newblock \bibinfo{journal}{Aerospace Science and Technology}
  \bibinfo{volume}{28}, \bibinfo{pages}{79--90}.
%Type = Article
\bibitem[{Triantafyllou and Howell(1994)}]{triantafyllou1994dynamic}
\bibinfo{author}{Triantafyllou, M.S.}, \bibinfo{author}{Howell, C.T.},
  \bibinfo{year}{1994}.
\newblock \bibinfo{title}{Dynamic response of cables under negative tension: an
  ill-posed problem}.
\newblock \bibinfo{journal}{Journal of Sound and Vibration}
  \bibinfo{volume}{173}, \bibinfo{pages}{433--447}.
%Type = Article
\bibitem[{Tzezana and Breuer(2019)}]{tzezana2019thrust}
\bibinfo{author}{Tzezana, G.A.}, \bibinfo{author}{Breuer, K.S.},
  \bibinfo{year}{2019}.
\newblock \bibinfo{title}{Thrust, drag and wake structure in flapping compliant
  membrane wings}.
\newblock \bibinfo{journal}{J. Fluid Mech.} \bibinfo{volume}{862},
  \bibinfo{pages}{871--888}.
%Type = Inproceedings
\bibitem[{Waldman and Breuer(2013)}]{waldman2013shape}
\bibinfo{author}{Waldman, R.M.}, \bibinfo{author}{Breuer, K.S.},
  \bibinfo{year}{2013}.
\newblock \bibinfo{title}{Shape, lift, and vibrations of highly compliant
  membrane wings}, in: \bibinfo{booktitle}{43rd AIAA Fluid Dynamics
  Conference}, p. \bibinfo{pages}{3177}.
%Type = Article
\bibitem[{Waldman and Breuer(2017)}]{waldman2017camber}
\bibinfo{author}{Waldman, R.M.}, \bibinfo{author}{Breuer, K.S.},
  \bibinfo{year}{2017}.
\newblock \bibinfo{title}{Camber and aerodynamic performance of compliant
  membrane wings}.
\newblock \bibinfo{journal}{J. Fluids and Struct.} \bibinfo{volume}{68},
  \bibinfo{pages}{390--402}.
%Type = Article
\bibitem[{Watanabe et~al.(2002)Watanabe, Suzuki, Sugihara and
  Sueoka}]{watanabe2002experimental}
\bibinfo{author}{Watanabe, Y.}, \bibinfo{author}{Suzuki, S.},
  \bibinfo{author}{Sugihara, M.}, \bibinfo{author}{Sueoka, Y.},
  \bibinfo{year}{2002}.
\newblock \bibinfo{title}{An experimental study of paper flutter}.
\newblock \bibinfo{journal}{J. Fluids and Struct.} \bibinfo{volume}{16},
  \bibinfo{pages}{529--542}.
%Type = Article
\bibitem[{Yang and Sultan(2016)}]{yang2016modeling}
\bibinfo{author}{Yang, S.}, \bibinfo{author}{Sultan, C.}, \bibinfo{year}{2016}.
\newblock \bibinfo{title}{Modeling of tensegrity-membrane systems}.
\newblock \bibinfo{journal}{International Journal of Solids and Structures}
  \bibinfo{volume}{82}, \bibinfo{pages}{125--143}.
%Type = Article
\bibitem[{Zhang et~al.(2000)Zhang, Childress, Libchaber and
  Shelley}]{zhang2000flexible}
\bibinfo{author}{Zhang, J.}, \bibinfo{author}{Childress, S.},
  \bibinfo{author}{Libchaber, A.}, \bibinfo{author}{Shelley, M.},
  \bibinfo{year}{2000}.
\newblock \bibinfo{title}{Flexible filaments in a flowing soap film as a model
  for one-dimensional flags in a two-dimensional wind}.
\newblock \bibinfo{journal}{Nature} \bibinfo{volume}{408},
  \bibinfo{pages}{835--839}.
%Type = Article
\bibitem[{Zhu and Peskin(2002)}]{zhu2002simulation}
\bibinfo{author}{Zhu, L.}, \bibinfo{author}{Peskin, C.S.},
  \bibinfo{year}{2002}.
\newblock \bibinfo{title}{Simulation of a flapping flexible filament in a
  flowing soap film by the immersed boundary method}.
\newblock \bibinfo{journal}{Journal of Computational Physics}
  \bibinfo{volume}{179}, \bibinfo{pages}{452--468}.

\end{thebibliography}

\end{document}